\documentclass[aps,prd,superscriptaddress,floatfix,nofootinbib,notitlepage,twocolumn]{revtex4-1}

\usepackage{graphicx}
\usepackage{amsmath}
\usepackage{longtable}
\usepackage{aas_macros}
\usepackage{xcolor}
\usepackage{color}
\usepackage{colortbl}

\graphicspath{{Figures/}}

\usepackage[sort&compress]{natbib}
\usepackage{hyperref}
\begin{document}

\title{Astrophysical interpretation of the anisotropies in the unresolved gamma-ray background}

\date{\today}

\author{Shin'ichiro Ando}
\email{s.ando@uva.nl}
\affiliation{GRAPPA, University of Amsterdam, Science Park, 1098 XH Amsterdam, The Netherlands}

\author{Mattia Fornasa}
\email{fornasam@gmail.com}
\affiliation{GRAPPA, University of Amsterdam, Science Park, 1098 XH Amsterdam, The Netherlands}

\author{Nicolao Fornengo}
\affiliation{Dipartimento di Fisica, Universit\'a di Torino, via P. Giuria 1, I-10125 Torino, Italy}
\affiliation{Istituto Nazionale di Fisica Nucleare, Sezione di Torino, via P. Giuria 1, I-10125 Torino, Italy}

\author{Marco Regis}
\affiliation{Dipartimento di Fisica, Universit\'a di Torino, via P. Giuria 1, I-10125 Torino, Italy}
\affiliation{Istituto Nazionale di Fisica Nucleare, Sezione di Torino, via P. Giuria 1, I-10125 Torino, Italy}

\author{Hannes-S. Zechlin}
\affiliation{Dipartimento di Fisica, Universit\'a di Torino, via P. Giuria 1, I-10125 Torino, Italy}
\affiliation{Istituto Nazionale di Fisica Nucleare, Sezione di Torino, via P. Giuria 1, I-10125 Torino, Italy}

\begin{abstract}
Recently, a new measurement of the auto- and cross-correlation angular power
spectrum (APS) of the isotropic gamma-ray background was performed, based on
81 months of data of the {\it Fermi} Large-Area Telescope (LAT). Here, we fit,
for the first time, the new APS data with a model describing the emission of 
unresolved blazars. These sources are expected to dominate the anisotropy 
signal. The model we employ in our analysis reproduces well the blazars 
resolved by {\it Fermi} LAT. When considering the APS obtained by masking the 
sources in the 3FGL catalogue, we find that unresolved blazars under-produce 
the measured APS below $\sim$1 GeV. Contrary to past results, this suggests the
presence of a new contribution to the low-energy APS, with a significance of, 
at least, $5\sigma$. The excess can be ascribed to a new class of faint 
gamma-ray emitters. If we consider the APS obtained by masking the sources in 
the 2FGL catalogue, there is no under-production of the APS below 1 GeV, but 
the new source class is still preferred over the blazars-only scenario (with 
a significance larger than $10\sigma$). The properties of the new source class 
and the level of anisotropies induced in the isotropic gamma-ray background 
are the same, independent of the APS data used. In particular, the new 
gamma-ray emitters must have a soft energy spectrum, with a spectral index 
ranging, approximately, from 2.7 to 3.2. This complicates their 
interpretation in terms of known sources, since, normally, star-forming and 
radio galaxies are observed with a harder spectrum. The new source class 
identified here is also expected to contribute significantly to the intensity 
of the isotropic gamma-ray background.
\end{abstract}

\maketitle

\pretolerance=10000

\section{Introduction}
In more than 8 years of operation, the {\it Fermi} Large Area Telescope (LAT) 
has significantly increased the census of known gamma-ray emitters: the most 
recent source catalogue (i.e., the {\it Fermi} LAT Third Source Catalogue, 
3FGL) contains 3033 objects, detected with a significance greater than 
$4\sigma$ between 100 MeV and 300 GeV~\cite{Acero:2015hja}. Gamma-ray sources 
that are too faint to be resolved individually by {\it Fermi} LAT contribute 
cumulatively to the so-called isotropic gamma-ray background (IGRB). See 
Ref.~\cite{Fornasa:2015qua} for a recent review.

The most recent measurement of the intensity energy spectrum of the IGRB was 
performed by {\it Fermi} LAT and it covers the energy range between 0.1 and 
820 GeV~\cite{Ackermann:2014usa}. By modeling known classes of gamma-ray 
emitters, Ref.~\cite{Ajello:2015mfa} showed that the measured energy spectrum 
of the IGRB can be explained by the concomitant emission of unresolved 
blazars, star-forming and radio galaxies. However, the exact composition of 
the IGRB is still unknown: reconstructing it would provide valuable insight 
on the characteristics of the contributing source classes.

Different populations of gamma-ray emitters are expected to induce different 
levels of anisotropies in the IGRB (see Refs.~\cite{Ando:2006cr,Ando:2006mt,
Ando:2009nk,SiegalGaskins:2010mp,Cuoco:2012yf,Harding:2012gk,Calore:2014oga,
DiMauro:2014wha} among others). Thus, a measurement of the gamma-ray angular 
power spectrum (APS) can constrain the nature of the IGRB in a complementary 
way with respect to the intensity energy spectrum. Other observables that can 
be employed to a similar goal are the 1-point photon count probability 
distribution of the IGRB~\cite{Malyshev:2011zi,Feyereisen:2015cea,
Zechlin:2015wdz,Lisanti:2016jub,Zechlin:2016pme}, the cross correlation of the 
IGRB with catalogues of resolved galaxies~\cite{Xia:2011ax,Xia:2015wka,
Cuoco:2015rfa,Regis:2015zka,Ando:2013xwa,Ando:2014aoa,Shirasaki:2015nqp,
Ando:2016ang} or of galaxy clusters~\cite{Branchini:2016glc}, the 
cross-correlation with the weak gravitational lensing of cosmic shear 
\cite{Camera:2012cj,Camera:2014rja,Shirasaki:2014noa,Shirasaki:2016kol,
Troster:2016sgf} or with the gravitational lensing of the cosmic microwave 
background \cite{Fornengo:2014cya}. 

The first detection of the IGRB anisotropy APS was reported by the {\it Fermi} 
LAT Collaboration in 2012, in the energy range between 1 and 50~GeV 
\cite{Ackermann:2012uf}. The signal was compatible with being due 
{\it entirely} to unresolved blazars \cite{Cuoco:2012yf,Harding:2012gk,
DiMauro:2014wha}.

An updated measurement of the anisotropy APS has been recently released, 
employing 81 months of {\it Fermi} LAT data, binned in 13 energy bins, from 
0.5 to 500 GeV~\cite{Fornasa:2016ohl}. Apart from the auto-correlation APS in 
each energy bin, the new analysis measured, for the first time, the 
cross-correlation APS between different energy bins. 
Ref.~\cite{Fornasa:2016ohl} suggested that the new APS data are due to more t
han one population of sources.

In this paper, we interpret the auto- and cross-correlation APS measured in 
Ref.~\cite{Fornasa:2016ohl} in terms of unresolved blazars. We employ a 
parametric model that was designed to describe the blazars observed by 
{\it Fermi} LAT~\cite{Collaboration:2010gqa,Ajello:2011zi,Ajello:2013lka,
Ajello:2015mfa}. The result of our fit to the new APS measurement will 
determine whether the latter can still be explained in terms of blazars alone 
or if an additional population of gamma-ray emitters is needed. Our analysis 
will also quantify the impact of the new APS measurement in constraining the 
properties of the blazar population, e.g., its redshift evolution and 
distribution in luminosity (especially at low luminosities). In general, this 
is the first work that takes advantage of the new anisotropy measurement to 
constrain the nature of the IGRB, employing physically-motivated models of 
astrophysical emitters.

The paper is organized as follows: in Sec.~\ref{sec:model}, we summarize the
model used to describe the blazar population. We also discuss how to compute 
the observables in which we are interested, i.e. the anisotropy APS of 
unresolved blazars, the source count distribution function of resolved blazars
and the intensity energy spectrum of all blazars (both resolved and 
unresolved). In Sec.~\ref{sec:data}, we present the data employed in the fit 
and we describe our fitting technique. The results are presented in 
Sec.~\ref{sec:results}, while Sec.~\ref{sec:discussion} and 
Sec.~\ref{sec:conclusions} are left for the discussion and the conclusions,
respectively.

\section{Modeling the blazar population}
\label{sec:model}
Refs.~\cite{Ajello:2011zi,Ajello:2013lka,Ajello:2015mfa} model the blazar
population by means of a parametric description of their gamma-ray luminosity 
function, of their redshift evolution and of their energy spectrum. 
Alternatively, in the ``blazar-sequence'' model, the whole spectral energy 
distribution of blazars is parametrized and their gamma-ray luminosity is 
related to the luminosity in X-rays~\cite{Inoue:2008pk,Abazajian:2010pc}. 
Here, we follow the former approach because we are interested in comparing our 
results directly to those of Ref.~\cite{Ajello:2015mfa}, in which the model 
parameters are constrained only by the properties of resolved blazars. Also, 
{\it Fermi} LAT has detected a significant number of sources to allow 
population studies to be performed entirely in gamma rays, without relying on 
a phenomenological correlation with X-rays.

The gamma-ray luminosity function of blazars, $\Phi(L_\gamma,z,\Gamma)$, is
defined as the number of sources per unit of luminosity $L_\gamma$ (defined in 
the rest frame of the source, for energies between 0.1 and 100 GeV), of 
comoving volume $dV$\footnote{We employ cosmological parameters in agreement 
with the observations by Planck \cite{Ade:2013zuv}.} and of spectral photon 
index $\Gamma$. The luminosity function at redshift $z=0$ is modeled as a 
double power law in $L_\gamma$, as follows:
\begin{eqnarray}
\Phi(L_\gamma,z=0,\Gamma) &=& \frac{dN}{dL_\gamma dV d\Gamma} \nonumber \\
&=& \frac{A}{\ln(10) L_\gamma}
\left[ \left( \frac{L_\gamma}{L_0} \right)^{\gamma_1} + 
\left( \frac{L_\gamma}{L_0} \right)^{\gamma_2} \right]^{-1} \nonumber \\
& & {} \times \exp \left[ - \frac{(\Gamma-\mu(L_\gamma))^2}{2\sigma^2} \right],
\label{eqn:luminosity_function}
\end{eqnarray}
where $\gamma_1$ and $\gamma_2$ are the indexes of the power laws and $L_0$ 
controls the transition between the two regimes. The factor $A$ sets the 
overall normalization and the exponential term describes a Gaussian 
distribution for the photon index $\Gamma$, with $\mu(L_\gamma)$ and $\sigma$ 
its mean and rms, respectively. 

In the literature, blazars are often divided 
into two subclasses, flat-spectrum radio quasars (FSRQs) and BL Lacertae 
objects (BL Lacs). The two subclasses populate different regimes of the same 
correlation between luminosity and spectral index: FSRQs are brighter with a 
softer spectrum but, as the luminosity decreases, sources become harder and it 
is more common to find BL Lacs than FSRQs~\cite{Ackermann:2011}. Allowing for 
the mean $\mu(L_\gamma)$ of the spectral-index distribution to depend on 
$L_\gamma$, one can reproduce the $L_\gamma$--$\Gamma$ correlation and, thus, 
describe both FSRQs and BL Lacs with the same model. In particular, we assume 
$\mu(L_\gamma)$ to be parametrized as follows \cite{Ghisellini:2009rb,
Meyer:2012jf,Ajello:2015mfa}:
\begin{equation}
\mu(L_\gamma) = \mu^\ast + \beta  
\left[\log \left(\frac{L_\gamma}{\mathrm{erg~s^{-1}}}\right) - 46\right].
\label{eqn:mu}
\end{equation}

The redshift evolution of the gamma-ray luminosity function is described
by the evolutionary factor $e(L_\gamma,z)$ \cite{Ajello:2011zi,Ajello:2015mfa},
\begin{equation}
\Phi(L_\gamma,z,\Gamma) = \Phi(L_\gamma,z=0,\Gamma) \, e(z,L_\gamma),
\label{eqn:LDDE}
\end{equation}
with 
\begin{eqnarray}
e(z,L_\gamma) &=&
\left[ \left( \frac{1+z}{1+z_c(L_\gamma)} \right)^{-p_1(L_\gamma)} \right. 
\nonumber \\
& & {} \left. + 
\left( \frac{1+z}{1+z_c(L_\gamma)} \right)^{-p_2(L_\gamma)} \right]^{-1}.
\label{eqn:e_factor}
\end{eqnarray}

The parameters in Eq.~(\ref{eqn:e_factor}) depend on the luminosity as follows:
\begin{eqnarray}
z_c(L_\gamma) &=& z_c^\ast 
\left( \frac{L_\gamma}{10^{48} \, \mbox{erg s}^{-1}} \right)^\alpha,
\label{eqn:z_c} \\
p_1(L_\gamma) &=& p_1^0 + \tau 
\left[ \log \left( \frac{L_\gamma}{\mathrm{erg~s^{-1}}} \right) - 44 \right],
\label{eqn:p1} \\
p_2(L_\gamma) &=& p_2^0 + \delta 
\left[ \log \left( \frac{L_\gamma}{\mathrm{erg~s^{-1}}} \right) - 44 \right].
\label{eqn:p2}
\end{eqnarray}

The indexes $p_1(L_\gamma)$ and $p_2(L_\gamma)$ control the redshift evolution 
of blazars, with a positive (negative) index corresponding to a positive 
(negative) evolution, i.e., the gamma-ray luminosity function increasing 
(decreasing) with $z$. The critical redshift $z_c$ determines the epoch of 
transition between the two evolutionary regimes.

Eqs.~(\ref{eqn:LDDE})--(\ref{eqn:p2}) describe the so-called 
luminosity-dependent density evolution (LDDE). Other evolutionary scenarios.
with different $e(z,L_\gamma)$ and modified versions of Eq.~(\ref{eqn:LDDE}) 
have been considered in the literature. Ref.~\cite{Ajello:2015mfa} employs 
them to describe a sample of 403 blazars detected by {\it Fermi} LAT with a 
test statistics larger than 50, at Galactic latitudes $|b|$ larger than 
$15^\circ$~\cite{Abdo:2010ge}. Even if the analysis of 
Ref.~\cite{Ajello:2015mfa} is not able to significantly prefer one 
evolutionary scheme over the other, the LDDE is the one yielding the best 
description of the sample of blazars. Thus, in this work, we restrict our 
analysis to the LDDE scheme.

Blazars are best described by a curved energy spectrum. Thus, we model their 
energy spectrum $dN_\gamma/dE$ as follows:
\begin{equation}
\frac{dN_\gamma}{dE_\gamma} \propto 
\left[ \left( \frac{E_\gamma}{E_b} \right)^{\gamma_a} +
\left( \frac{E_\gamma}{E_b} \right)^{\gamma_b} \right]^{-1} 
\exp[-\tau(E_\gamma,z)]
\label{eqn:ES}
\end{equation}
and we assume that $E_b$ correlates with $\Gamma$ according to 
$\log(E_b/\mbox{GeV}) = 9.25 - 4.11 \Gamma$ \cite{Ajello:2015mfa}. The factor 
$\exp[-\tau(E_\gamma,z)]$ accounts for the absorption of gamma rays due to pair 
conversion with the extragalactic background light. We model it following 
Ref.~\cite{Dominguez:2010bv}.

According to the model defined above, the differential source count 
distribution of blazars $dN/dF$ (i.e. the number of sources per unit solid 
angle and per unit flux, measured in $\mbox{cm}^{2}~\mbox{s}~\mbox{deg}^{-2}$) 
can be written as follows:
\begin{equation}
\frac{dN}{dF} = \int_{0.01}^{5.0} dz \int_1^{3.5} d\Gamma \, 
\Phi[ L_\gamma(F_{\rm E},z,\Gamma),z,\Gamma] \, \frac{dV}{dz} \, 
\frac{dL_\gamma}{dF},
\label{eqn:source_count}
\end{equation}
where $F$ denotes the flux above 100 MeV and $F_{\rm E}$ indicates the 
{\it energy} flux, as opposed to the {\it number} flux $F$. The quantity 
$L_\gamma(F_{\rm E},z,\Gamma)$ is the luminosity associated with a source with 
flux $F_{\rm E}$ at a redshift $z$ and with spectral index $\Gamma$. The bounds 
of the integration in $\Gamma$ in Eq.~(\ref{eqn:source_count}) are chosen to 
properly sample the distribution of $\Gamma$, while those in redshift probe 
the region where the majority of the emission comes from. In particular, we 
assume that there are no blazars below $z=0.01$~\cite{Ajello:2015mfa}.

The cumulative intensity energy spectrum $dI/dE_\gamma$ of {\it all} blazars 
(i.e. resolved and unresolved) can be computed (in units of 
$\mbox{cm}^{-2}~\mbox{s}^{-1}~\mbox{sr}^{-1}~\mbox{GeV}^{-1}$) as follows:
\begin{eqnarray}
\frac{dI}{dE_\gamma} &=& \int_{0.01}^{5.0} dz \frac{dV}{dz} \int_1^{3.5} d\Gamma 
\int_{L_{\rm min}}^{L_{\rm max}} dL_\gamma  \Phi(L_\gamma,z,\Gamma) \nonumber\\
& & {} \times F(L_\gamma,z,\Gamma) \frac{dN_\gamma(\Gamma,z,E_\gamma)}{dE_\gamma}.
\label{eqn:intensity}
\end{eqnarray}
Similarly to Eq.~(\ref{eqn:source_count}), $F(L_\gamma,z,\Gamma)$ is the
flux (between 0.1 and 100 GeV) produced by a source with luminosity
$L_\gamma$, spectral index $\Gamma$ and at a redshift $z$, and 
$dN_\gamma/dE_\gamma$ is the energy spectrum from Eq.~(\ref{eqn:ES}), properly
normalized so that $F \, dN_\gamma/dE$ provides the differential flux of 
the source. The upper bound in the integration in $L_\gamma$ is fixed at 
$10^{52}~\mbox{erg~s}^{-1}$ and the lower one at $10^{43}~\mbox{erg~ s}^{-1}$. 
Their precise value is not particularly important as the integrand in 
Eq.~(\ref{eqn:intensity}) drops quickly at low and high luminosities.

Finally, for the APS $C_{\rm P}^{i,j}$ between energy bins $i$ and $j$ ($i=j$
for the auto-correlation APS and otherwise for the cross-correlation), we 
assume that blazars are point-like and that their APS is dominated by their 
so-called 1-halo term~\cite{Cooray:2002dia,Ando:2006cr}. This is a good 
assumption if the sources producing the anisotropy signal are relatively 
bright and not numerous, which is the case for unresolved 
blazars~\cite{Ando:2006cr,Ando:2006mt}. In that case, the APS is Poissonian, 
i.e., independent of angular multipoles. It can be computed as follows (in 
units of $\mbox{cm}^{-4}~\mbox{s}^{-2}~ \mbox{sr}^{-1}$):
\begin{eqnarray}
C_{\rm P}^{i,j} &=& \int_{0.01}^{5.0} dz\frac{dV}{dz} \int_1^{3.5} d\Gamma 
\int_{L_{\rm min}}^{L_{\rm max}} dL_\gamma \Phi(L_\gamma,z,\Gamma) \nonumber \\ 
& & {} \times F_i(L_\gamma,z,\Gamma) \, F_j(L_\gamma,z,\Gamma) \nonumber \\
& & {} \times \left[ 1-\Omega(F_{\rm E}(L_\gamma,z,\Gamma),\Gamma) \right] .
\label{eqn:CP}
\end{eqnarray}

\begin{table*}
\tiny
\caption{\label{tab:CP_3FGL} Value of $C_{\rm P,cat}^{i,j}$ (in units of $\mbox{cm}^{-4}~\mbox{s}^{-2}~\mbox{sr}^{-1}$) for all the independent combinations of the first 10 energy bins considered in Ref.~\cite{Fornasa:2016ohl} and for the 3FGL catalogue. The last row contains the values of $F_{\rm thr}$ (in units of $10^{-10}~\mbox{cm}^{-2}~\mbox{s}^{-1}$) employed to compute $C_{\rm P,cov=1}^{i,j}$.}
\begin{tabular}{c|c|c|c|c|c|c|c|c|c|c}
\hline
Energy $[ \mbox{GeV} ]$ & 0.50-0.72 & 0.72-1.04 & 1.04-1.38 & 1.38-1.99 & 1.99-3.15 & 3.15-5.00 & 5.00-7.23 & 7.23-10.45 & 10.45-21.83 & 21.83-50.00 \\
\hline
0.50-0.72 & $1.15 \times 10^{-17}$ & & & & & & & & & \\
0.72-1.04 & $6.51 \times 10^{-18}$ & $4.23 \times 10^{-18}$ & & & & & & & \\
1.04-1.38 & $1.42 \times 10^{-18}$ & $8.84 \times 10^{-19}$ & $4.41 \times 10^{-19}$ & & & & & & \\ 
1.38-1.99 & $1.75 \times 10^{-18}$ & $1.10 \times 10^{-18}$ & $3.94 \times 10^{-19}$ & $4.85 \times 10^{-19}$ & & & & & & \\ 
1.99-3.15 & $1.59 \times 10^{-18}$ & $1.02 \times 10^{-18}$ & $3.00 \times 10^{-19}$ & $3.62 \times 10^{-19}$ & $3.91 \times 10^{-19}$ & & & & & \\
3.15-5.00 & $1.08 \times 10^{-18}$ & $6.89 \times 10^{-19}$ & $1.79 \times 10^{-19}$ & $2.11 \times 10^{-19}$ & $2.23 \times 10^{-19}$ & $2.05 \times 10^{-19}$ & & & & \\
5.00-7.23 & $4.48 \times 10^{-19}$ & $2.88 \times 10^{-19}$ & $8.22 \times 10^{-20}$ & $9.67 \times 10^{-20}$ & $1.02 \times 10^{-19}$ & $8.06 \times 10^{-20}$ & $3.96 \times 10^{-20}$ & & & \\
7.23-10.45 & $2.78 \times 10^{-19}$ & $1.78 \times 10^{-19}$ & $5.25 \times 10^{-20}$ & $6.09 \times 10^{-20}$ & $6.38 \times 10^{-20}$ & $5.12 \times 10^{-20}$ & $2.45 \times 10^{-20}$ & $1.76 \times 10^{-20}$ & & \\
10.45-21.83 & $2.54 \times 10^{-19}$ & $1.63 \times 10^{-19}$ & $4.92 \times 10^{-20}$ & $5.71 \times 10^{-20}$ & $5.83 \times 10^{-20}$ & $4.52 \times 10^{-20}$ & $2.19 \times 10^{-20}$ & $1.51 \times 10^{-20}$ & $1.70 \times 10^{-20}$ & \\
21.83-50.00 & $1.06 \times 10^{-19}$ & $6.75 \times 10^{-20}$ & $1.98 \times 10^{-20}$ & $2.26 \times 10^{-20}$ & $2.33 \times 10^{-20}$ & $1.87 \times 10^{-20}$ & $8.59 \times 10^{-21}$ & $5.96 \times 10^{-21}$ & $6.32 \times 10^{-21}$ & $3.51 \times 10^{-21}$ \\
\hline
$F_{\rm thr}$ & 5.0 & 3.0 & 1.0 & 1.0 & 0.9 & 0.7 & 0.3 & 0.2 & 0.2 & 0.1 \\
\hline
\end{tabular}
\end{table*}

\begin{table*}
\tiny
\caption{\label{tab:CP_2FGL} The same as Table~\ref{tab:CP_3FGL} but for the 2FGL catalogue.}
\begin{tabular}{c|c|c|c|c|c|c|c|c|c|c}
\hline
Energy $[ \mbox{GeV} ]$ & 0.50-0.72 & 0.72-1.04 & 1.04-1.38 & 1.38-1.99 & 1.99-3.15 & 3.15-5.00 & 5.00-7.23 & 7.23-10.45 & 10.45-21.83 & 21.83-50.00 \\
\hline
0.50-0.72 & $2.42 \times 10^{-17}$ & & & & & & & & & \\
0.72-1.04 & $1.08 \times 10^{-17}$ & $6.86 \times 10^{-18}$ & & & & & & & & \\
1.04-1.38 & $8.56 \times 10^{-19}$ & $5.64 \times 10^{-19}$ & $2.96 \times 10^{-19}$ & & & & & & & \\
1.38-1.99 & $1.38 \times 10^{-18}$ & $8.72 \times 10^{-19}$ & $2.75 \times 10^{-19}$ & $4.06 \times 10^{-19}$ & & & & & & \\
1.99-3.15 & $1.21 \times 10^{-18}$ & $7.53 \times 10^{-19}$ & $1.99 \times 10^{-19}$ & $2.81 \times 10^{-19}$ & $2.63 \times 10^{-19}$ & & & & & \\
3.15-5.00 & $1.13 \times 10^{-18}$ & $6.79 \times 10^{-19}$ & $1.21 \times 10^{-19}$ & $1.69 \times 10^{-19}$ & $1.51 \times 10^{-19}$ & $1.47 \times 10^{-19}$ & & & & \\
5.00-7.23 & $3.94 \times 10^{-19}$ & $2.32 \times 10^{-19}$ & $4.55 \times 10^{-20}$ & $6.47 \times 10^{-20}$ & $5.93 \times 10^{-20}$ & $4.72 \times 10^{-20}$ & $2.36 \times 10^{-20}$ & & & \\
7.23-10.45 & $5.10 \times 10^{-19}$ & $2.92 \times 10^{-19}$ & $4.40 \times 10^{-20}$ & $5.73 \times 10^{-20}$ & $4.91 \times 10^{-20}$ & $4.60 \times 10^{-20}$ & $1.53 \times 10^{-20}$ & $2.63 \times 10^{-20}$ & & \\
10.45-21.83 & $4.42 \times 10^{-19}$ & $2.50 \times 10^{-19}$ & $3.58 \times 10^{-20}$ & $4.81 \times 10^{-20}$ & $4.13 \times 10^{-20}$ & $4.03 \times 10^{-20}$ & $1.47 \times 10^{-20}$ & $2.23 \times 10^{-20}$ & $2.26 \times 10^{-20}$ & \\
21.83-50.00 & $1.68 \times 10^{-19}$ & $9.27 \times 10^{-20}$ & $1.12 \times 10^{-20}$ & $1.60 \times 10^{-20}$ & $1.35 \times 10^{-20}$ & $1.41 \times 10^{-20}$ & $4.87 \times 10^{-21}$ & $6.82 \times 10^{-21}$ & $7.19 \times 10^{-21}$ & $3.39 \times 10^{-21}$ \\
\hline
$F_{\rm thr}$ & 10.0 & 5.0 & 1.3 & 1.3 & 1.0 & 0.7 & 0.3 & 0.3 & 0.3 & 0.13 \\
\hline
\end{tabular}
\end{table*}

The quantity $\Omega(F,\Gamma)$ is the so-called ``sky coverage'' and it 
describes the probability of {\it Fermi} LAT to resolve a source characterized 
by $(F,\Gamma)$. It accounts for the fact that the telescope has a lower 
(i.e. better) sensitivity for harder sources. Note that the number fluxes $F_i$ 
and $F_j$ are integrated inside energy bins $i$ and $j$, respectively, while 
the energy flux $F_E$ in the sky coverage is integrated between 0.1 and 100 
GeV. However, we ignore the exact behaviour of $\Omega$ and the estimated 
$C_{\rm P}^{i,j}$ is very sensitive to how one models the transition between 
resolved and unresolved sources. Therefore, we decide to follow the procedure 
adopted in Ref.~\cite{Zechlin:2016pme} and we compute $C_{\rm P}^{i,j}$ as the 
difference of two terms, i.e. $C_{\rm P}^{i,j}=C_{\rm P,cov=1}^{i,j}-C_{\rm P,cat}^{i,j}$. 
Here, $C_{\rm P,cov=1}^{i,j}$ is the APS produced by all sources up to a certain 
threshold $F_{\rm thr}$ large enough that $\Omega(F_{\rm thr},\Gamma)=1$, for all 
values of $\Gamma$. This threshold depends on the energy bin considered and 
$C_{\rm P,cov=1}^{i,j}$ is computed as in Eq.~(\ref{eqn:CP}) but replacing the 
factor $(1-\Omega)$ with $\Theta(F^i_{\rm thr}-F^i) \Theta(F^j_{\rm thr}-F^j)$, 
where $\Theta(x)$ is the Heaviside step function. The second term 
$C_{\rm P,cat}^{i,j}$ is the APS of the sources resolved in the catalogue. It is 
computed directly from the catalogue for all sources fainter than $F_{\rm thr}$
and located at $|b|>30^\circ$. The fluxes $F^i$ and $F^j$ employed to compute 
$C_{\rm P,cat}^{i,j}$ are obtained by integrating the best-fit spectral model of 
each source, as provided in the catalogue. For the {\it Fermi} LAT Second 
Source Catalogue (2FGL) \cite{Fermi-LAT:2011yjw} and 3FGL catalogues, it is 
known that those fits are not reliable for energies larger than few tens of 
GeV and, therefore, we only consider the APS $C_{\rm P}^{i,j}$ for energies below 
50~GeV. By explicitly subtracting the contribution of resolved blazars, our 
method estimates the APS of the unresolved ones without having to assume 
a specific shape for the sky coverage. Ref.~\cite{Zechlin:2016pme} tested that 
this procedure does not depend on the value chosen for $F_{\rm thr}$, as long as 
it lies in a region with coverage equal to 1. 

Table~\ref{tab:CP_3FGL} shows the values of $C_{\rm P,cat}^{i,j}$, for all the 
independent combinations of the first 10 energy bins considered in 
Ref.~\cite{Fornasa:2016ohl} and in the case of the 3FGL catalogue. The last 
row shows the values of $F_{\rm thr}$ employed for the different energy bins. 
Table~\ref{tab:CP_2FGL} contains the same information but for the 2FGL 
catalogue. 

To conclude, we note that, contrary to $C_{\rm P}^{i,j}$, the way we compute the
source count distribution in Eq.~(\ref{eqn:source_count}) and the intensity
energy spectrum in Eq.~(\ref{eqn:intensity}) is independent of the telescope 
sensitivity and it is not associated with a specific analysis or source 
catalogue.

\section{Data and fitting technique}
\label{sec:data}
The auto- and cross-correlation APS are taken from Ref.~\cite{Fornasa:2016ohl}. 
The APS is computed from flux sky maps obtained after subtracting a model for 
the Galactic foreground, i.e., the emission induced by the interaction of 
cosmic rays with the interstellar medium and radiation fields. Also, the 
regions of the sky where resolved sources and the Galactic foreground are 
largely dominated are masked. Two masks are considered separately in 
Ref.~\cite{Fornasa:2016ohl}: they both mask the Galactic plane (i.e., the 
region with $|b|<30^\circ$) and few extended sources. Then, the so-called 
``3FGL mask'' excludes the region around each of the sources in the 3FGL 
catalogue, while the ``2FGL mask'' cuts all objects in the 2FGL catalogue. 
Inside the multipole range considered in Ref.~\cite{Fornasa:2016ohl} (i.e., 
between $\ell=49$ and 706), both the auto- and cross-correlation APS are found 
compatible with being Poissonian. We only consider the best-fit APS 
$C_{\rm P}^{i,j}$ for the first 10 energy bins, i.e. below 50~GeV. The 55 
independent $C_{\rm P}^{i,j}$ are taken from Tables~I and II of 
Ref.~\cite{Fornasa:2016ohl}.

The data on the source count distribution, in the case of the {\it Fermi} LAT
First Source catalog (1FGL), are taken from Fig.~14 of 
Ref.~\cite{Collaboration:2010gqa} (data set labelled ``all blazars'') and 
they refer to emitters associated with blazars and detected in the energy 
range between 100~MeV and 100~GeV. Note that these data have already been 
corrected for the sky coverage, so that we can directly compare them with the 
model prediction computed in Eq.~(\ref{eqn:source_count}). For the 2FGL and 
3FGL catalogue, the $dN/dF$ is computed directly from the list of sources as 
discussed in Appendix B of Ref.~\cite{Zechlin:2015wdz}\footnote{Here, we 
consider sources detected with a significance larger than $6\sigma$, at 
Galactic latitudes $|b|>30^\circ$ and associated with blazars, i.e. classes 
\texttt{bzb}, \texttt{bzq} and \texttt{agu} for the 2FGL catalogue and classes 
\texttt{bll}, \texttt{fsrq} and \texttt{bcu} for the 3FGL one.}. We restrict 
ourselves to a regime in flux in which the catalogues are complete, i.e. above 
$1.98 \times 10^{-8} ~\mbox{cm}^{-2} ~\mbox{s}^{-1}$ for 2FGL and above 
$1.34 \times 10^{-8} ~\mbox{cm}^{-2} ~\mbox{s}^{-1}$ for 3FGL.\footnote{Both 
fluxes are computed above 100 MeV.}

Finally, we will also consider the intensity energy spectrum of the so-called 
extragalactic gamma-ray background (EGB), i.e. the residual emission observed 
by {\it Fermi} LAT after subtracting a model of the Galactic 
foreground~\cite{Ackermann:2014usa}. The EGB is interpreted as the emission of 
all sources (both resolved and unresolved), as opposed to the IGRB that only 
includes unresolved ones. The measurement in Ref.~\cite{Ackermann:2014usa} is 
based on 50 months of data and it covers the energy range between 100~MeV and 
820~GeV. We consider the entries from Table~3 of Ref.~\cite{Ackermann:2014usa}, 
in the case of its ``model A'' for the Galactic foreground. Note that, in the 
following, we will not include the data of the EGB intensity energy spectrum 
in our fits. However, we will compare them to our prediction for the total 
emission associated with blazars.

The fits are performed by scanning over the parameters defining the blazar
population and computing the likelihood function. Scans are performed with 
PyMultiNest\footnote{http://johannesbuchner.github.io/PyMultiNest/index.html}
\cite{Buchner:2014nha}, based on MultiNest 
v3.10\footnote{https://ccpforge.cse.rl.ac.uk/gf/project/multinest/} 
\cite{Feroz:2007kg,Feroz:2008xx,Feroz:2013hea}. The tolerance is fixed at
0.5 with 5000 live points. This guarantees a good sampling of the likelihood
so that results of the scans can be interpreted both in a Bayesian and 
frequentist framework.

We assume that all the data are independent\footnote{As discussed in 
Ref.~\cite{Fornasa:2016ohl}, the covariances between two measured 
$C_{\rm P}^{i,j}$ are negligible. Also, we decide to work with the differential
$dN/dF$, instead of the cumulative one $N(>F)$, so that we can neglect the 
covariance between the source count distribution in different flux bins.} and 
they arise from a Gaussian probability distribution. Thus, the logarithm of 
the likelihood function $\ln \mathcal{L}(\mathbf{\Theta})$ is proportional to 
$-\sum_i [\bar{D}_i - D_i(\mathbf{\Theta})]^2 / 2 \bar{\sigma}_i^2$, where 
the sum runs over all the data considered. $\bar{D}_i$ is the measured value
for data point $i$ and $\bar{\sigma}_i$ its estimated error, while 
$D_i(\mathbf{\Theta})$ is the value of the same observable estimated for point 
$\mathbf{\Theta}$ in the parameter space. Different scans will feature 
different parameter spaces and different data sets. 

We start by considering a parameter space that is 4-dimensional, comprised of 
parameters $A$, $\gamma_1$, $L_0$, and $p_2^0$ defined in 
Eqs.~(\ref{eqn:luminosity_function}) and (\ref{eqn:p2}). The choice is made in 
order to guarantee a significant freedom and variability for the theoretical 
predictions, without having to deal with too many free parameters. In 
particular, $\gamma_1$ controls the gamma-ray luminosity function in the 
low-luminosity regime, while $A$ and $L_0$ its overall normalization. The 
parameter $p_2^0$ determines the evolution of blazars with redshift. All the 
other parameters in Eqs.~(\ref{eqn:luminosity_function}), (\ref{eqn:mu}), 
(\ref{eqn:z_c})-(\ref{eqn:ES}) are fixed to their median values from 
Ref.~\cite{Ajello:2015mfa} (in the case of the LDDE evolutionary scheme). In 
particular, $\gamma_2=1.83$, $\mu^\ast=2.22$, $\beta=0.10$, $\sigma=0.28$, 
$z_c^\ast=1.25$, $\alpha=7.23$, $p_1^0=3.39$, $\tau=3.16$, $\delta=0.64$, 
$\gamma_a=1.7$ and $\gamma_b=2.6$. We implement log-priors on $A$ and $L_0$ 
(between $10^{-6}$ and $10^{2}~\mbox{Gpc}^{-3}$ and between $10^{45}$ and 
$10^{53}~\mbox{erg~ s}^{-1}$, respectively) and linear priors on $\gamma_1$ and 
$p_1^0$ (between 0 and 5, and between $-20$ and 0, respectively). Different 
parameter spaces and prior distributions will be defined in the following 
sections.

For all the parameters $\Theta_i$ we compute the marginalized probability 
distribution function and the Profile Likelihood (PL). The former is obtained 
by integrating the posterior probability distribution over all parameters 
except $\Theta_i$, while the PL is computed by maximizing over
them~\cite{Trotta:2008bp,Feroz:2011bj}. We also derive two-dimensional 
marginalized posterior distribution functions and PL for certain combinations 
of parameters. In all the cases considered, the frequentist (i.e., 
maximization of the likelihood) and the Bayesian (i.e., marginalization of the
posterior distribution) approach yield similiar results. Thus, in the 
following sections, we show only the frequentist case.

\section{Results}
\label{sec:results}

\begin{figure*}
\includegraphics[width=0.49\textwidth]{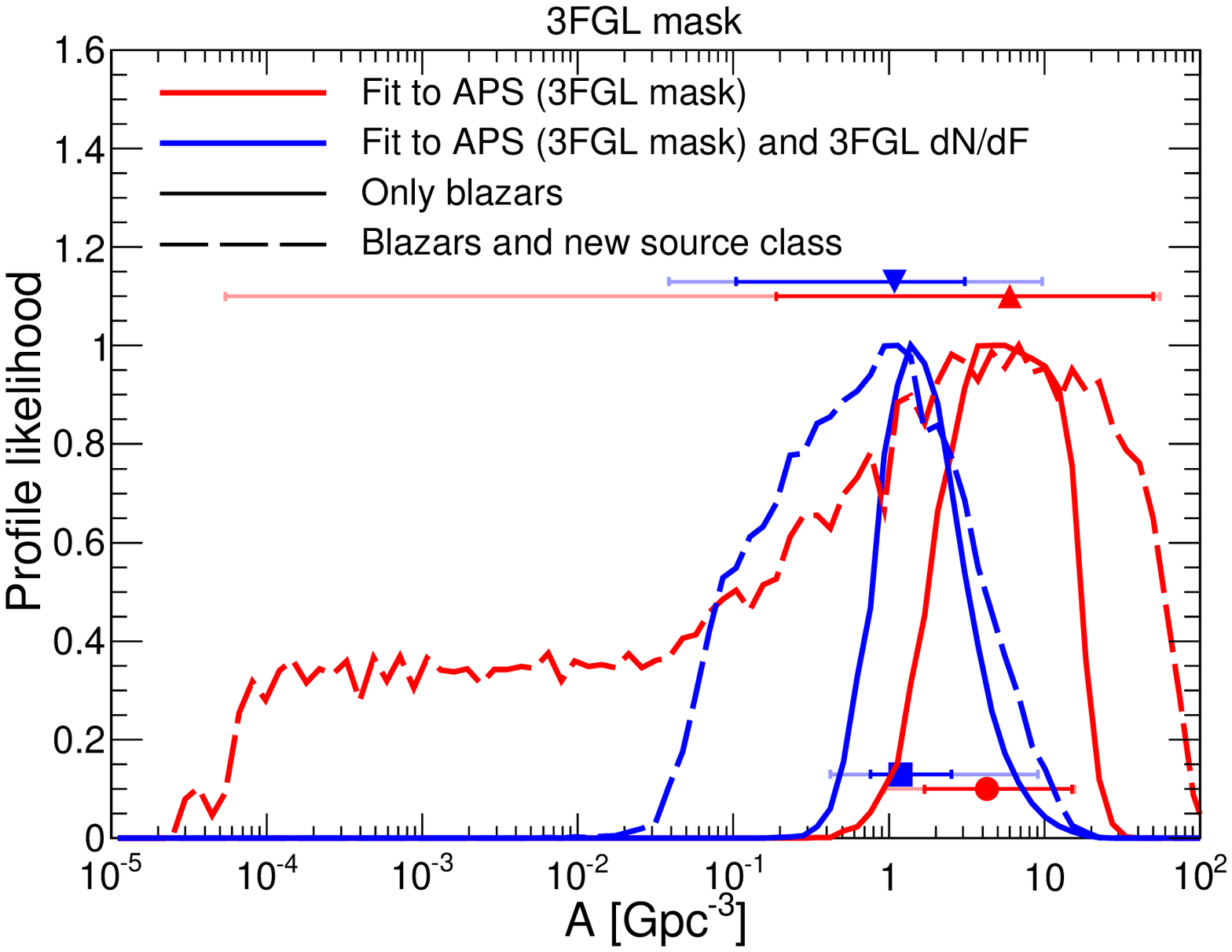}
\includegraphics[width=0.49\textwidth]{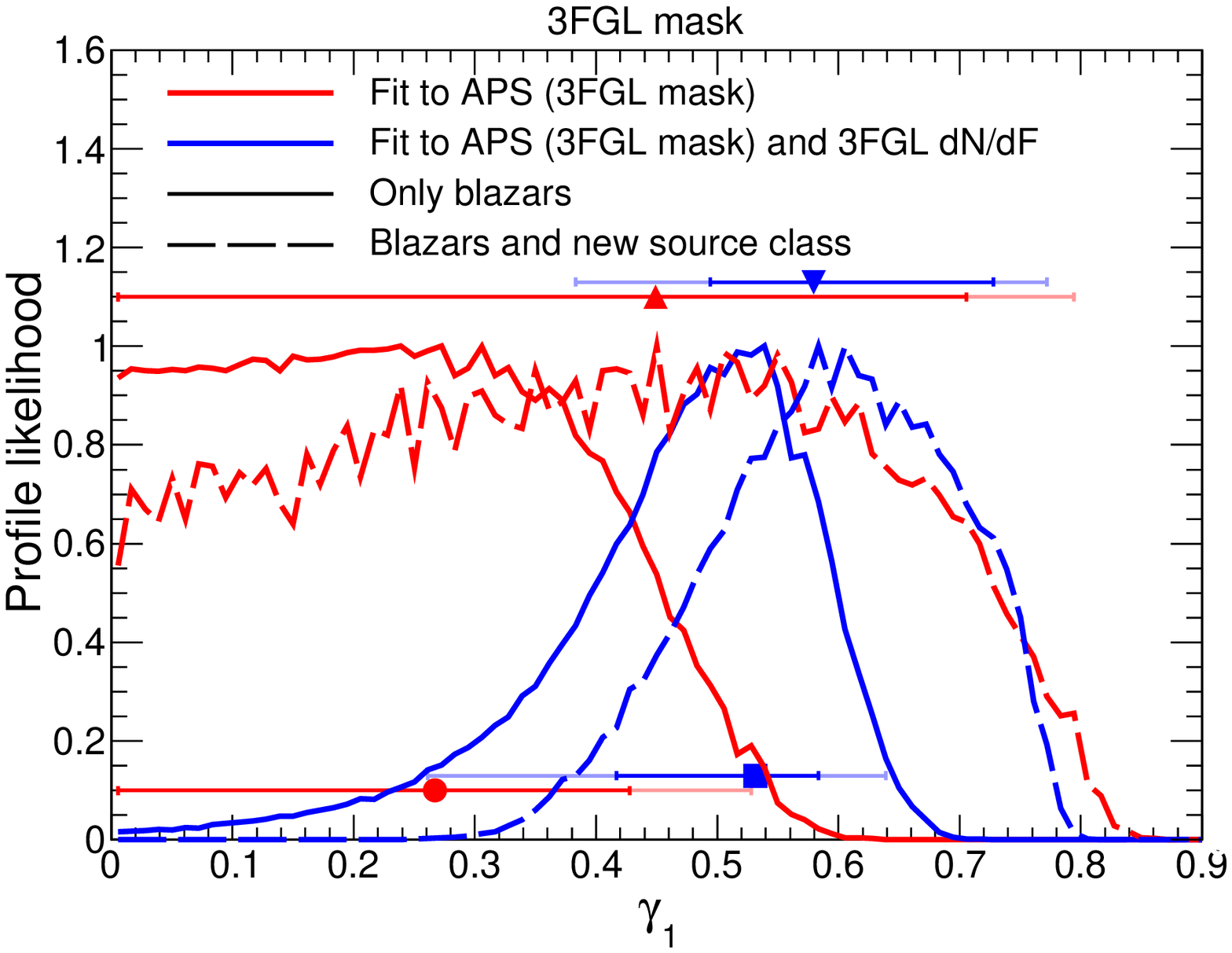}
\includegraphics[width=0.49\textwidth]{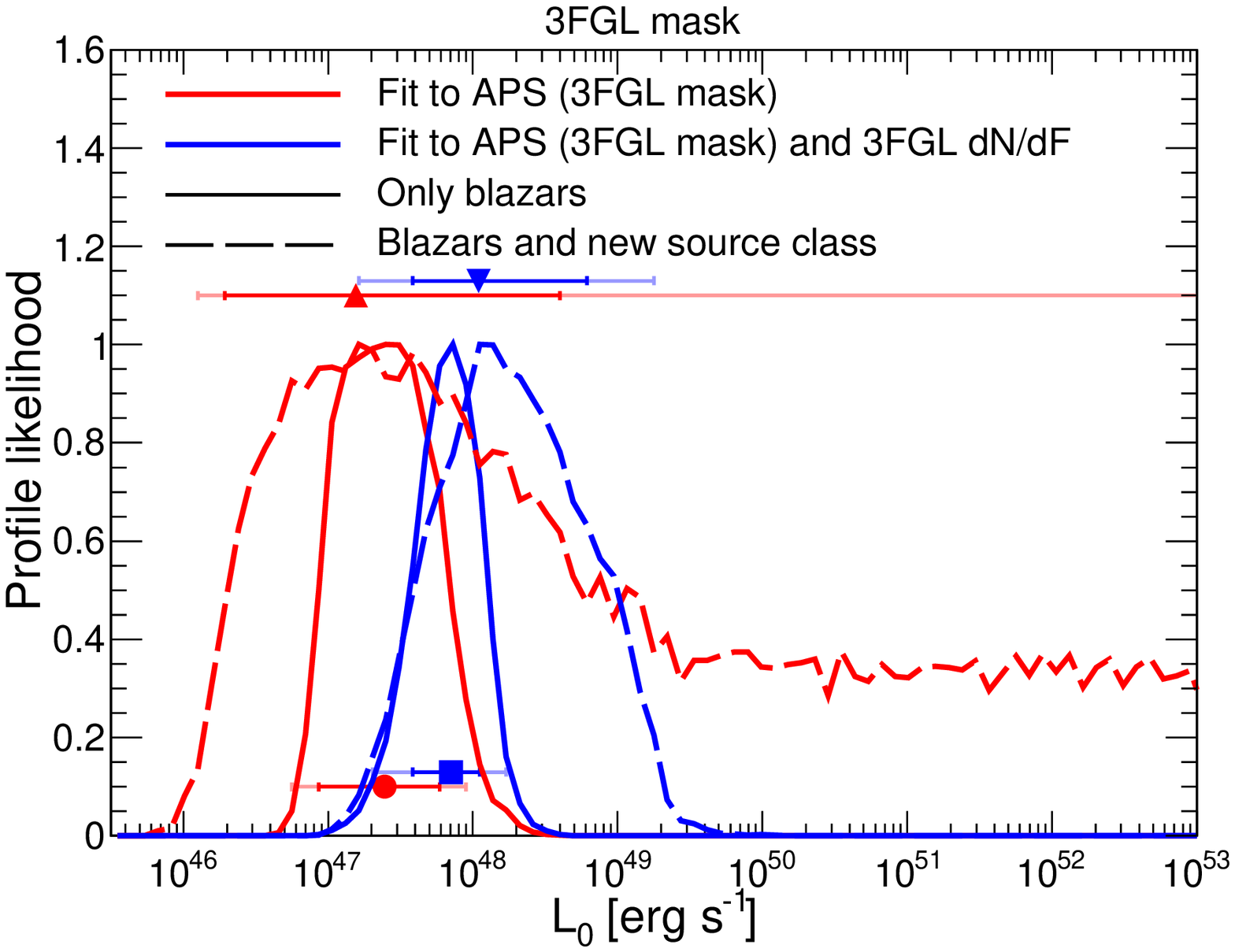}
\includegraphics[width=0.49\textwidth]{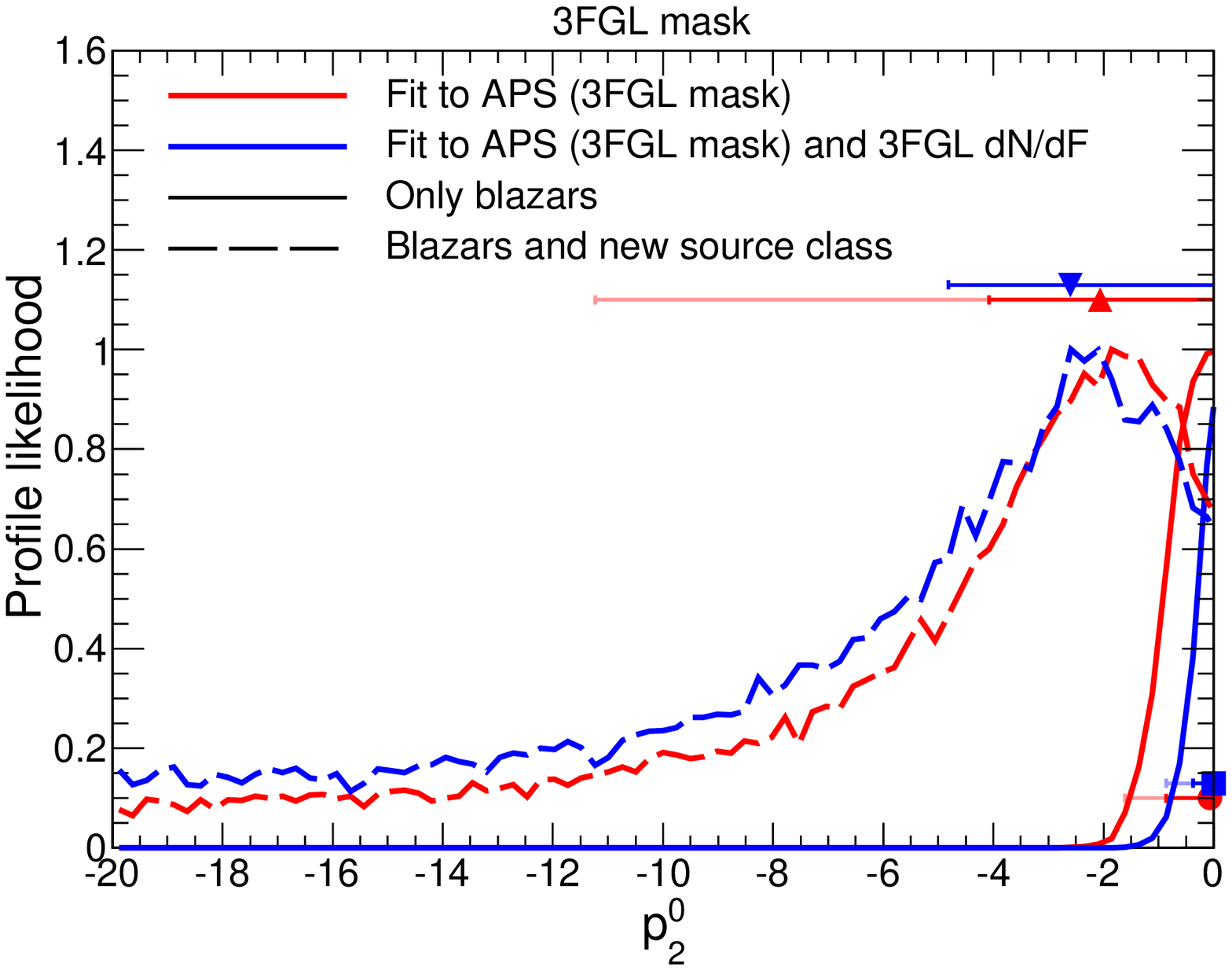}
\caption{\label{fig:1D_3FGL} One-dimensional profile-likelihood distribution for parameters $A$, $\gamma_1$, $L_0$ and $p_2^0$ (see text for details). The red lines refer to the scans performed by fitting only the auto- and cross-correlation APS from Ref.~\cite{Fornasa:2016ohl} in the case of the 3FGL mask, while the blue lines are for the fits to the APS and the source count distribution $dN/dF$ from the 3FGL catalogue. The solid lines refer to fits in which model predictions are computed only in terms of blazars (described by the LDDE scheme from Sec.~\ref{sec:model}), while for the dashed lines we include an additional population of sources (see text). Squares, circles and triangles indicate the best-fit solutions. The sets of lines near the bottom of the panels are for the scans performed only with blazars and the ones near the top (below the legend) are for the scans including the additional source class. The horizontal lines indicate the 68\% (red or dark blue) and 95\% CL region (pink or light blue).}
\end{figure*}

\begin{figure*}
\includegraphics[width=0.32\textwidth]{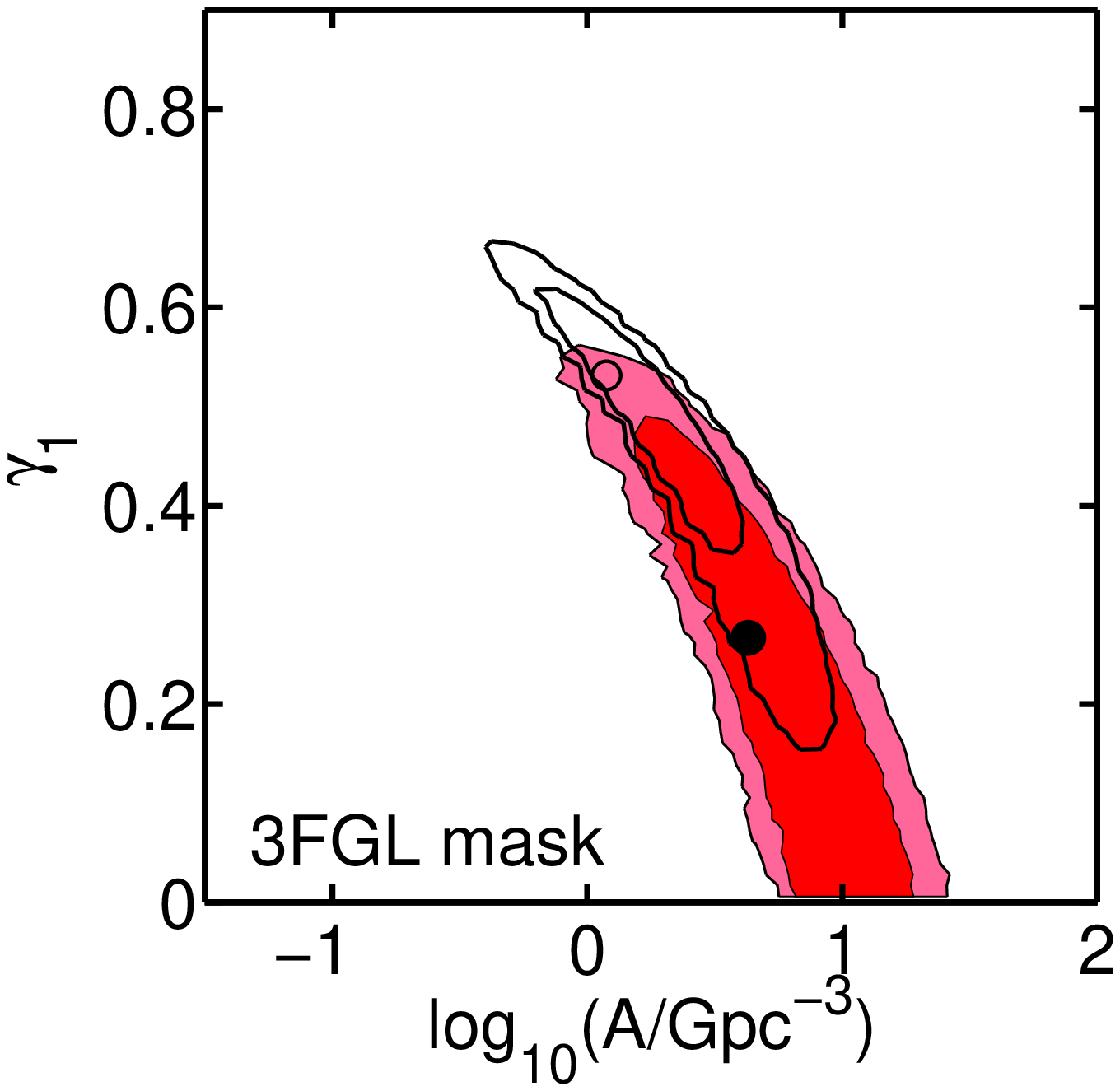}
\includegraphics[width=0.32\textwidth]{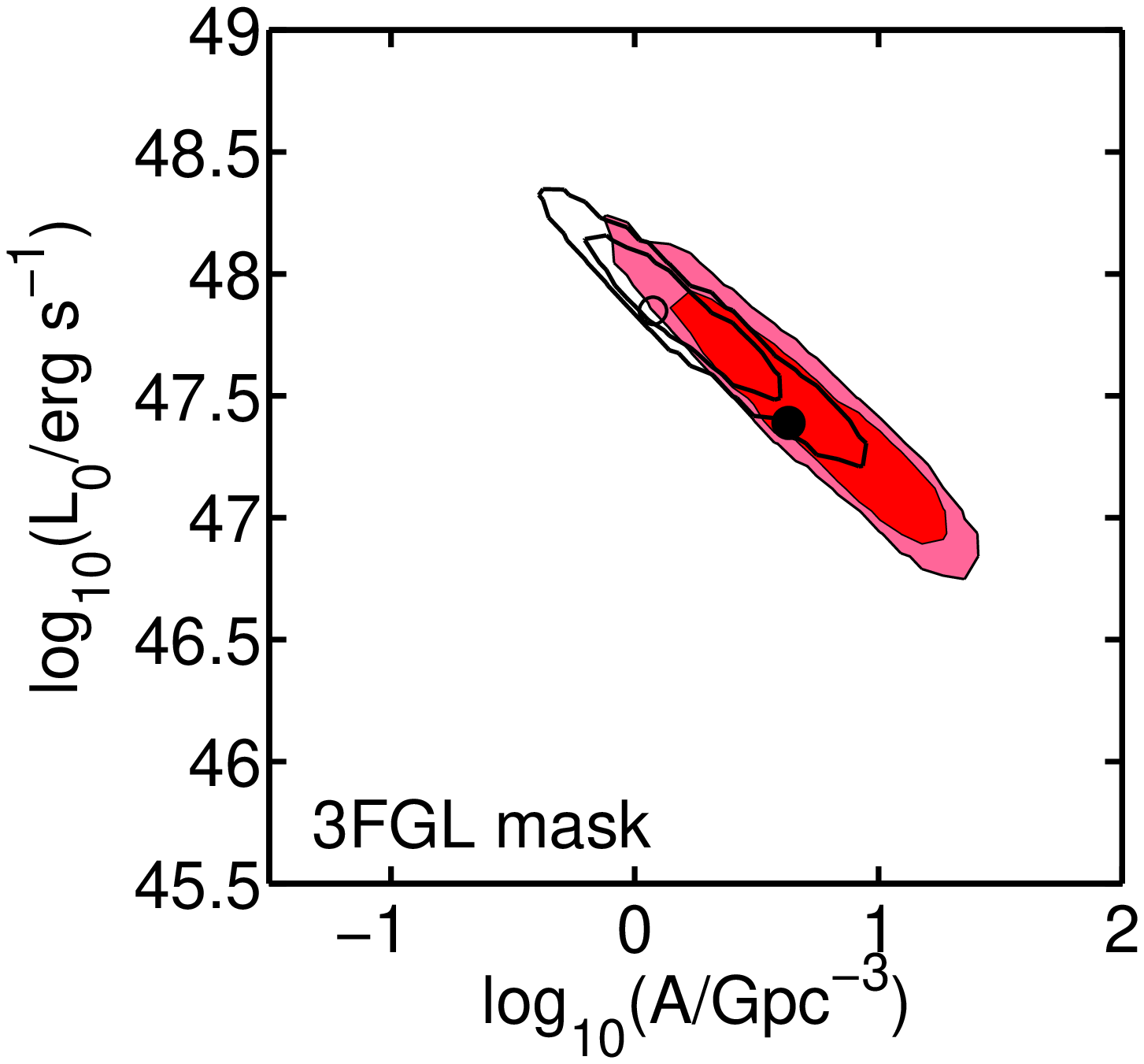}
\includegraphics[width=0.32\textwidth]{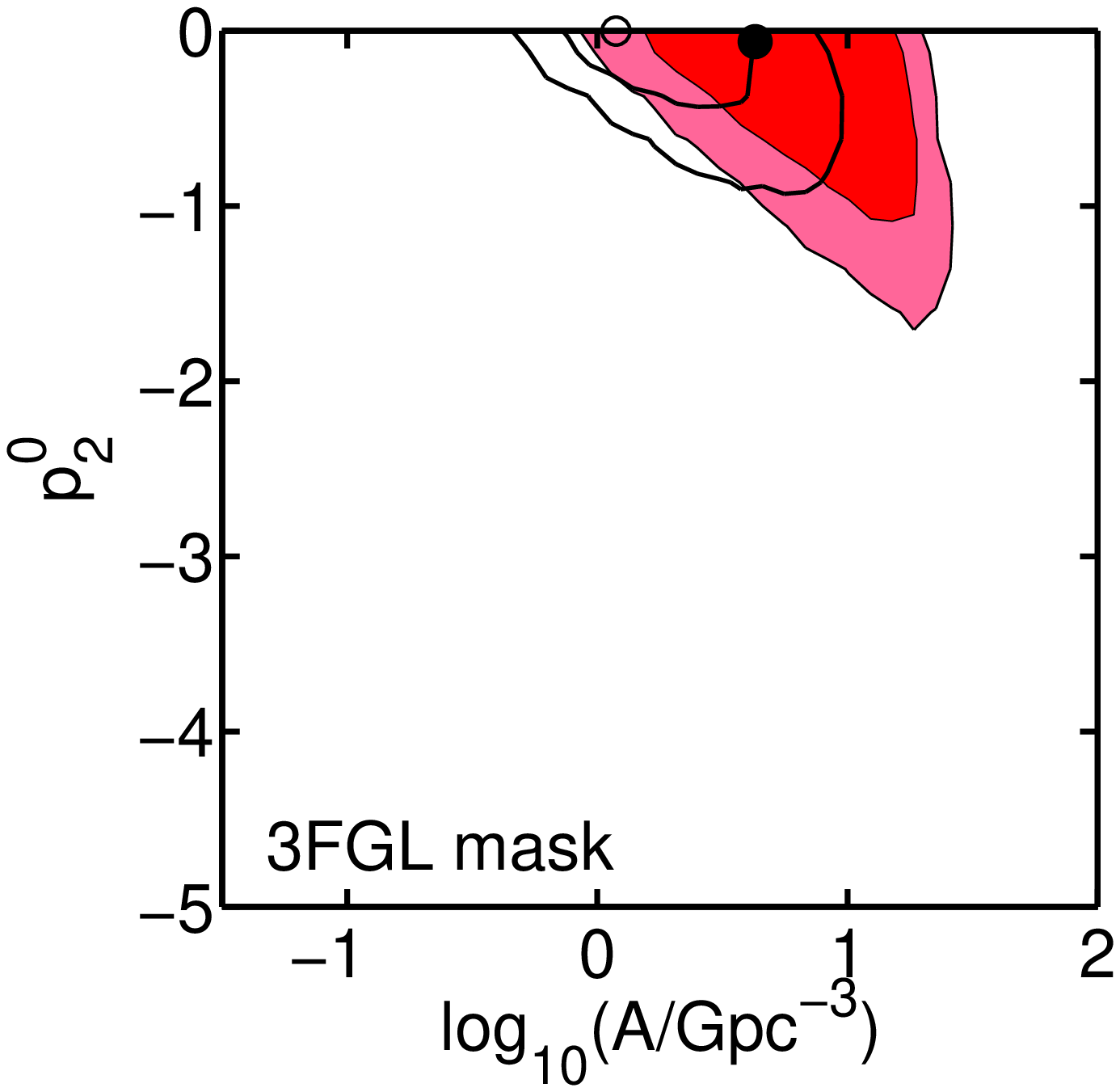}
\includegraphics[width=0.32\textwidth]{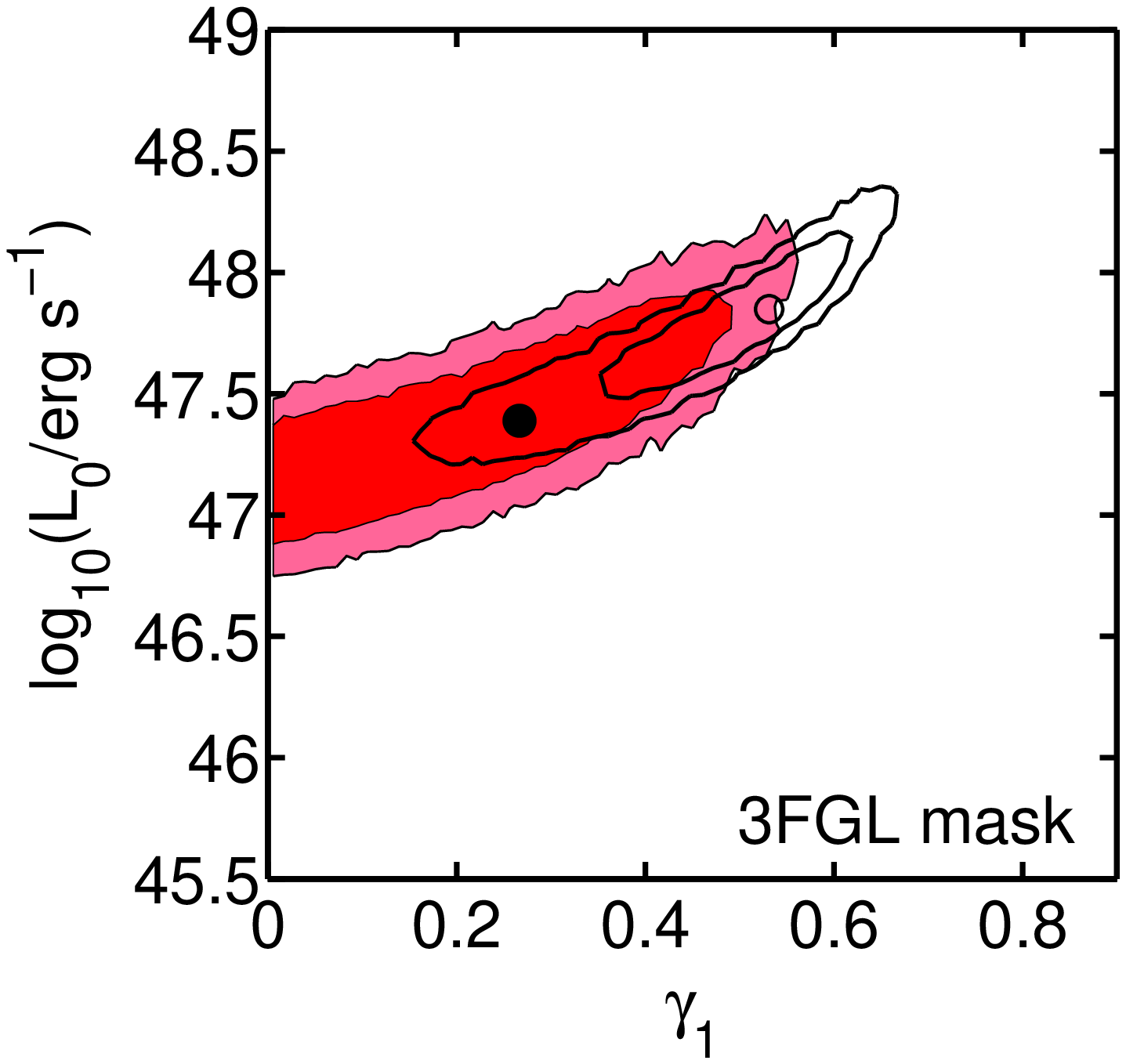}
\includegraphics[width=0.32\textwidth]{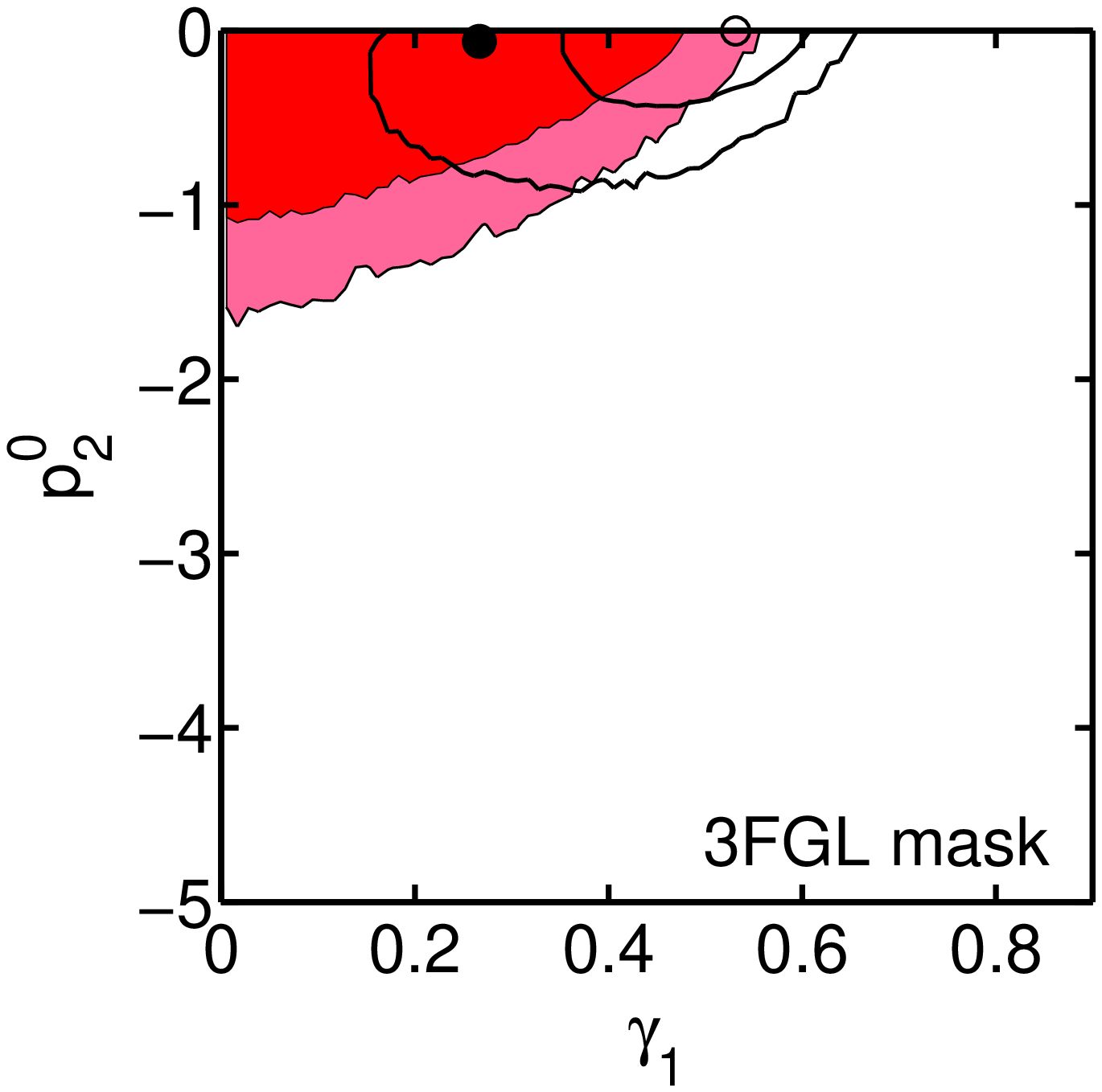}
\includegraphics[width=0.32\textwidth]{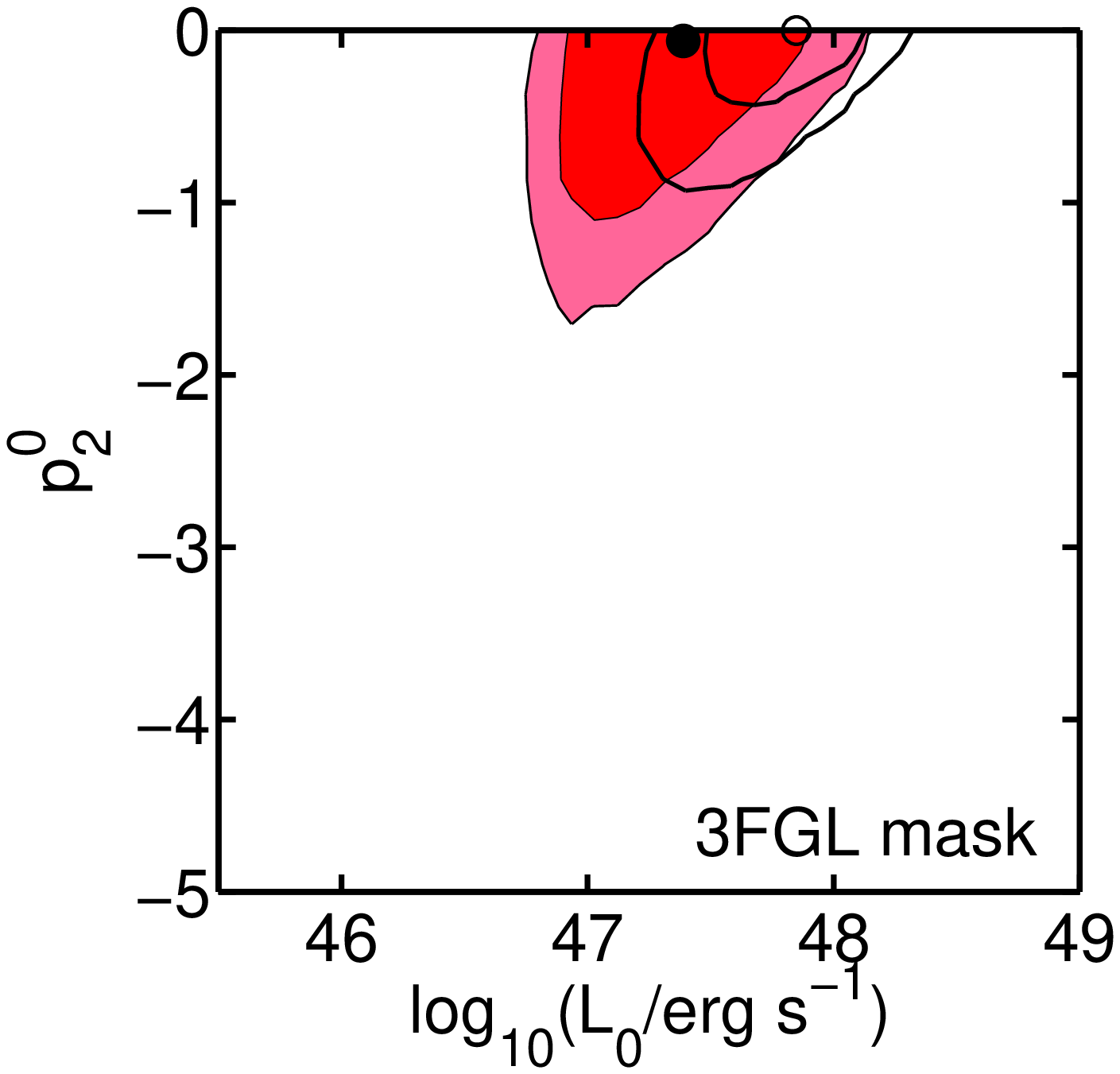}
\caption{\label{fig:2D_3FGL_onlyblazars} Two-dimensional profile-likelihood contour plots for all the combinations of parameters $A$, $\gamma_1$, $L_0$ and $p_2^0$. Filled contours and filled black dots refer to the scans performed by fitting only the auto- and cross-correlation APS from Ref.~\cite{Fornasa:2016ohl} in the case of the 3FGL mask, while empty contours and empty circles are for the fits to the APS and the source count distribution $dN/dF$ from the 3FGL catalogue. Inner contours mark the 68\% CL region and outer ones the 95\% CL region. The model predictions in the scans are computed only in terms of blazars.}
\end{figure*}

\begin{figure*}
\includegraphics[width=0.49\textwidth]{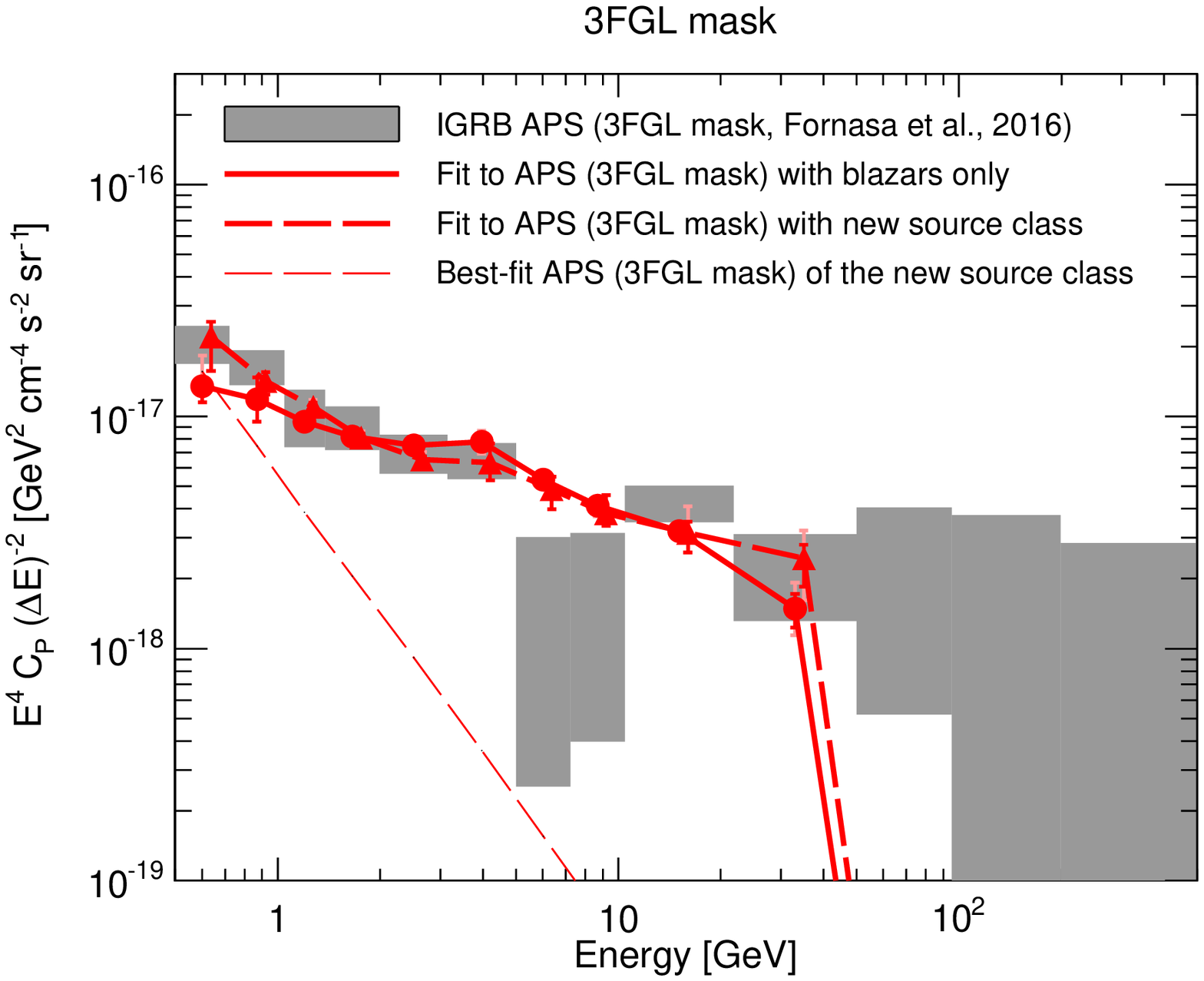}
\includegraphics[width=0.49\textwidth]{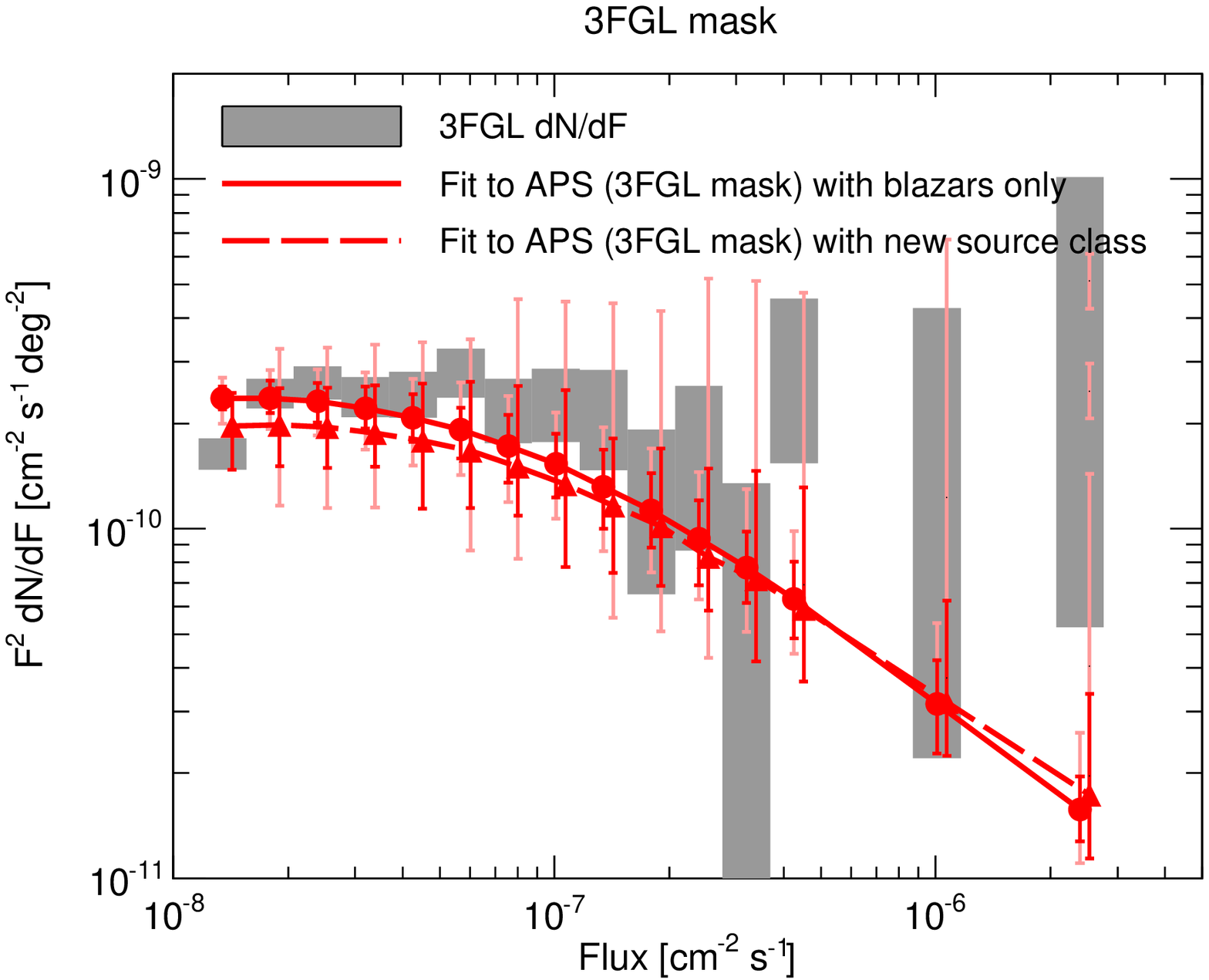}
\caption{\label{fig:bestfit_APS_dNdF_3FGL_onlyAPS} The gray boxes indicate the data points used in the scans, for the auto-correlation APS with the 3FGL mask (left panel) and for the 3FGL $dN/dF$ (right panel). The solid red lines and red circles indicate the best-fit solution for the scan performed by fitting only the auto- and cross-correlation APS from Ref.~\cite{Fornasa:2016ohl} (3FGL mask) with a model including only blazars. The thicker dashed red lines and the red triangles denote the best-fit solution for a scan performed fitting the same APS data but including a new source class (see text for details). The thinner red dashed line in the left panel shows the best-fit auto-correlation APS for the new class source, separately. Around each red circle/triangle, the red (pink) vertical line shows the 68\% (95\%) CL uncertainty. Circles and triangles are slightly shifted with respect to each other to increase readibility.}
\end{figure*}

\subsection{Fitting only the auto- and cross-correlation angular power spectrum (3FGL mask)}
\label{sec:APS_only_blazars_only}
We start by performing a scan over $A$, $\gamma_1$, $L_0$ and $p_2^0$, and by 
computing the likelihood only in terms of the auto- and cross-correlation APS 
from Ref.~\cite{Fornasa:2016ohl} for the 3FGL mask. The solid red lines in 
Fig.~\ref{fig:1D_3FGL} shows the derived one-dimensional PL for the four model 
parameters. The red circles near the bottom of the panels indicate the 
best-fit points and the corresponding red (pink) horinzontal lines the 68\% 
(95\%) confidence-level (CL) region. On the other hand, the red regions in 
Fig.~\ref{fig:2D_3FGL_onlyblazars} show the two-dimensional PL for different 
combinations of the model parameters. The inner contours denote the 68\% CL 
region and the outer ones the 95\% CL. The black circles are the best fits. It 
appears that the reconstruction of the model parameters is still affected by 
significant degeneracies. For example, the upper middle panel of 
Fig.~\ref{fig:2D_3FGL_onlyblazars} shows a degeneracy between $A$ and $L_0$ 
since, for a fixed value of $\gamma_1$, they both control the normalization of 
the gamma-ray luminosity function. Also, from the upper left panel, increasing 
(decreasing) $\gamma_1$ corresponds to a lower (larger) $A$, since making 
$dN/dF$ steeper (shallower) increases (decreases) the total number of 
low-luminosity sources so that a large (smaller) normalization is needed to 
reproduce the measured APS. Another way of increasing the abundance of 
low-flux sources is to increase $p_2^0$ and, thus, shifting blazars to larger 
redshifts. Therefore, iso-PL lines in the lower middle panel are diagonal,
from the lower left corner to the upper right corner of the panel.

\begin{figure*}
\includegraphics[width=0.49\textwidth]{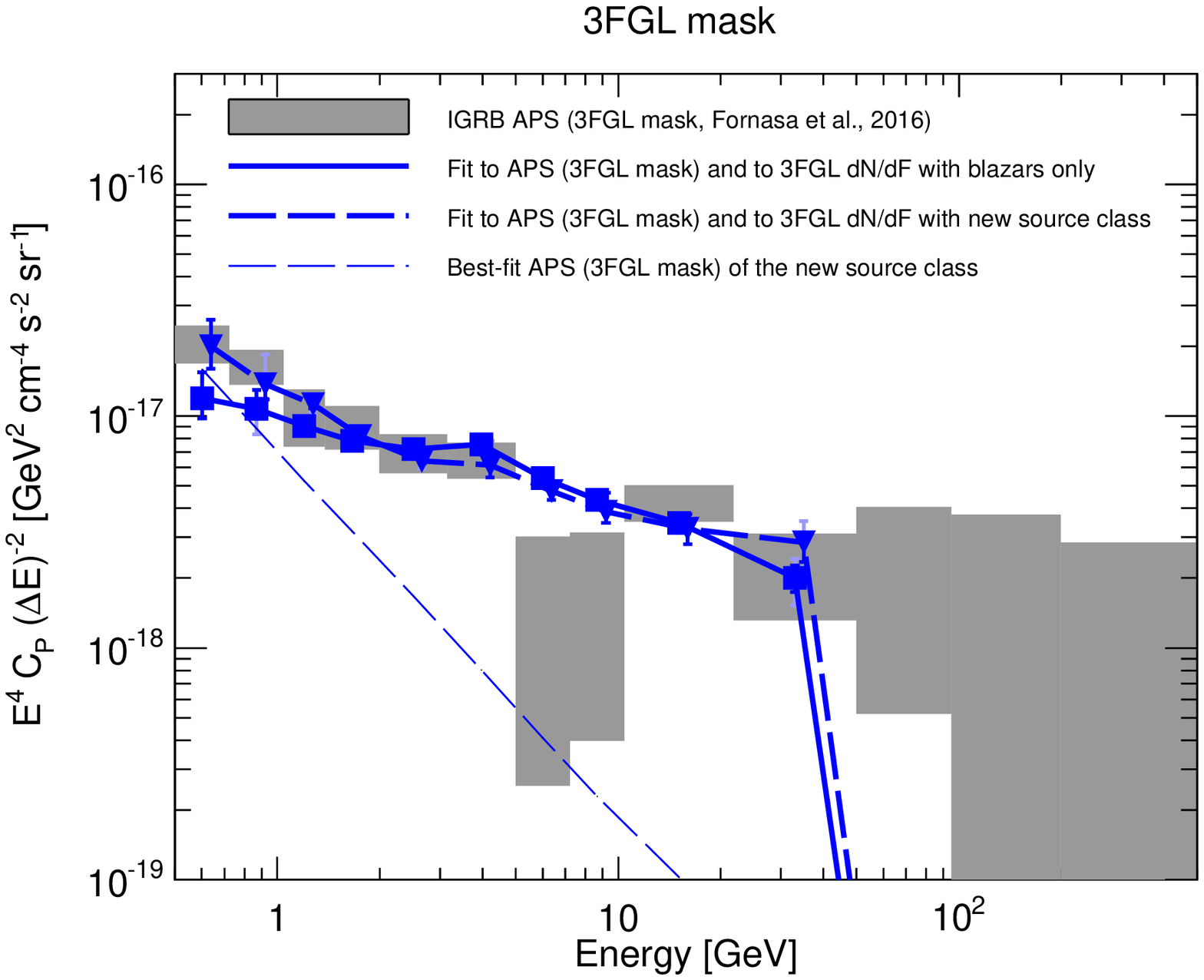}
\includegraphics[width=0.49\textwidth]{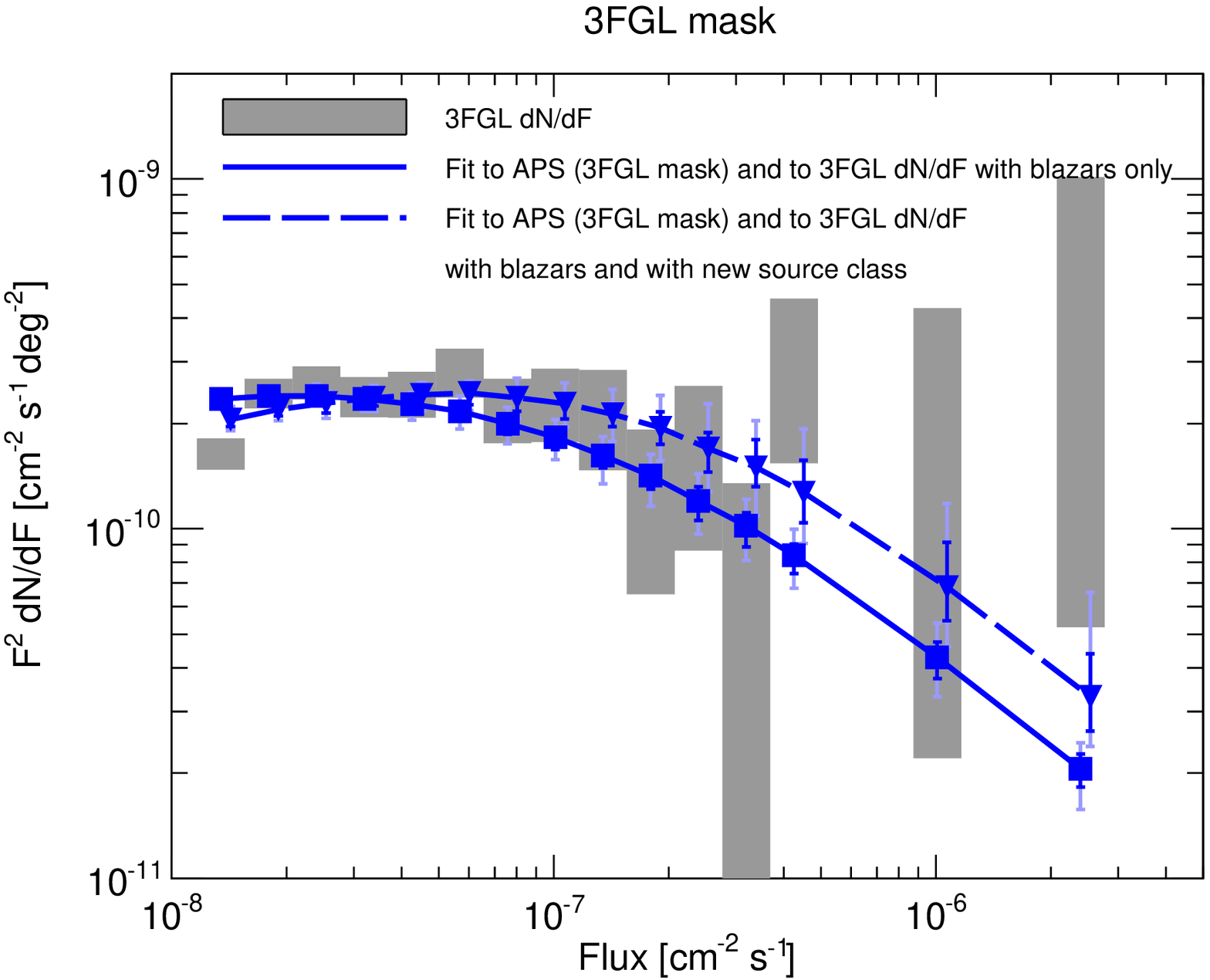}
\caption{\label{fig:bestfit_APS_dNdF_3FGL_APS_and_dNdF} The gray boxes indicate the data points used in the scans, for the auto-correlation APS with the 3FGL mask (left panel) and the 3FGL $dN/dF$ (right panel). The solid blue lines and blue squares indicate the best-fit solution for the scan performed by fitting the auto- and cross-correlation APS from Ref.~\cite{Fornasa:2016ohl} (3FGL mask) and the 3FGL $dN/dF$, with a model including only blazars. The thicker dashed blue lines and the blue triangles denote the best-fit solution for a scan performed fitting the same APS and $dN/dF$ data but including a new source class (see text for details). The thinner dashed blue line in the left panel shows the auto-correlation APS of the new source class, separately. Around each blue square/triangle, the dark blue (light blue) vertical line shows the 68\% (95\%) CL uncertainty. Squares and triangles are slightly shifted with respect to each other to increase readibility.}
\end{figure*}

Fig.~\ref{fig:bestfit_APS_dNdF_3FGL_onlyAPS} shows the predicted 
auto-correlation APS (left panel) and the 3FGL $dN/dF$ (right panel), in 
comparison with the data points used in the fit (gray boxes). The best-fit 
solution is represented by the full red circles and the solid red line, 
embedded in the 68\% (red) and 95\% CL (pink) error bar. We note that, even if 
we do not include the measured source count distribution in our fit, the 
best-fit solution describes the data reasonably well (right panel of 
Fig.~\ref{fig:bestfit_APS_dNdF_3FGL_onlyAPS}). On the other hand, the best fit 
underpredicts the auto-correlation APS in the first energy bin, but it is 
compatible with the measured APS at 95\% CL\footnote{Similarly, the best-fit 
solution overestimates most of the cross-correlation data points involving the 
first two energy bins.}. This discrepancy is responsible for the best-fit 
$\chi^2=-2\ln\mathcal{L}$ of 80.88, corresponding to a $\chi^2$ per degree of 
freedom of 1.47 and $p$-value of 0.005. Note also that the best-fit solution 
predicts negligible anisotropies above 50~GeV. We remind that, in our 
likelihood, we are neglecting any measured APS above that energy. The drop of 
the red line above 50~GeV is compatible with the fact that, at high energies, 
the measured APS has a detection significance lower than 
$3\sigma$~\cite{Fornasa:2016ohl}.

From Figs.~\ref{fig:1D_3FGL} and \ref{fig:2D_3FGL_onlyblazars} we also note 
that both $\gamma_1$ and $p_2^0$ prefer values that are at the edge of the 
prior range considered. This suggests that increasing the range may lead to 
a solution with a better $\chi^2$. We do not consider the possibility of 
negative $\gamma_1$ since there is no indication of this in the analysis of 
resolved blazars from Ref.~\cite{Ajello:2015mfa}, but we perform another scan 
with a prior range for $p_2^0$ that extends up to 20.0. Even if a positive 
evolution of blazars for $z>z_{\rm c}$ is probably unphysical and against the 
findings of Ref.~\cite{Ajello:2015mfa}, we consider this possibility as it 
may increase the abundance of blazars at low energies and, thus, potentially 
improve the agreement with the APS below 1 GeV. However, the best-fit value 
for $p_2^0$ results to be $0.23_{-0.59}^{+1.72}$ and the best-fit APS still 
underproduces the low-energy APS.

\subsection{Adding the source count distribution (3FGL mask)}
\label{sec:APS_and_dNdF_blazars_only}
A realistic model of blazars needs to reproduce the number of sources observed 
by {\it Fermi} LAT. Thus, we perform a new scan over the same four parameters 
as in the previous section but including the observed source count 
distribution $dN/dF$ in the likelihood. We only show results for the $dN/dF$ 
of the 3FGL catalogue but we performed separate scans for the 1FGL and 2FGL 
$dN/dF$. Results are qualititatively similar to what is presented in the 
following.

The solid blue lines in Fig.~\ref{fig:1D_3FGL} show the one-dimensional PL 
for the four free parameters. The blue squares at the bottom of the panels 
mark the best-fit values and the surrounding horizontal blue (light blue) 
lines the 68\% (95\%) CL uncertainty. The two-dimensional PL is shown in 
Fig.~\ref{fig:2D_3FGL_onlyblazars} by means of empty contours and the empty 
circle indicates the best-fit point. Overall, the precision in the estimation
of the parameters has improved and the best-fit solution is not very different 
than what obtained when fitting only the auto- and cross-correlation APS: the 
preferred regions are located along the same degeneracies as before, with a 
slight shift towards smaller normalisations $A$ and, therefore, a larger 
$\gamma_1$ and $L_0$. The PL distribution for $p_2^0$ is still clustered 
towards the upper edge of its prior range, while, this time, $\gamma_1$ is 
different from zero at 95\% CL. Solutions with a $\gamma_1 \sim 0$ would 
underestimate the source count distribution and, thus, are now excluded. In 
Fig.~\ref{fig:bestfit_APS_dNdF_3FGL_APS_and_dNdF}, the solid blue line and 
the blue squares indicate the auto-correlation APS (left panel) and the 3FGL 
$dN/dF$ (right panel) corresponding to the best fit, in comparison with the 
data employed in the likelihood (gray boxes). The dark blue (light blue) 
vertical bars show the 68\% and 95\% CL uncertainty. Similar to the previous 
section, we note an underestimation of the auto-correlation APS below 1~GeV. 
The best-fit solution has a $\chi^2$ of 112.41, corresponding to a $\chi^2$ 
per degree of freedom of 1.70 and a $p$-value of $3 \times 10^{-4}$. Allowing 
$p_2^0$ to have positive values leads to a best-fit $p_2^0$ that is different 
than zero at 95\% CL (i.e. $p_2^0=2.48^{+0.81}_{-0.73}$). In that case, the 
best-fit solution has a $\chi^2$ value of 106.10, but the reconstructed APS 
still undereproduces the measured one below 1 GeV. The results of these first 
two scans suggest that blazars alone are not able to reproduce the measured 
APS below the GeV scale. This agrees with the findings of 
Ref.~\cite{Fornasa:2016ohl}.

We also tested this result by performing another scan in which we fit the
APS with the 3FGL mask and the 3FGL $dN/dF$ but the source count distribution
is computed with all sources in the catalogue, not only blazars. In this 
case, the best-fit solution predicts approximately 10\% more APS below 1 
GeV and, thus, the best-fit $\chi^2$ is smaller (i.e. $\chi^2=93.78$) than 
with the blazars $dN/dF$. This suggests that taking unassociated sources into
account could improve the agreement with the measured APS. We comment more 
about this in Sec.~\ref{sec:discussion}.

We note that our best-fit values from Figs.~\ref{fig:1D_3FGL} and 
\ref{fig:2D_3FGL_onlyblazars} are in agreement (at the 95\% CL) with the
results of Ref.~\cite{Ajello:2015mfa}. In that paper, the authors employ the 
same LDDE scheme used here to describe 403 blazars observed by {\it Fermi} 
LAT, but they used a larger parameter space and they constrained their model 
by fitting the blazars' flux and spectral index distributions, taking also
advantage of some redshift estimates. 

\label{sec:with_new_source_class}
\begin{figure*}
\includegraphics[width=0.32\textwidth]{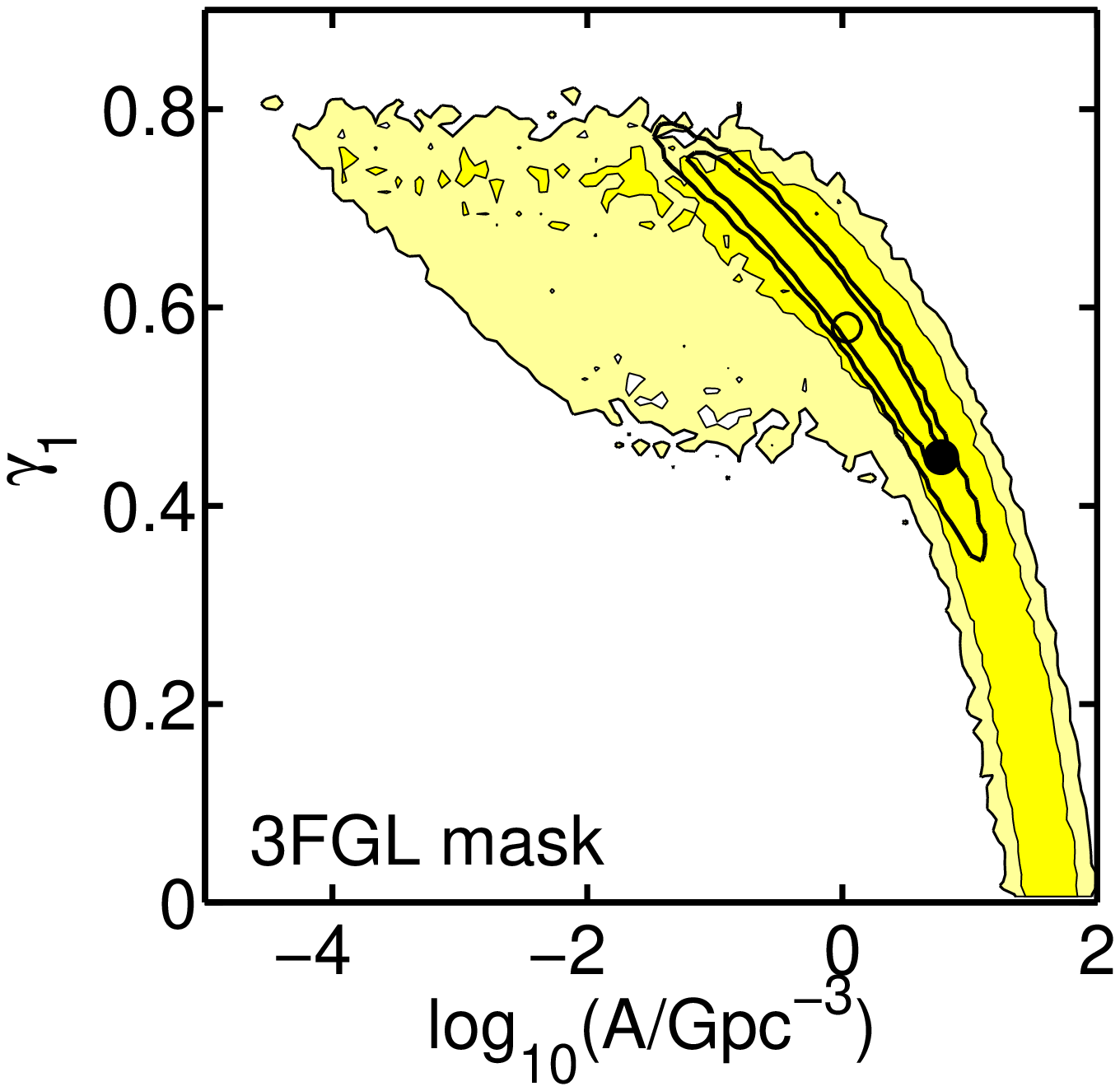}
\includegraphics[width=0.32\textwidth]{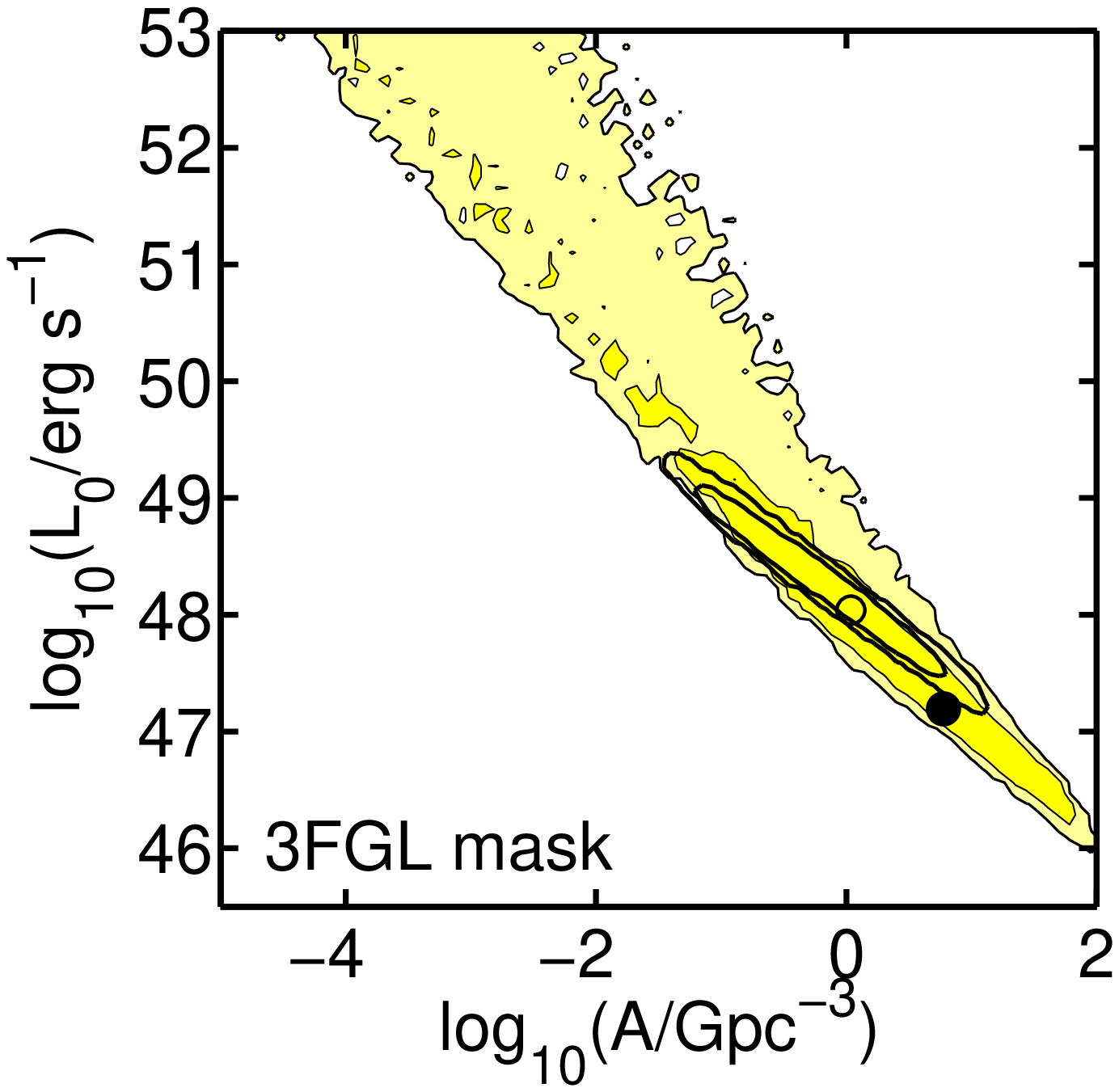}
\includegraphics[width=0.32\textwidth]{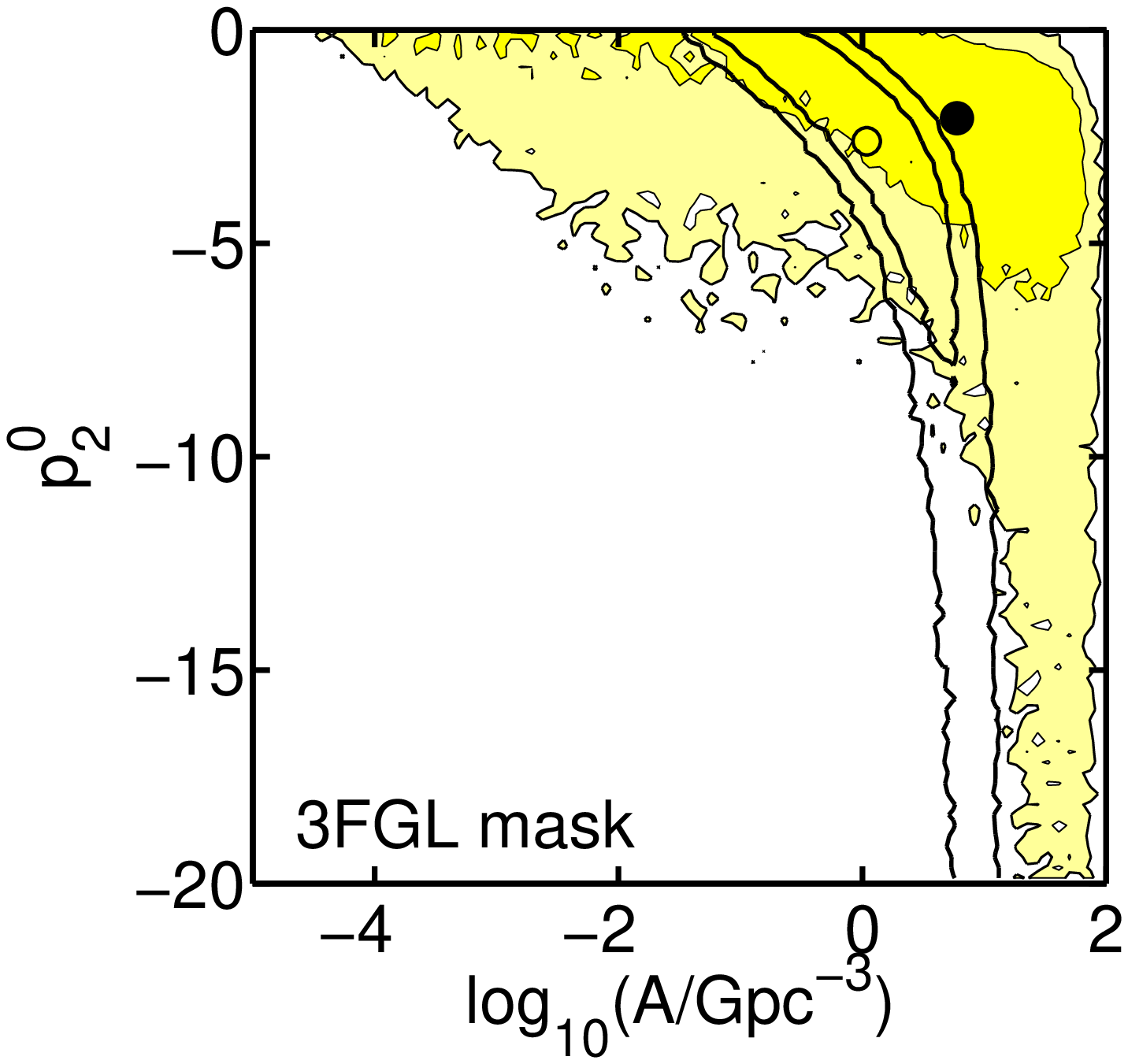}
\includegraphics[width=0.32\textwidth]{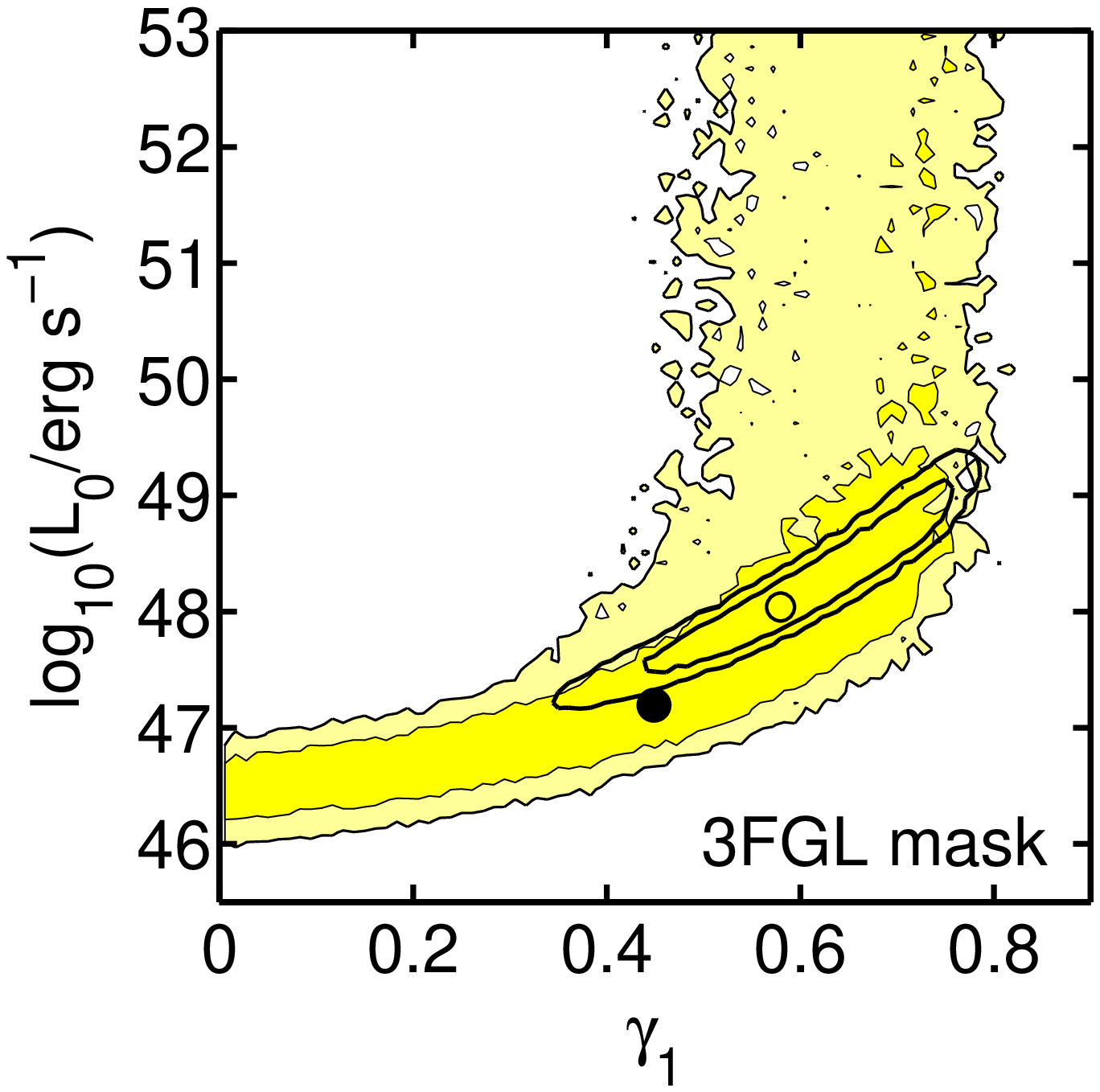}
\includegraphics[width=0.32\textwidth]{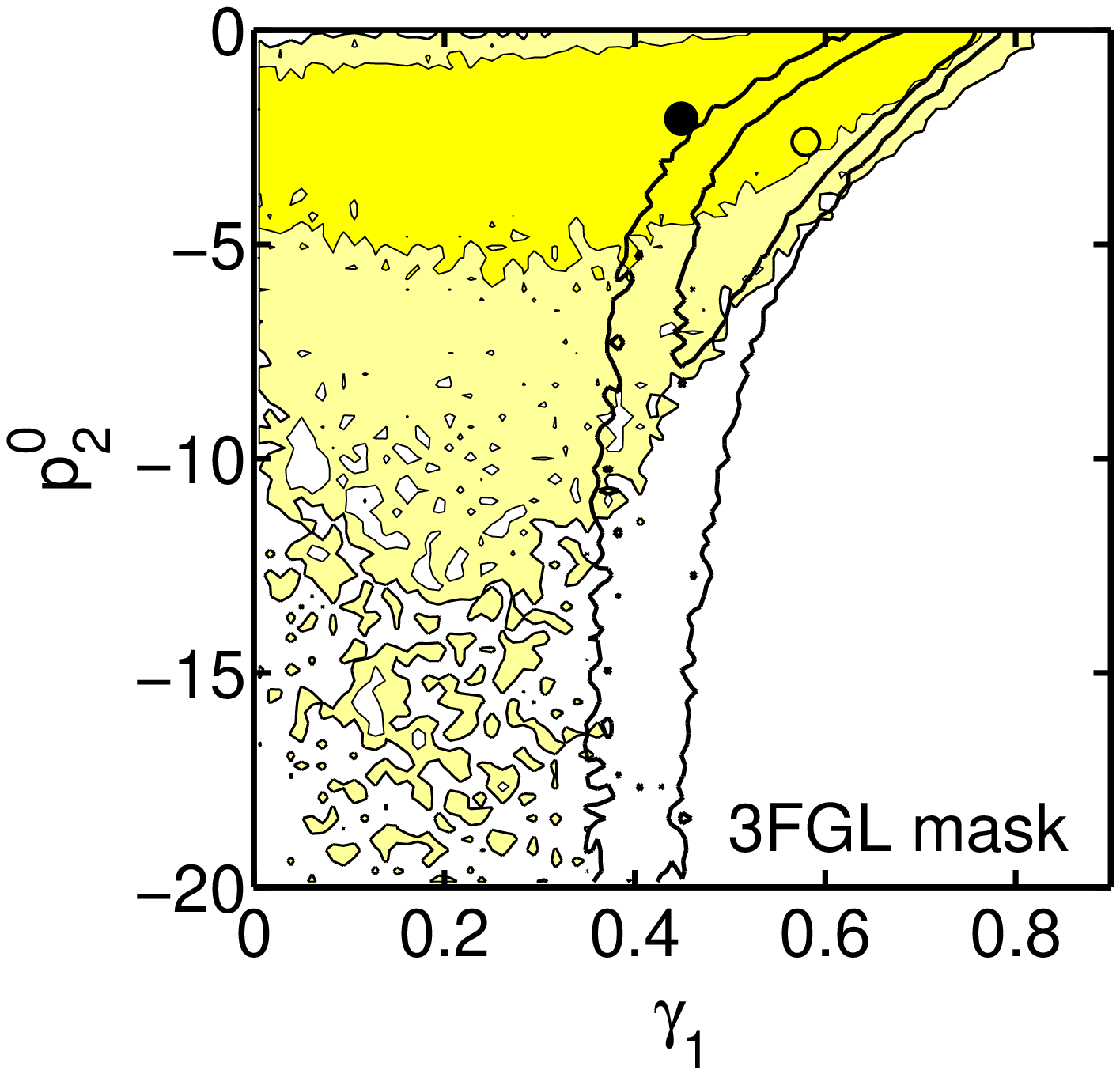}
\includegraphics[width=0.32\textwidth]{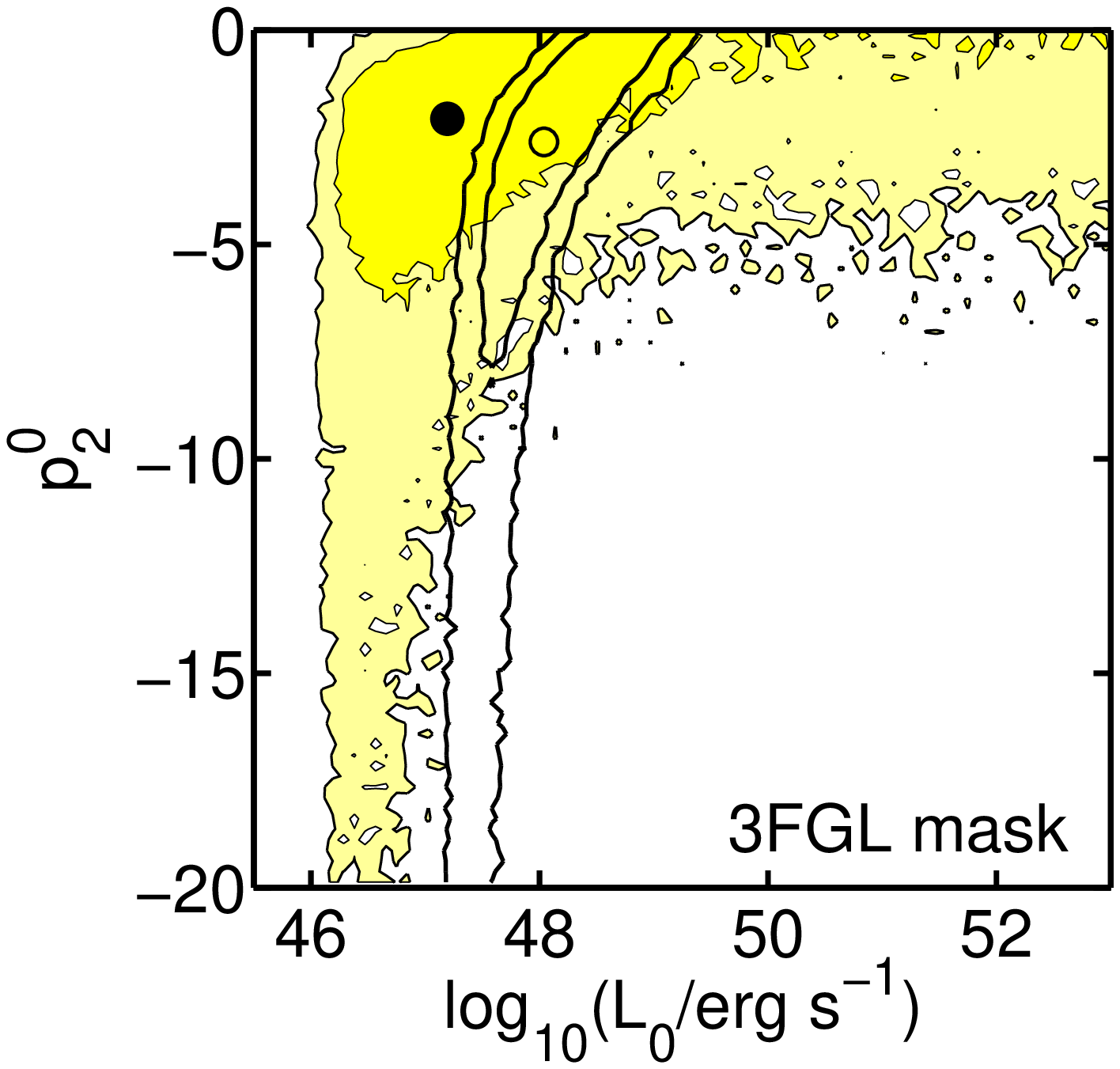}
\caption{\label{fig:2D_3FGL_new_class} Two-dimensional profile-likelihood contour plots for all the combinations of parameters $A$, $\gamma_1$, $L_0$ and $p_2^0$ (see text for details). Filled contours and filled black dots refer to the scans performed by fitting only the auto- and cross-correlation APS from Ref.~\cite{Fornasa:2016ohl} in the case of the 3FGL mask, while empty contours and empty circles are for the fits to the APS and the source count distribution $dN/dF$ from the 3FGL catalogue. Inner contours mark the 68\% CL region and outer ones the 95\% CL region. The model predictions in the scans are computed with blazars and a new class of gamma-ray emitters (see text).}
\end{figure*}

\subsection{Including a new class of sources (3FGL mask)}
Motivated by the results of the previous section, we expand our theoretical 
model by including one additional population of sources. We adopt a 
phenomenological description and we assume that the new sources emit with a 
power-law energy spectrum ($\propto E^{-\Gamma_{\rm new}}$) and that they are 
point-like and unclustered. Thus, their auto- and cross-correlation can be 
written as follows:
\begin{equation}
C_{\rm P}^{i,j} = C_{\rm P}^{0,0} 
\frac{[E_{{\rm max},i}^{1-\Gamma_{\rm new}} - E_{{\rm min},i}^{1-\Gamma_{\rm new}}]
[E_{{\rm max},j}^{1-\Gamma_{\rm new}} - E_{{\rm min},j}^{1-\Gamma_{\rm new}}]}
{[E_{{\rm max},0}^{1-\Gamma_{\rm new}} - E_{{\rm min},0}^{1-\Gamma_{\rm new}}]^2},
\label{eqn:APS_newpop}
\end{equation}
where $C_{\rm P}^{0,0}$ is the auto-correlation APS of the new population in the 
first energy bin, between $E_{{\rm min,0}}$ and $E_{{\rm max},0}$, while 
$E_{{\rm min},i}$ and $E_{{\rm max},i}$ indicate the lower and upper edge of the 
$(i+1)$-th energy bin.

We perform one additional scan fitting the auto- and cross-correlation APS
from Ref.~\cite{Fornasa:2016ohl} for the 3FGL mask but adding the 
contribution of the new source class. $\Gamma_{\rm new}$ and $C_{\rm P}^{0,0}$ are 
added to the list of the free parameters in the scan. We assume flat priors 
for $\Gamma_{\rm new}$ between 2.2 and 3.4 and log priors for $C_{\rm P}^{0,0}$ 
between $10^{-20}$ and $10^{-15}~\mbox{cm}^{-4}~\mbox{s}^{-2}~\mbox{sr}^{-1}$.

The one-dimensional PL distributions for $A$, $\gamma_1$, $L_0$ and $p_2^0$ for 
the new scan are plotted as red dashed lines in Fig.~\ref{fig:1D_3FGL}. The
best fits are marked by the red triangles, within their 68\% and 95\% CL 
uncertainty denoted by the red and pink horinzontal lines. The two-dimensional 
68\% and 95\% CL regions are also shown in Fig.~\ref{fig:2D_3FGL_new_class} 
by the dark yellow and light yellow areas, respectively. The full black 
circles indicate the best fit. We note that the size of the contours has 
increased significantly: apart from the same degeneracy present in 
Fig.~\ref{fig:2D_3FGL_onlyblazars} (i.e., the diagonal band between $A$ and 
$\gamma_1$, for $A > 1~\mbox{Gpc}^{-3}$), a new region appears with 
$A < 1~\mbox{Gpc}^{-3}$, $\gamma_1 > 0.5$ and $L_0 > 10^{48}~\mbox{erg s}^{-1}$.
There is also another new region for large $A$ and very negative $p_2^0$. 
Models in both of these regions significantly underproduce the measured auto- 
and cross-APS at low energies, and therefore, they would be excluded if 
blazars were the only class of gamma-ray emitters in the fit. 

\begin{figure*}
\includegraphics[width=0.49\textwidth]{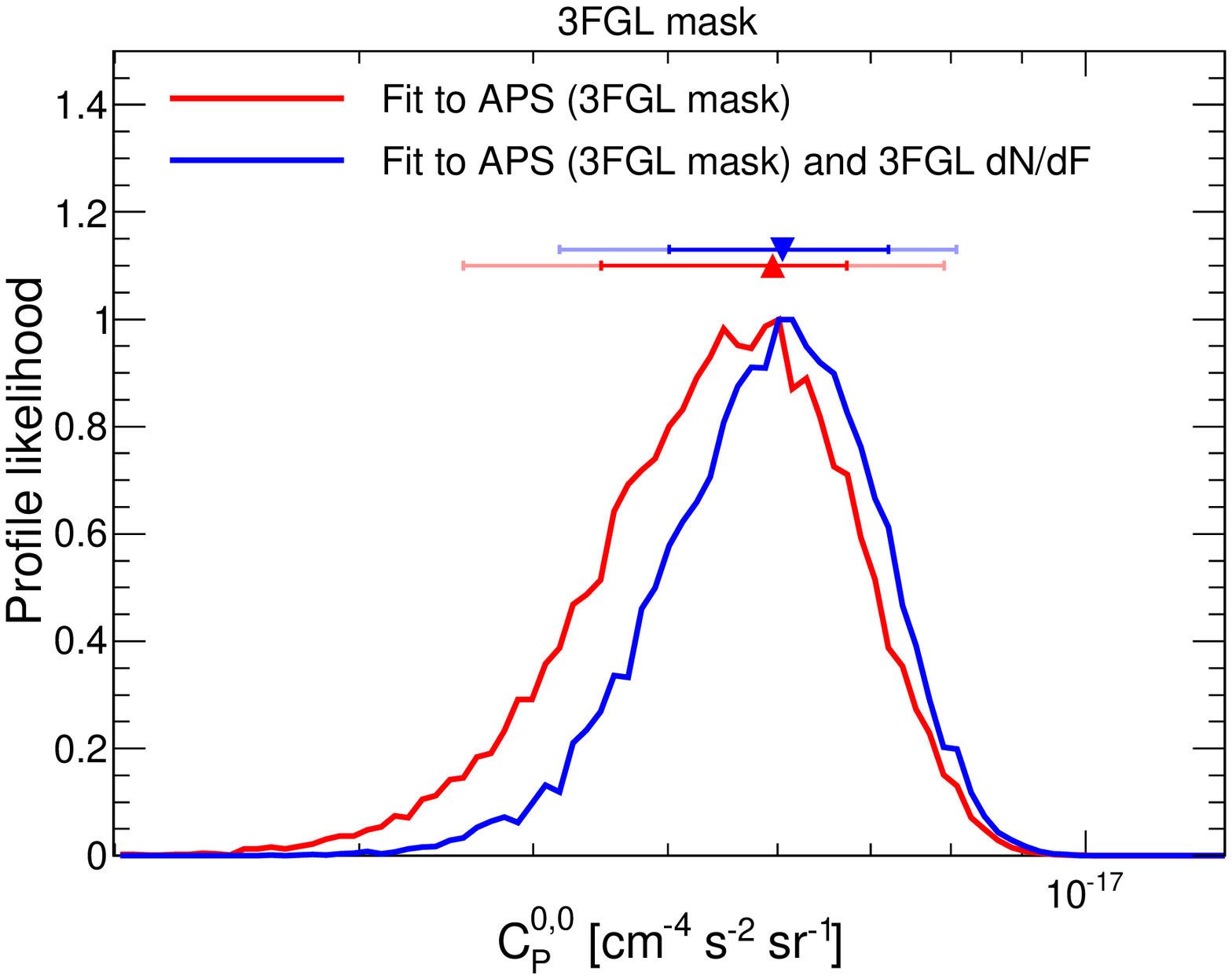}
\includegraphics[width=0.49\textwidth]{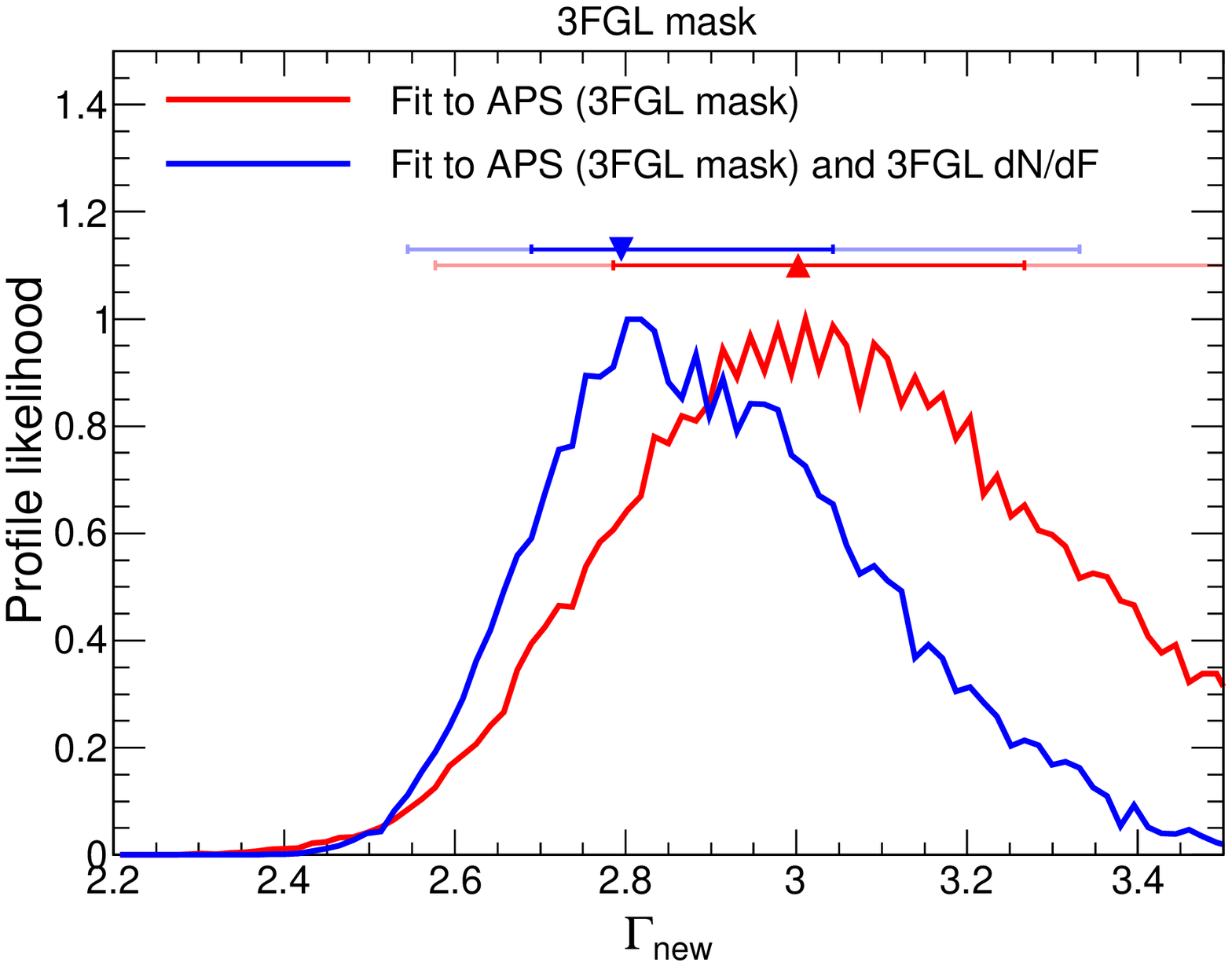}
\caption{\label{fig:1D_3FGL_new_class} As Fig.~\ref{fig:1D_3FGL}, but for the parameters describing the new source class.}
\end{figure*}

\begin{figure}
\includegraphics[width=0.49\textwidth]{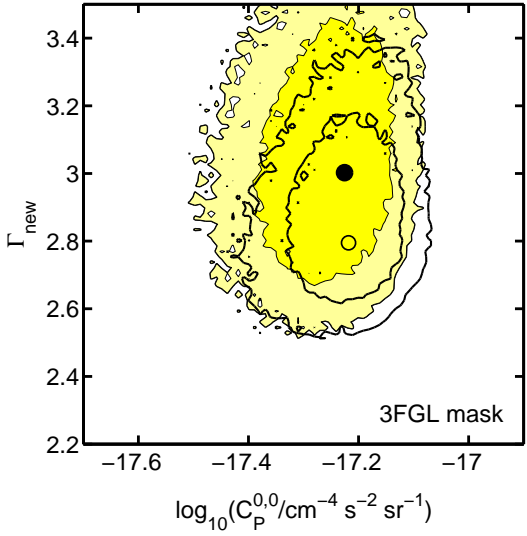}
\caption{\label{fig:2D_3FGL_new_params} Two-dimensional profile-likelihood contour plot for the parameters describing the new source class, i.e. $C_{\rm P}^{0.0}$ and $\Gamma_{\rm new}$. Filled contours and filled black dots refer to the scans performed by fitting only the auto- and cross-correlation APS from Ref.~\cite{Fornasa:2016ohl} in the case of the 3FGL mask, while empty contours and empty circles are for the fits to the APS and the source count distribution $dN/dF$ from the 3FGL catalogue. Inner contours mark the 68\% CL region and outer ones the 95\% CL one.}.
\end{figure}

The new source population takes care of improving the agreement with the
low-energy APS. We can see this in 
Fig.~\ref{fig:bestfit_APS_dNdF_3FGL_onlyAPS}, where the thick dashed red 
lines and the red triangles show the predicted best-fit auto-correlation APS 
(left panel) and 3FGL $dN/dF$ (right panel). As before, the vertical red 
(pink) lines indicate the 68\% (95\%) CL uncertainty. The auto-correlation APS
of the new component alone is shown separately by the thinner dashed red line.
The predicted $dN/dF$ (dashed red line in the right panel) is much more 
uncertain than with blazars only and it systematically underproduces the
3FGL source count distribution. 

Red lines in Fig.~\ref{fig:1D_3FGL_new_class} show the one-dimensional PL 
distributions for the new parameters in the scans, i.e., $C_{\rm P}^{0,0}$ (left) 
and $\Gamma_{\rm new}$ (right). In Fig.~\ref{fig:2D_3FGL_new_params}, the filled
contours indicate the 68\% and 95\% CL regions on the 
($C_{\rm P}^{0.0}, \Gamma_{\rm new}$) plane. The PL of $C_{\rm P}^{0,0}$ is quite 
peaked and the reconstruction has a precision of 15\%. The reconstruction is 
less precise for $\Gamma_{\rm new}$ and values between 2.79 and 3.27 are allowed 
at 68\% CL.

The best-fit $\chi^2$ is significantly better than without the new source 
class, i.e. $\chi^2=54.29$, corresponding to a $\chi^2$ per degree of freedom
of 1.11 and a $p$-value of 0.28. We can perform a likelihood-ratio test by
defining $\Delta\chi^2$, i.e., the difference between the best-fit $\chi^2$ of 
the simpler model (i.e. without the new source class, see 
Sec.~\ref{sec:APS_only_blazars_only}) and the best-fit $\chi^2$ of the
one with the new source class. Since the simpler model is located on the 
boundary of the more complex one, we apply Chernoff's 
theorem~\cite{Chernoff:1954,Shapiro:1988} and obtain a $p$-value of 
$5 \times 10^{-7}$, indicating that the model with the additional source class 
is preferred over the interpretation in terms of only blazars at $5\sigma$.

Alternatively, following Bayesian statistics, the comparison between models 
can be performed by computing the Bayes factor $B$, defined as the ratio of 
the evidences of the two scans. In our scan $\ln B$ is 12.37, a value 
suggesting a strong preference for the model with the new source class,
according to Jeffrey's scale~\cite{Trotta:2008qt}.

Finally, we perform one additional scan including the new source class but
fitting both the APS (with the 3FGL mask) and the 3FGL $dN/dF$. The 
one-dimensional PL for $A$, $\gamma_1$, $L_0$ and $p_2^0$ are shown as dashed
blue lines (with blue triangles) in Fig.~\ref{fig:1D_3FGL}, while the 
two-dimensional PL are denoted as empty contours in 
Fig.~\ref{fig:2D_3FGL_new_class}. The most relevant difference with respect to
the fit to the APS data only (filled yellow regions in 
Fig.~\ref{fig:2D_3FGL_new_class}) is the fact that values of $\gamma_1$ 
smaller than 0.3 are disfavoured at 95\% CL, as well as 
$A<10^{-2}~\mbox{Gpc}^{-3}$. Solutions in those regions would underpredict
or overpredict the source count distribution, respectively. 
Fig.~\ref{fig:bestfit_APS_dNdF_3FGL_APS_and_dNdF} compares the auto-correlation
APS and $dN/dF$ of the best-fit solution (dashed thick blue lines and blue 
triangles) to the data (grey boxes). As for the previous case, including the 
new source improves the agreement with the APS data below 1~GeV. The 
auto-correlation APS of the new class alone is shown by the thin blue dashed 
line in the left panel of Fig.~\ref{fig:bestfit_APS_dNdF_3FGL_APS_and_dNdF}. 
The one-dimensional PL for $C_{\rm P}^{0,0}$ and $\Gamma_{\rm new}$ are shown in 
Fig.~\ref{fig:1D_3FGL_new_class} by blue lines and their two-dimensional PL by 
empty contours in Fig.~\ref{fig:2D_3FGL_new_params}. Results are compatible
with the fit to the APS alone, confirming that the properties of the new 
source class do not change if we include the $dN/dF$ data. In this case, the 
best-fit $\chi^2$ is 70.19, corresponding to $\chi^2$ per degree of freedom of 
1.10 and a $p$-value of 0.28. The likelihood-ratio test yields a $p$-value of 
$2 \times 10^{-10}$, corresponding to more than 10$\sigma$. The presence of the 
new source class is strongly favoured also in a Bayesian framework as the 
Bayes factor $\ln B$ is 19.13.

\begin{figure*}
\includegraphics[width=0.49\textwidth]{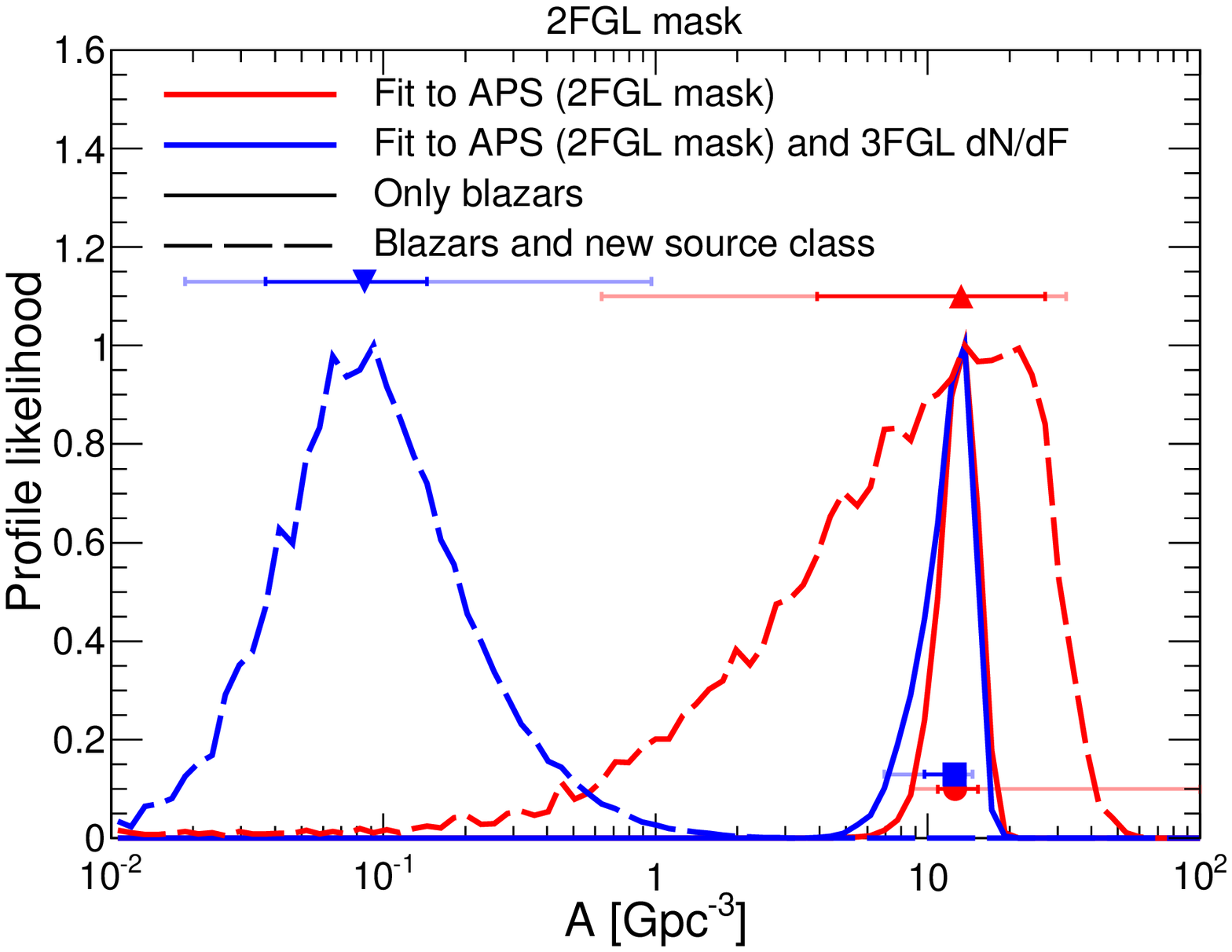}
\includegraphics[width=0.49\textwidth]{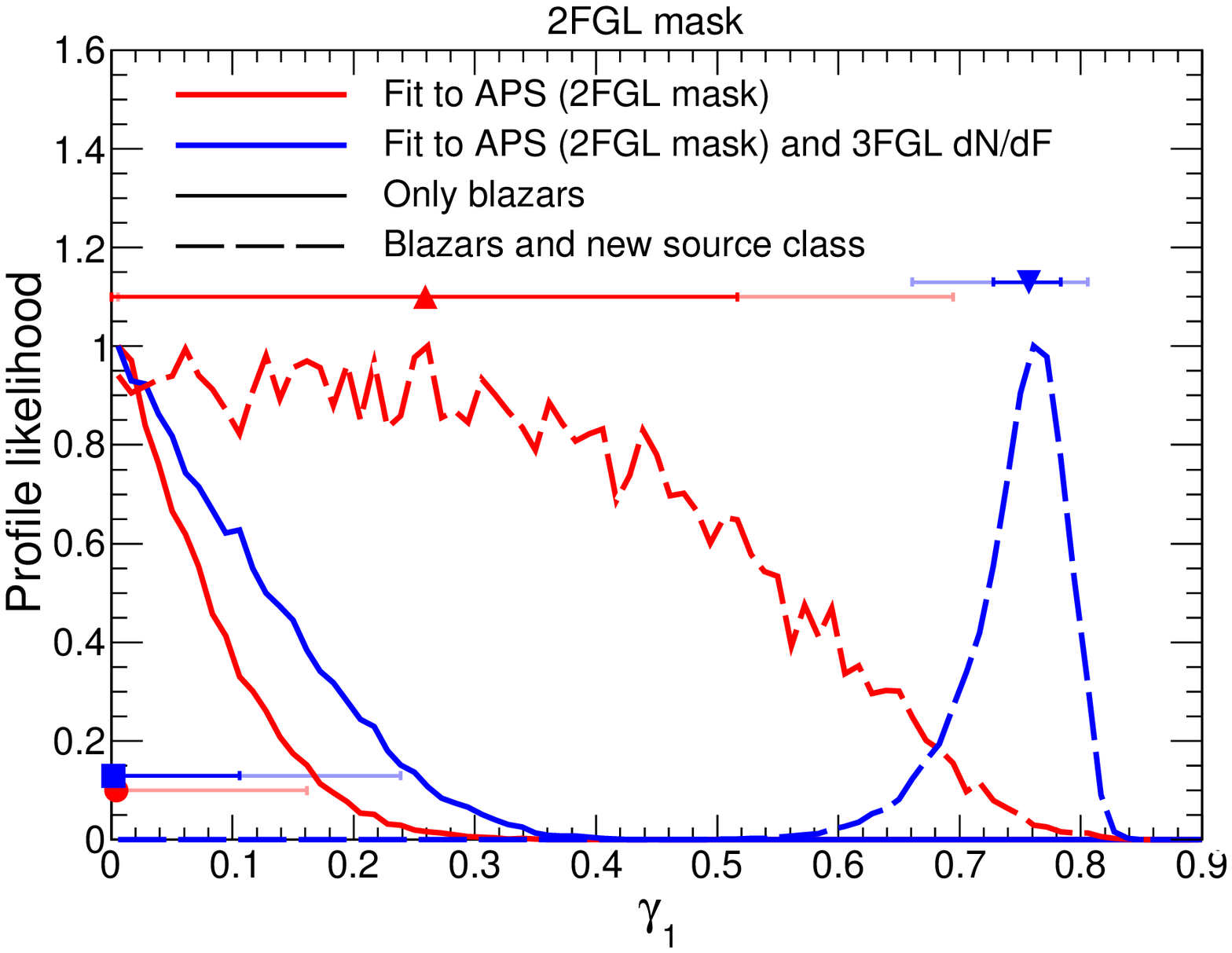}
\includegraphics[width=0.49\textwidth]{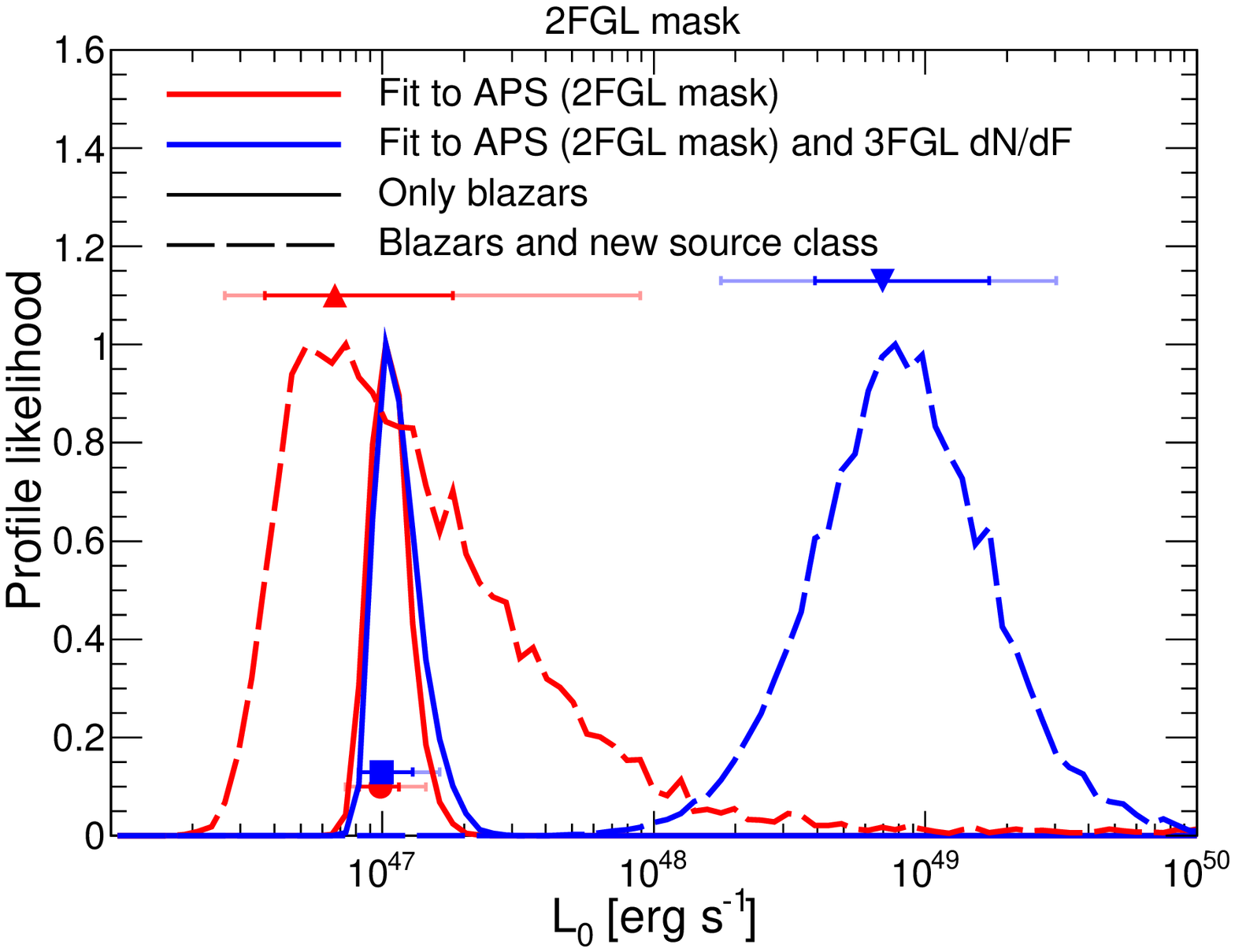}
\includegraphics[width=0.49\textwidth]{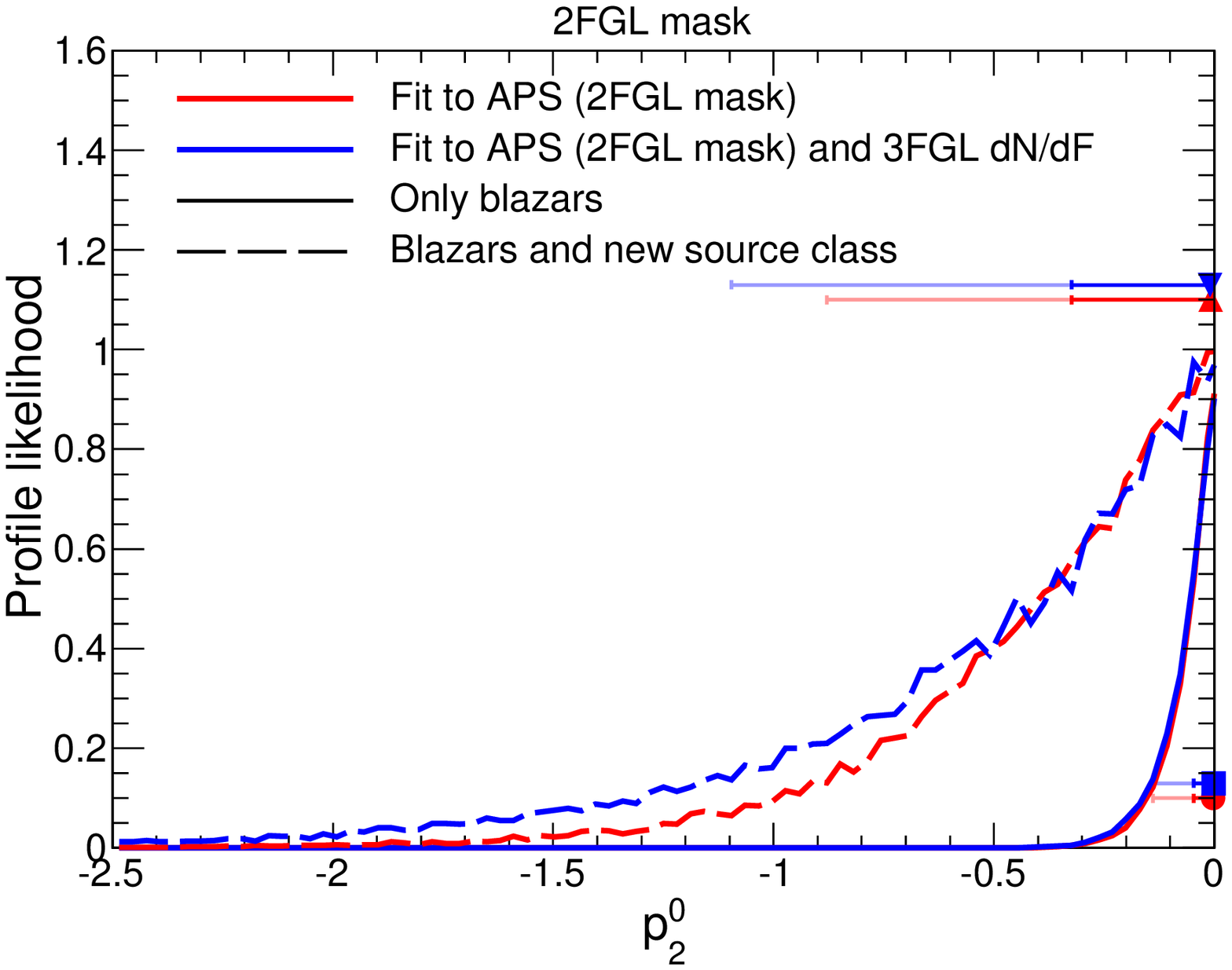}
\includegraphics[width=0.49\textwidth]{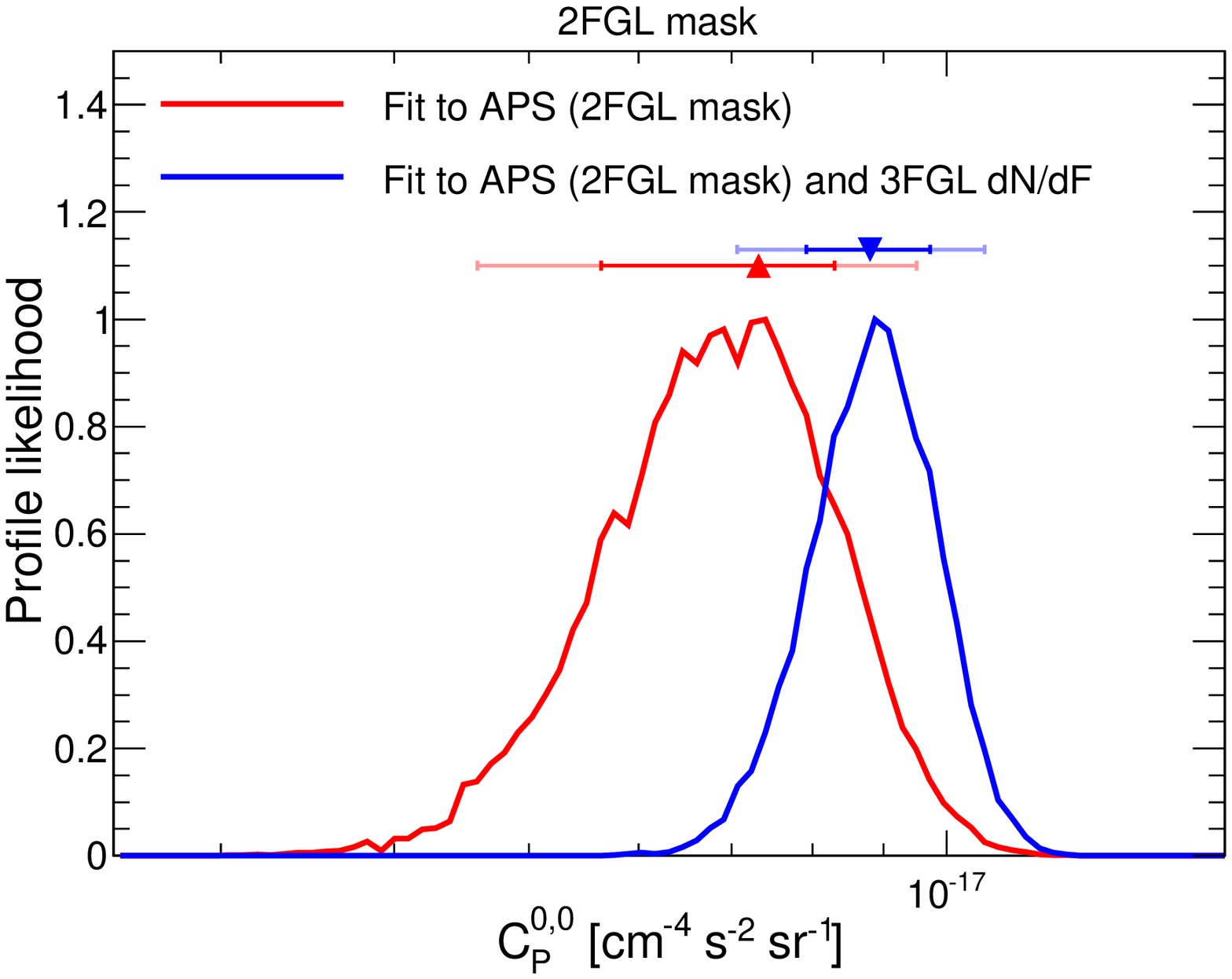}
\includegraphics[width=0.49\textwidth]{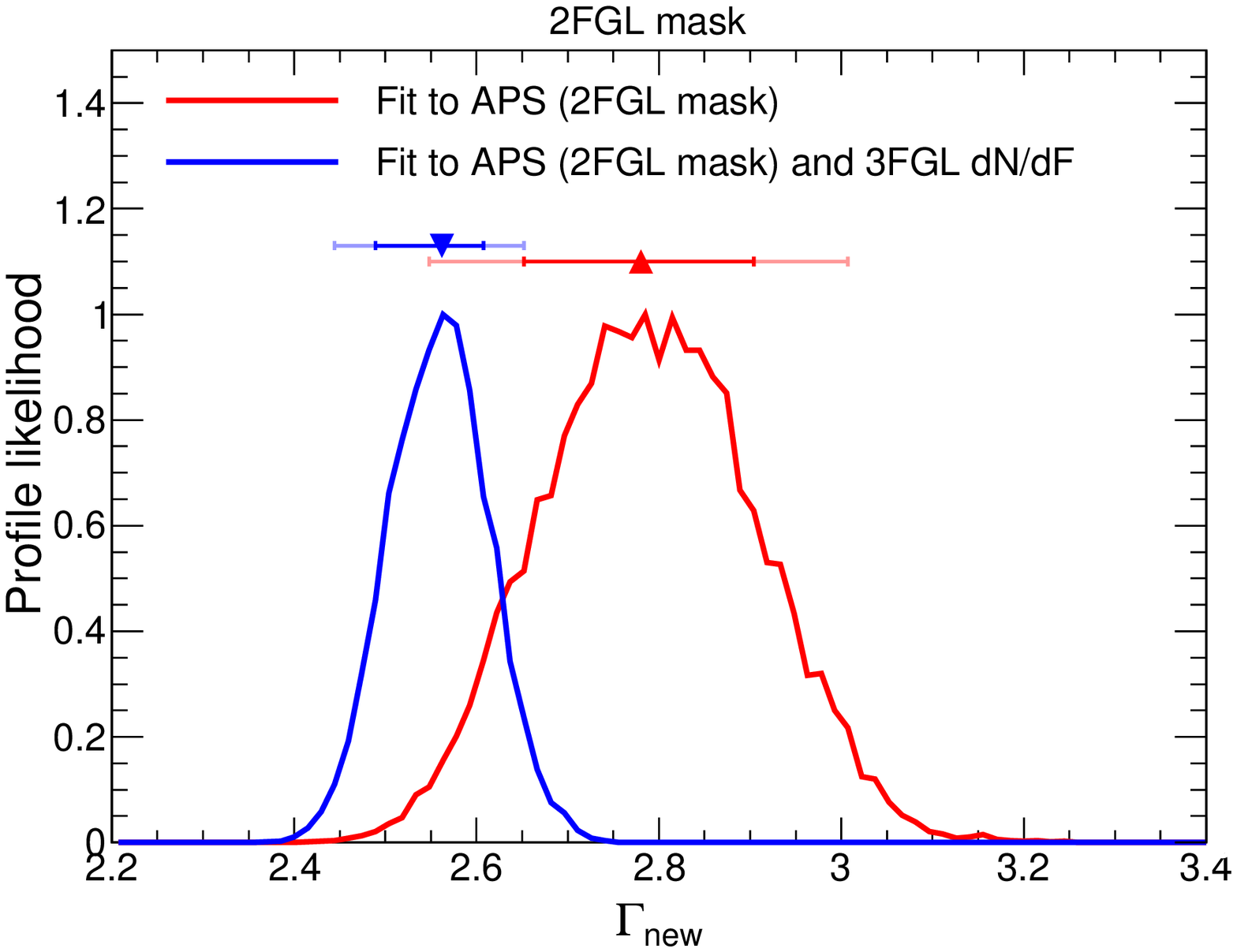}
\caption{\label{fig:1D_2FGL} One-dimensional profile-likelihood distribution for parameters $A$, $\gamma_1$, $L_0$, $p_2^0$, $C_{\rm P}^{0,0}$ and $\Gamma_{\rm new}$ (see text for details). The red lines refer to the scans performed by fitting only the auto- and cross-correlation APS from Ref.~\cite{Fornasa:2016ohl} in the case of the 2FGL mask, while the blue lines are for the fits to the APS and the source count distribution $dN/dF$ from the 3FGL catalogue. The solid lines refer to fits in which model predictions are computed only in terms of blazars (described by the LDDE scheme from Sec.~\ref{sec:model}), while for the dashed lines we include an additional population of sources (see text). Squares, circles and triangles indicate the best-fit solutions. The sets of lines near the bottom of the panels are for the scans performed only with blazars and the ones near the top (i.e. below the legend) are for the scans including the additional source class. The red/blue (pink/ligher blue) horizontal lines indicate the 68\% (95\%) CL region.}
\end{figure*}

\begin{figure*}
\includegraphics[width=0.32\textwidth]{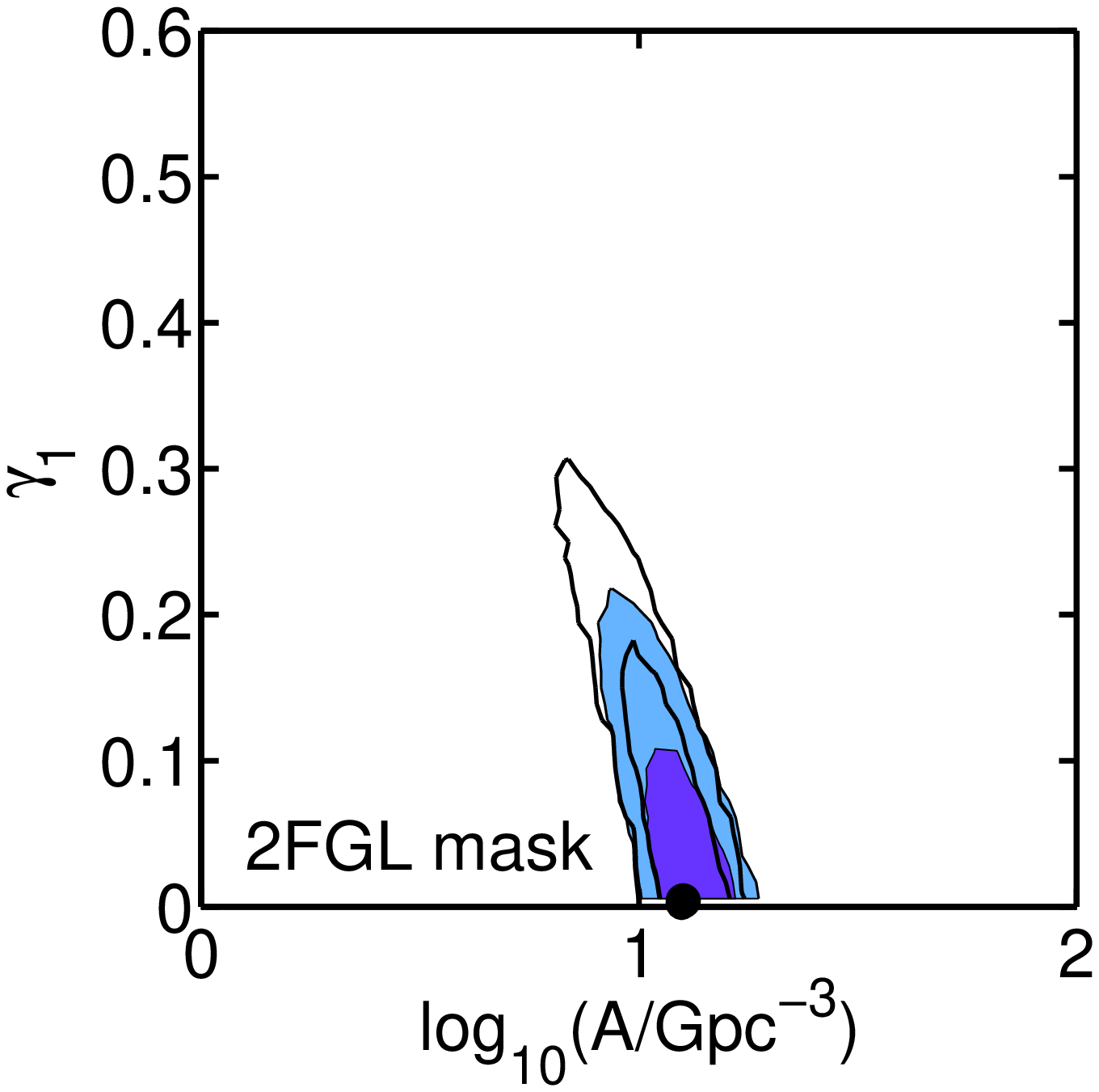}
\includegraphics[width=0.32\textwidth]{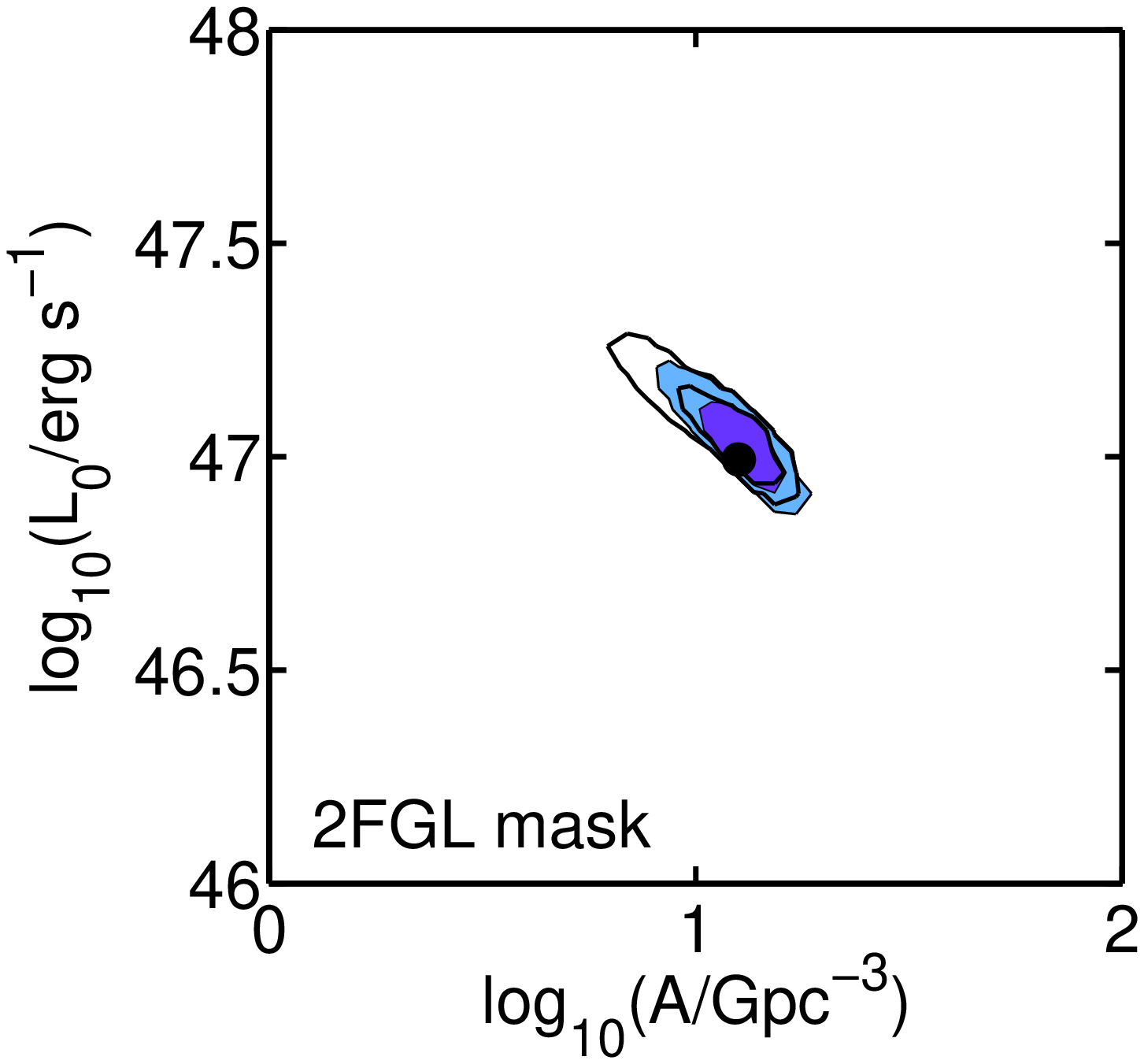}
\includegraphics[width=0.32\textwidth]{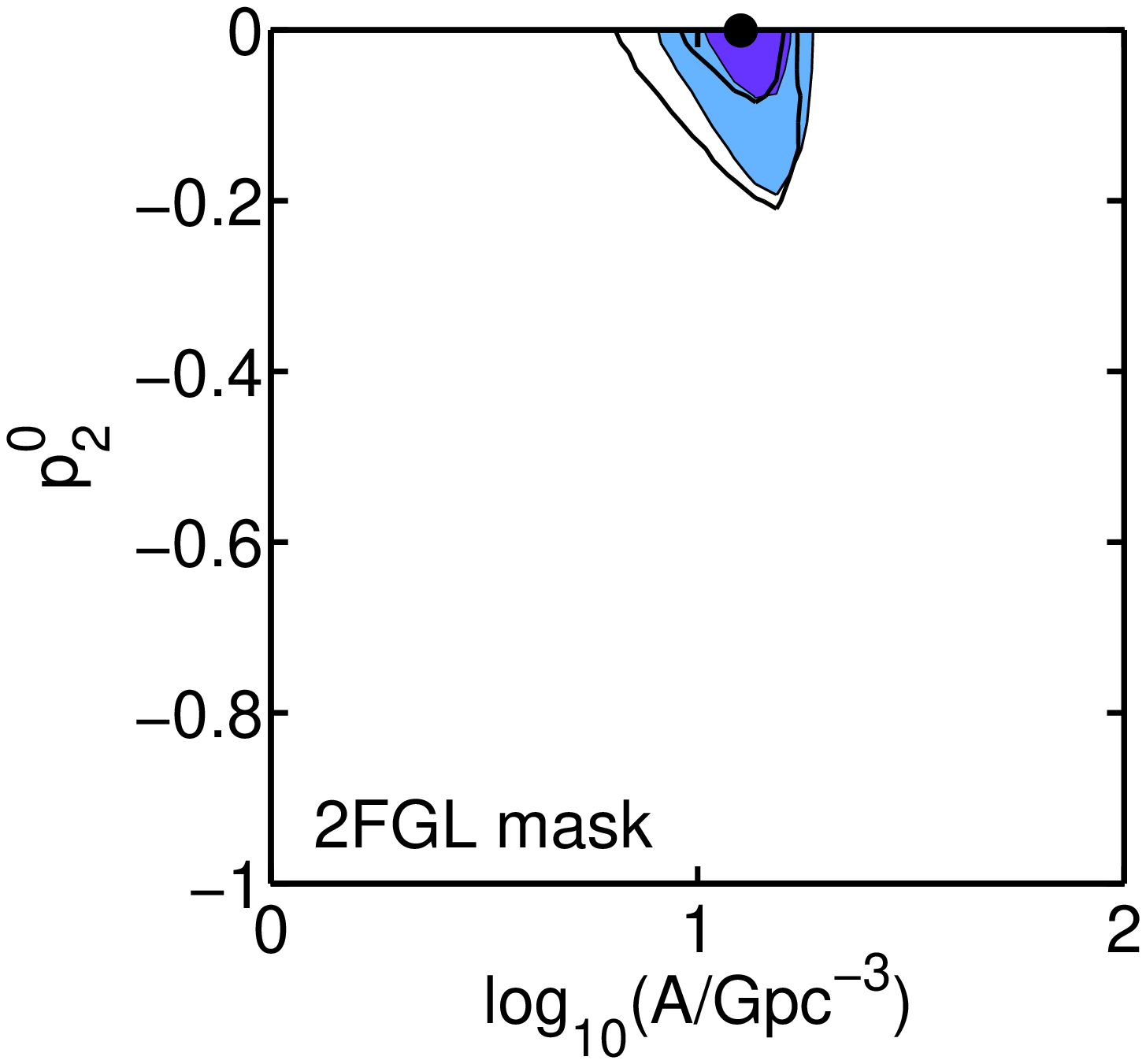}
\includegraphics[width=0.32\textwidth]{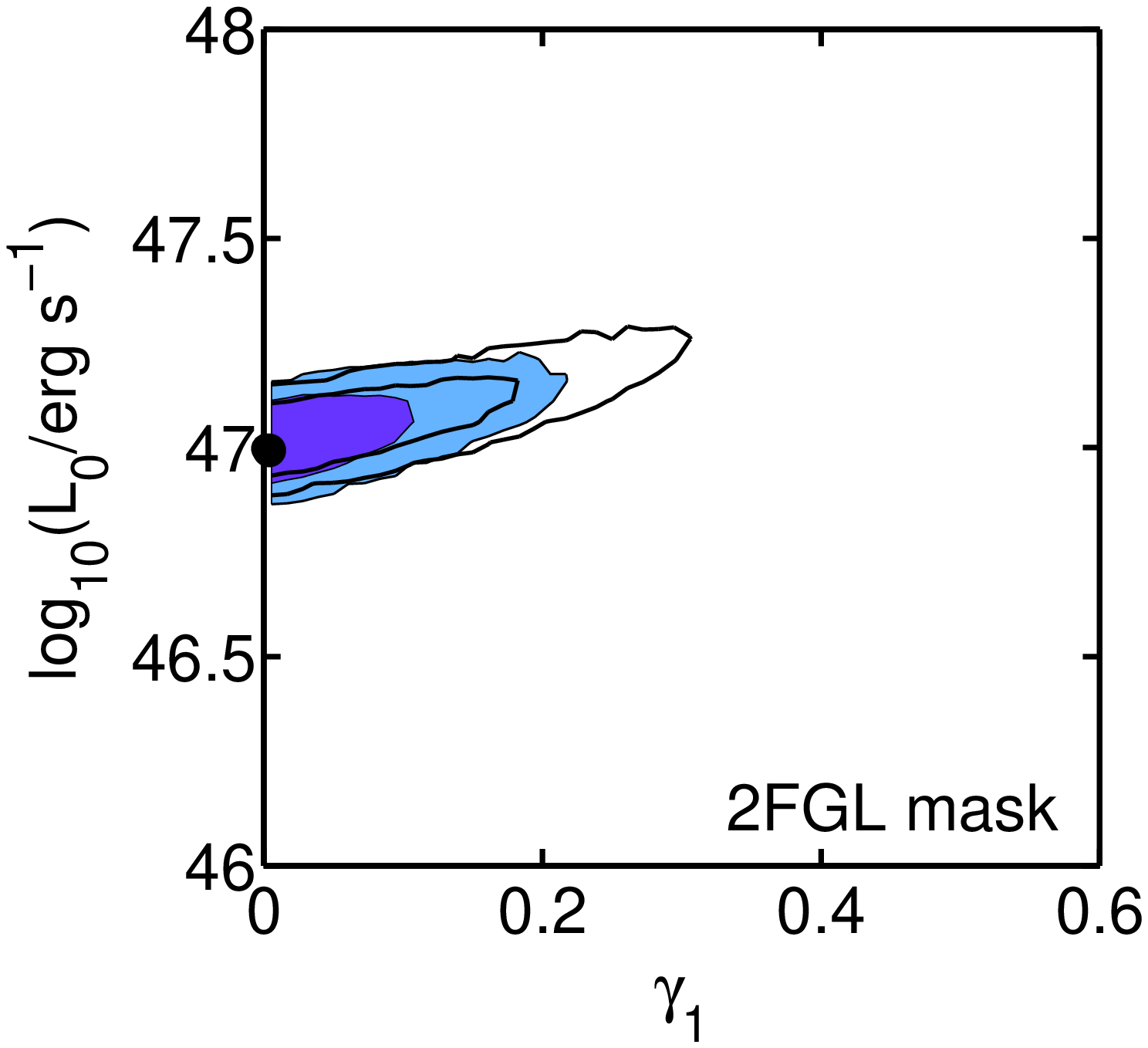}
\includegraphics[width=0.32\textwidth]{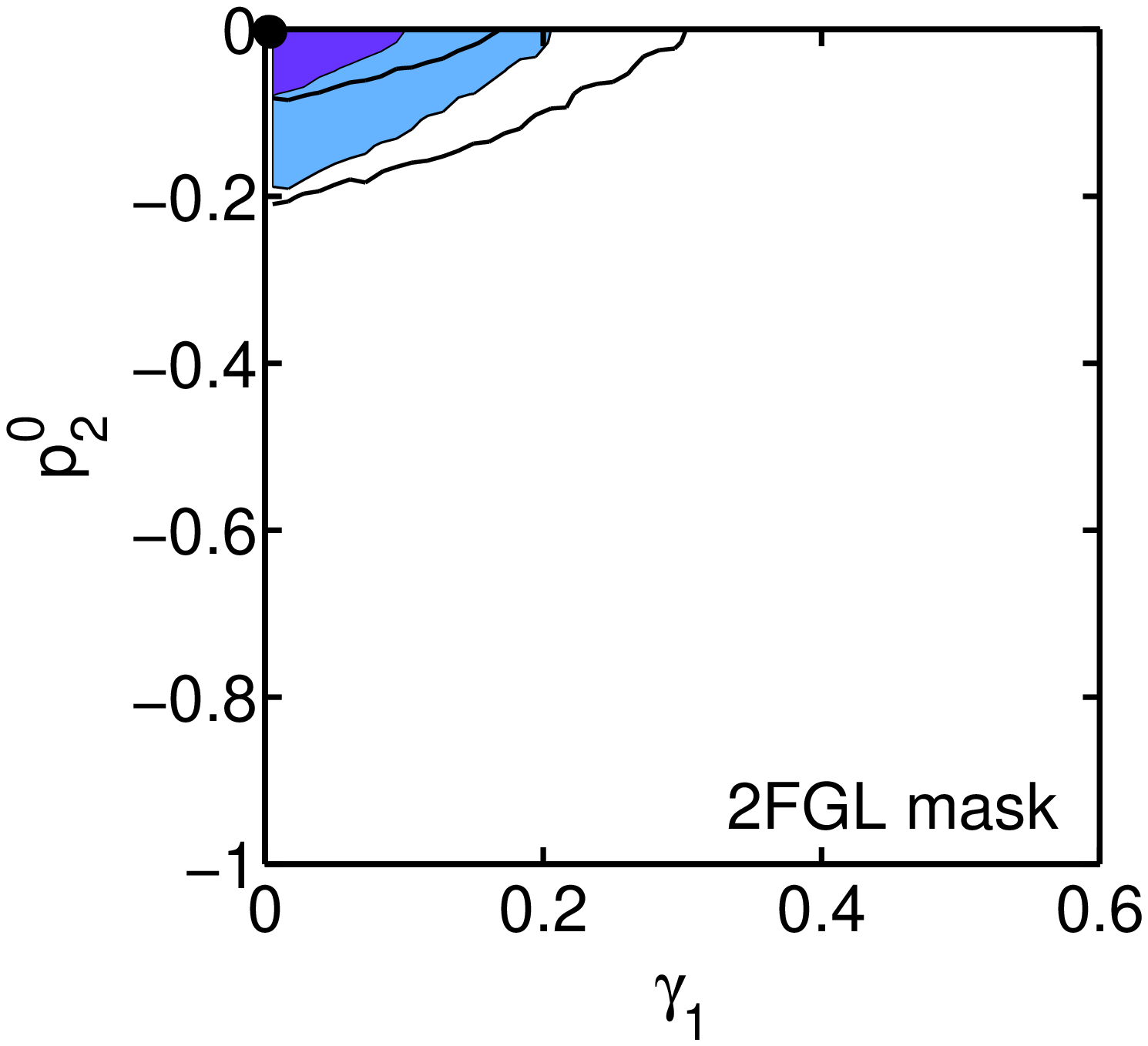}
\includegraphics[width=0.32\textwidth]{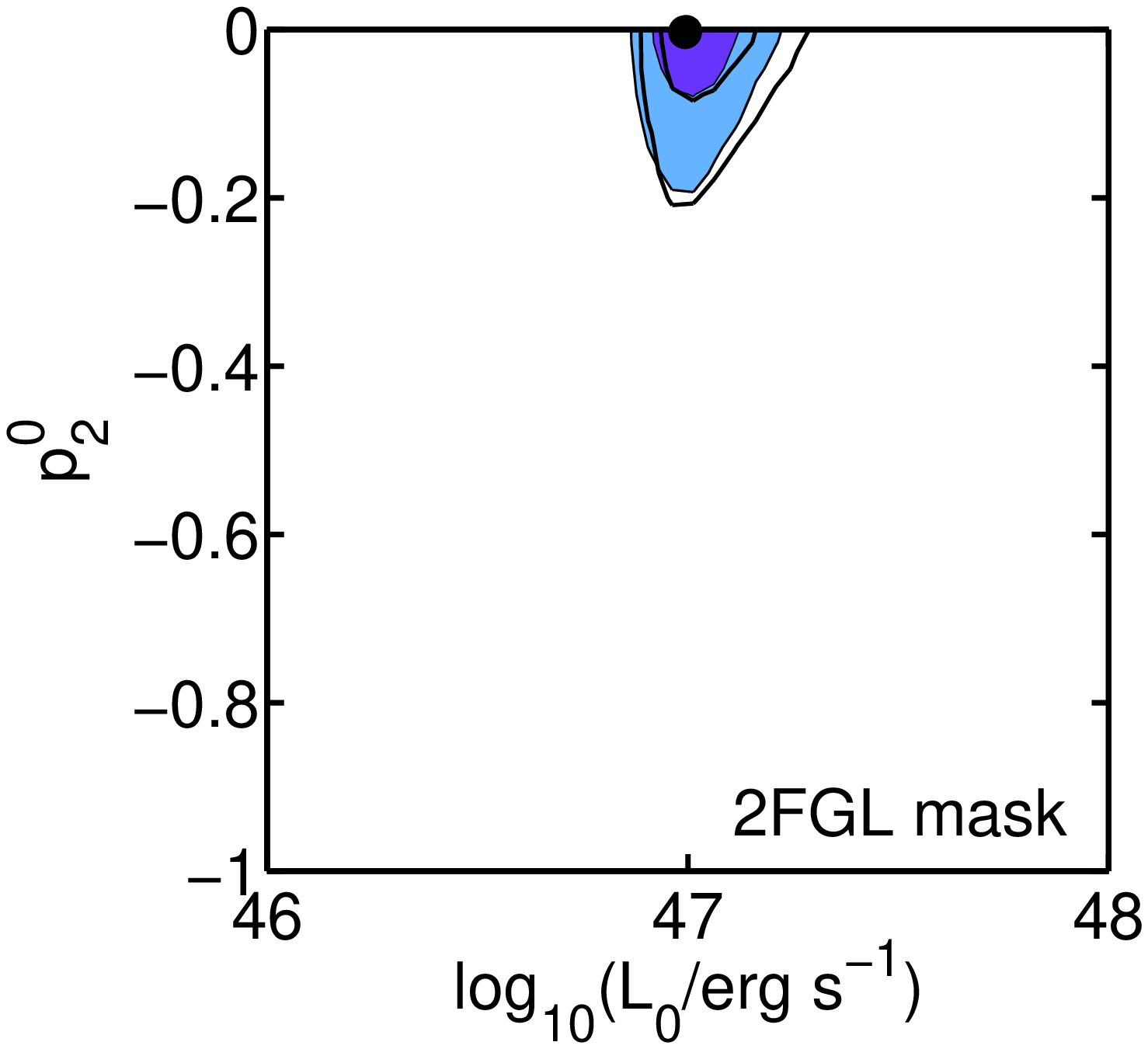}
\caption{\label{fig:2D_2FGL_onlyblazars} Two-dimensional profile-likelihood contour plots for all the combinations of parameters $A$, $\gamma_1$, $L_0$ and $p_2^0$. Filled contours and filled black dots refer to the scans performed by fitting only the auto- and cross-correlation APS from Ref.~\cite{Fornasa:2016ohl} in the case of the 2FGL mask, while empty contours and empty circles are for the fits to the APS and the source count distribution $dN/dF$ from the 3FGL catalogue. Inner contours mark the 68\% CL region and outer ones the 95\% CL region. The model predictions in the scans are computed including only blazars.}
\end{figure*}

\begin{figure*}
\includegraphics[width=0.32\textwidth]{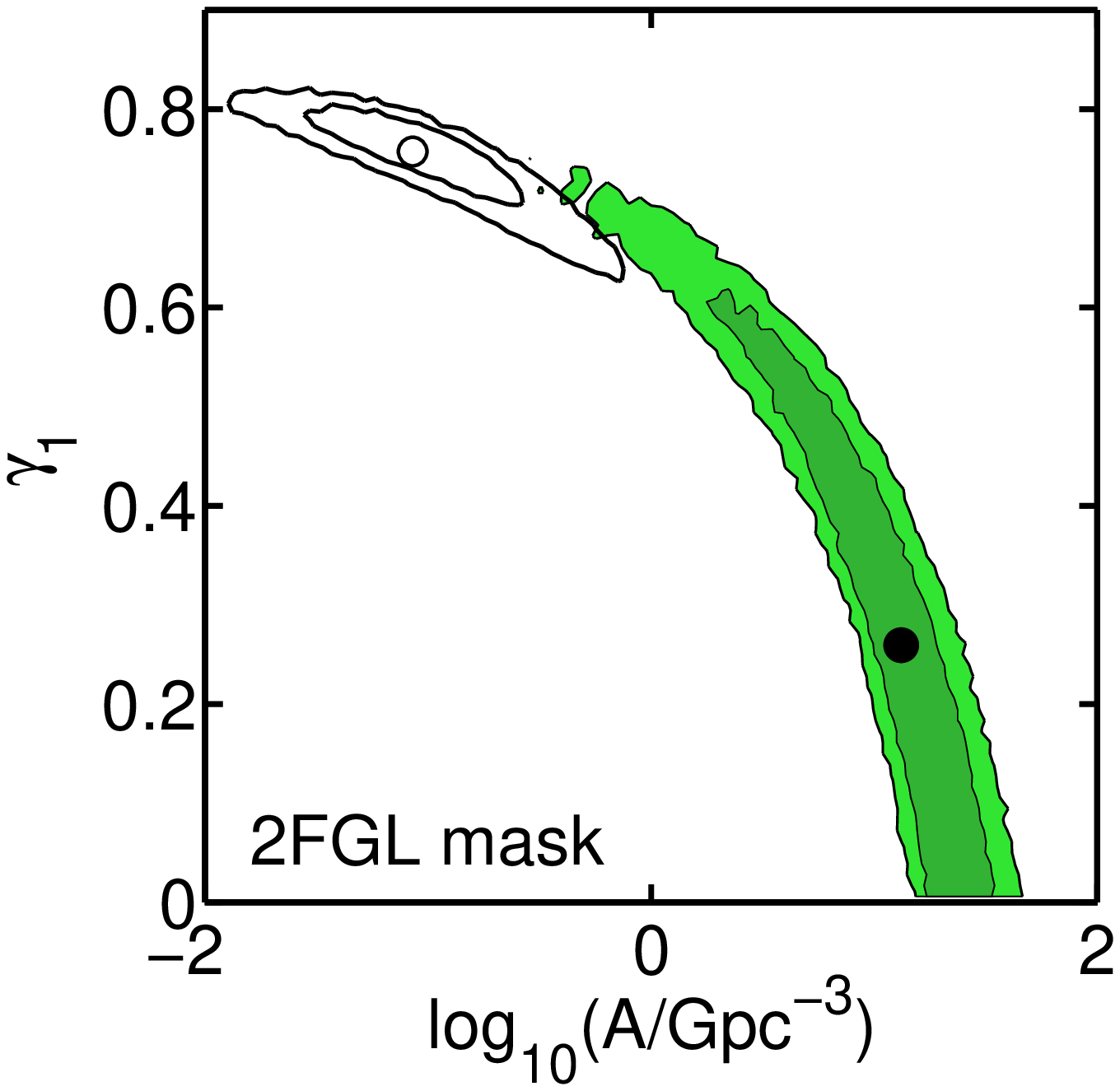}
\includegraphics[width=0.32\textwidth]{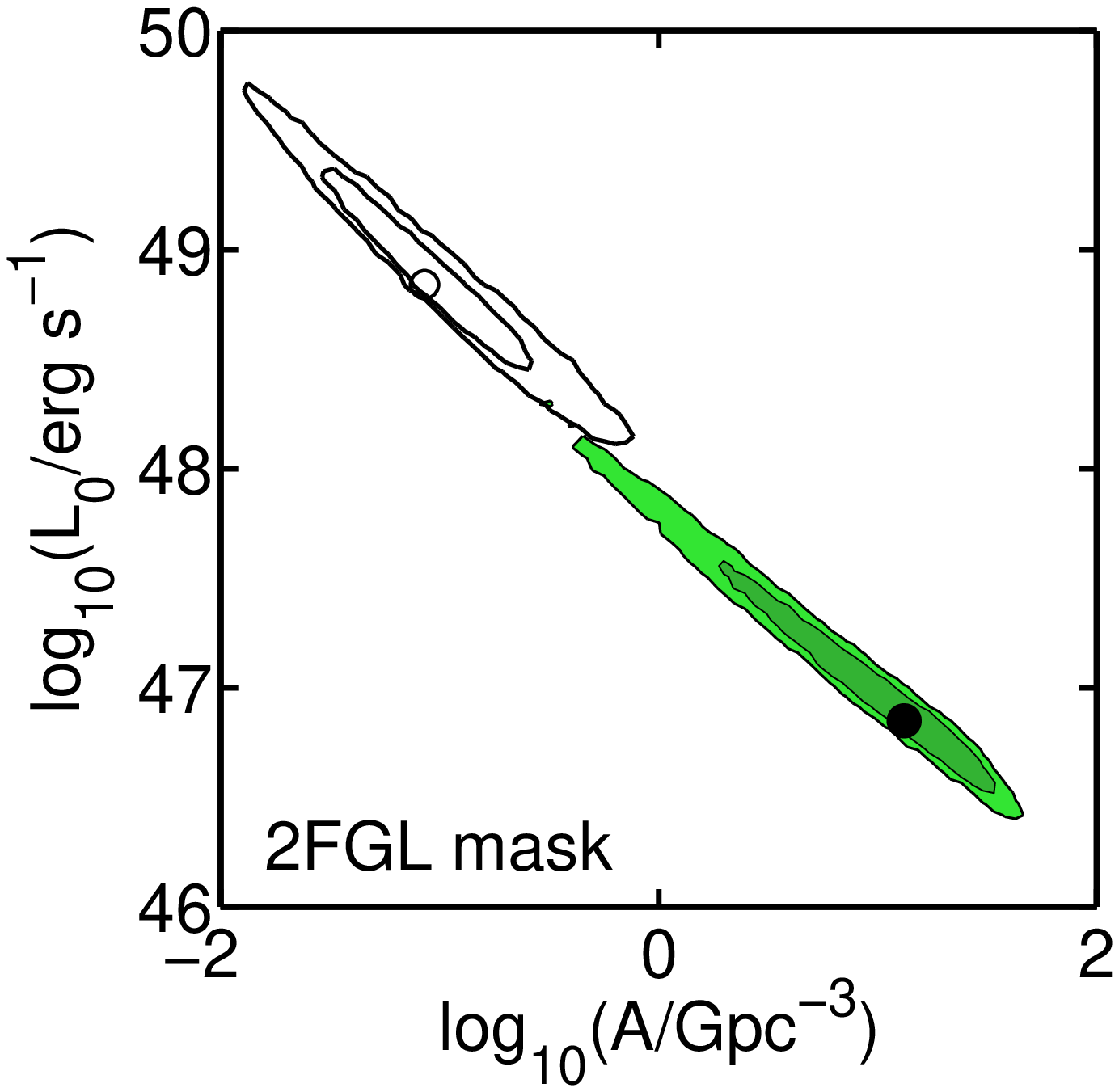}
\includegraphics[width=0.32\textwidth]{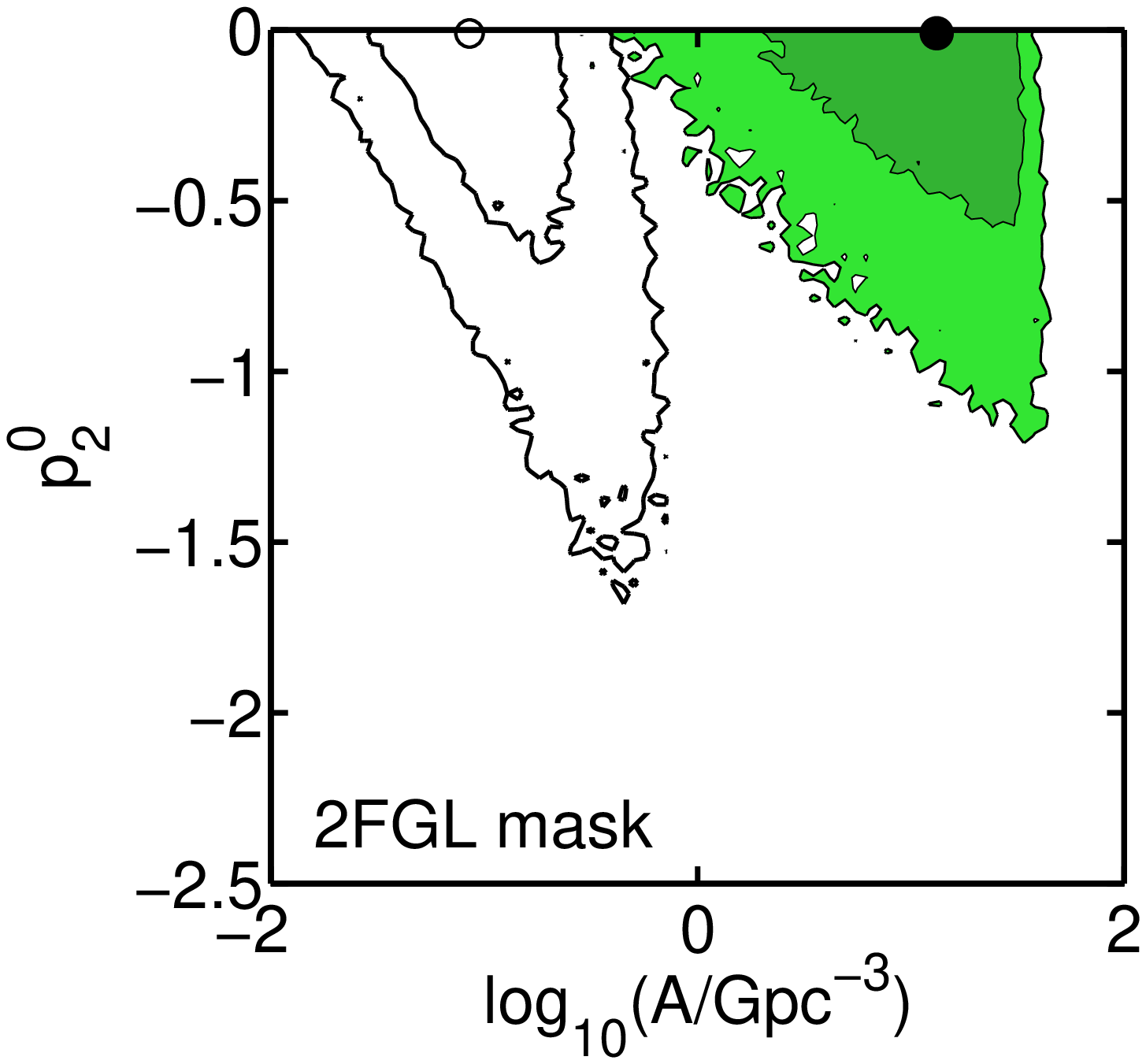}
\includegraphics[width=0.32\textwidth]{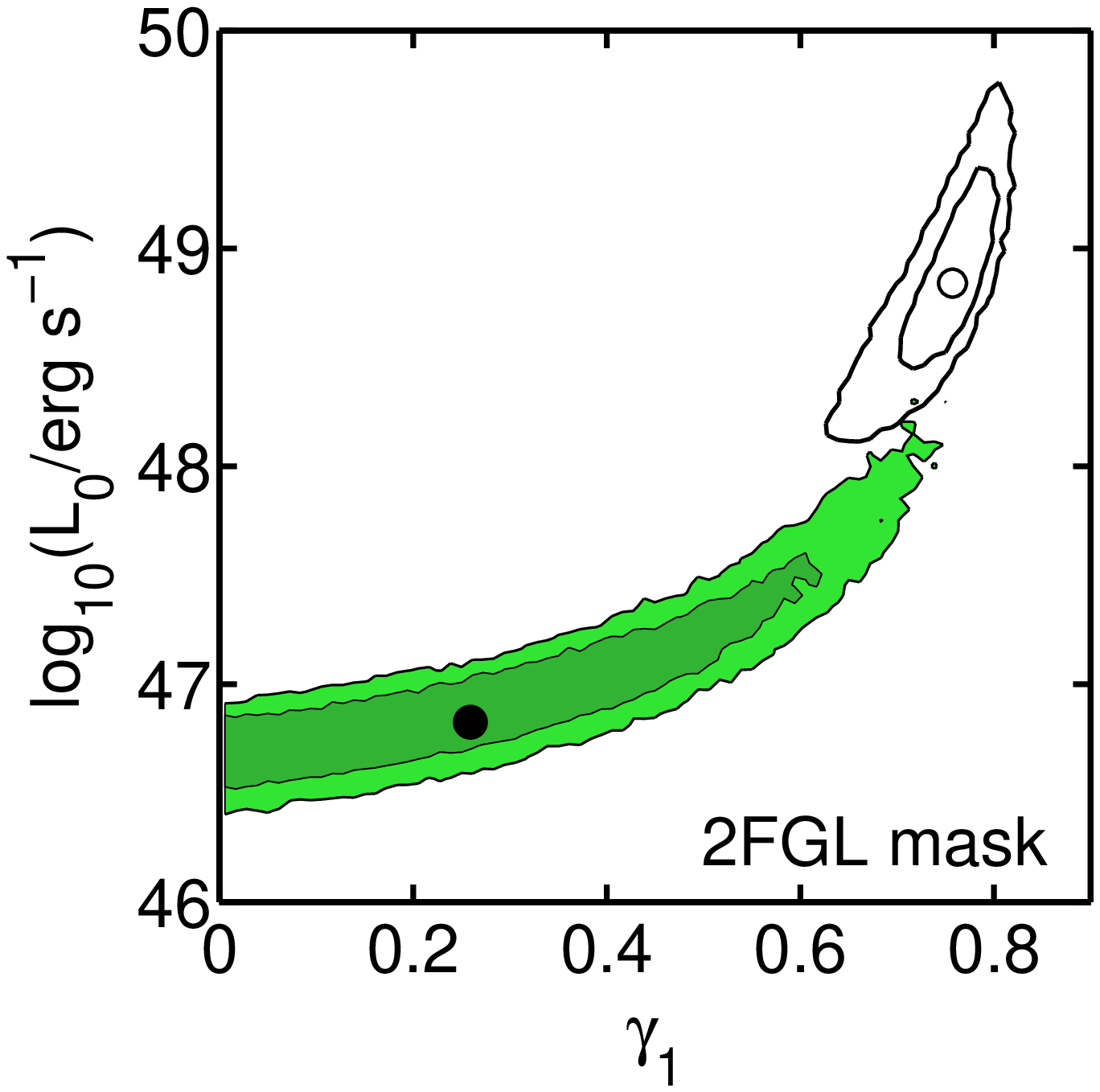}
\includegraphics[width=0.32\textwidth]{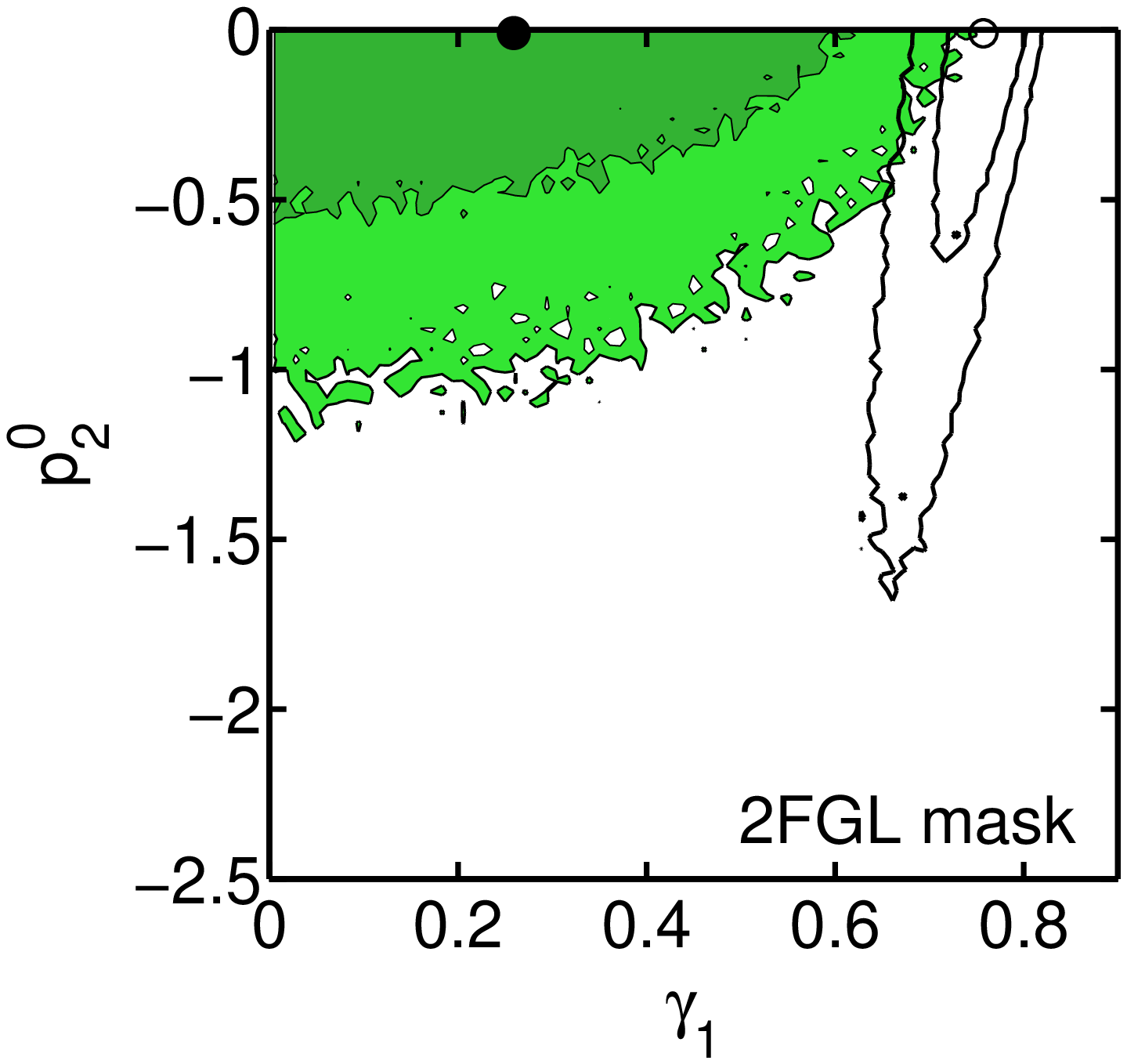}
\includegraphics[width=0.32\textwidth]{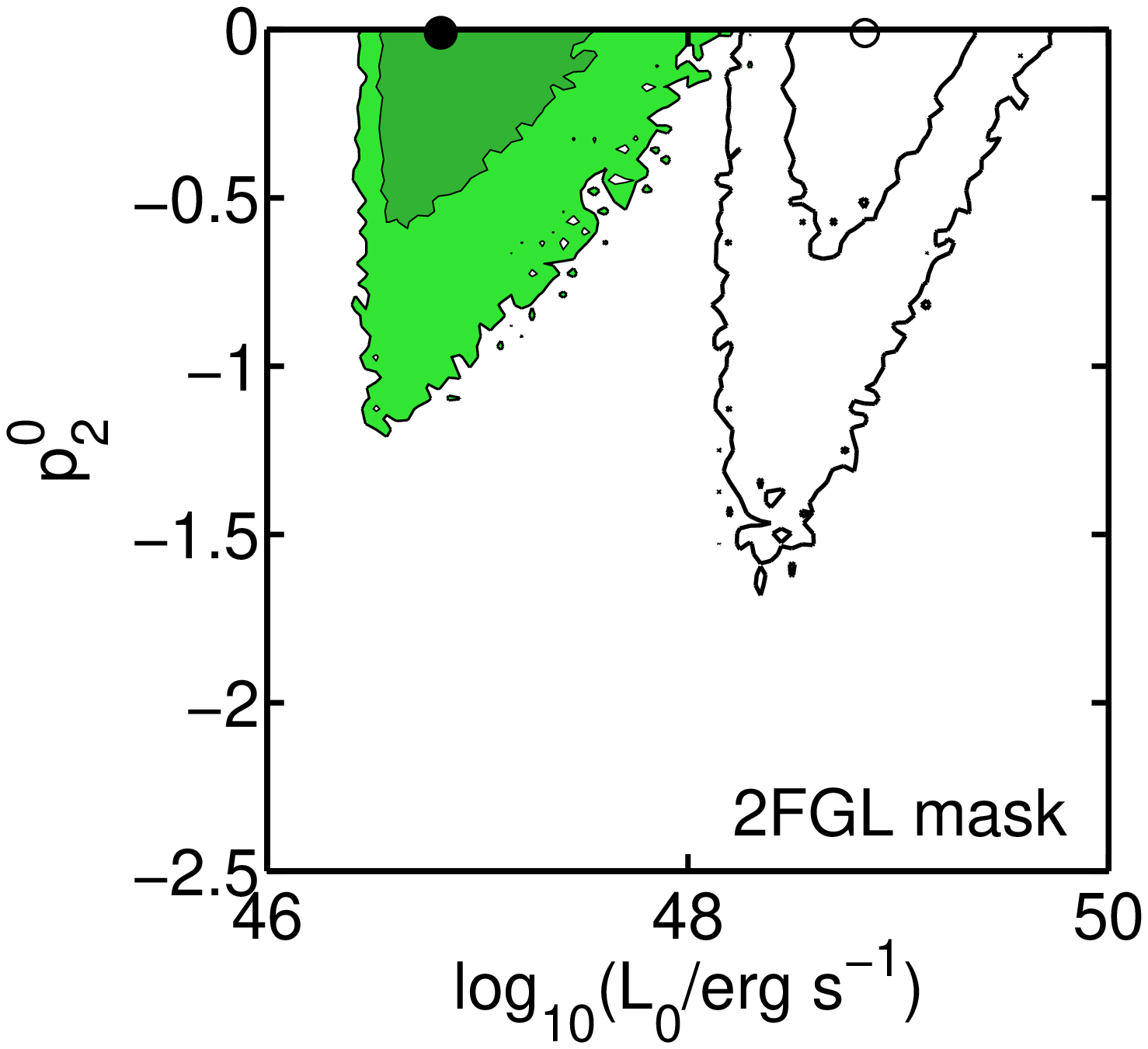}
\includegraphics[width=0.32\textwidth]{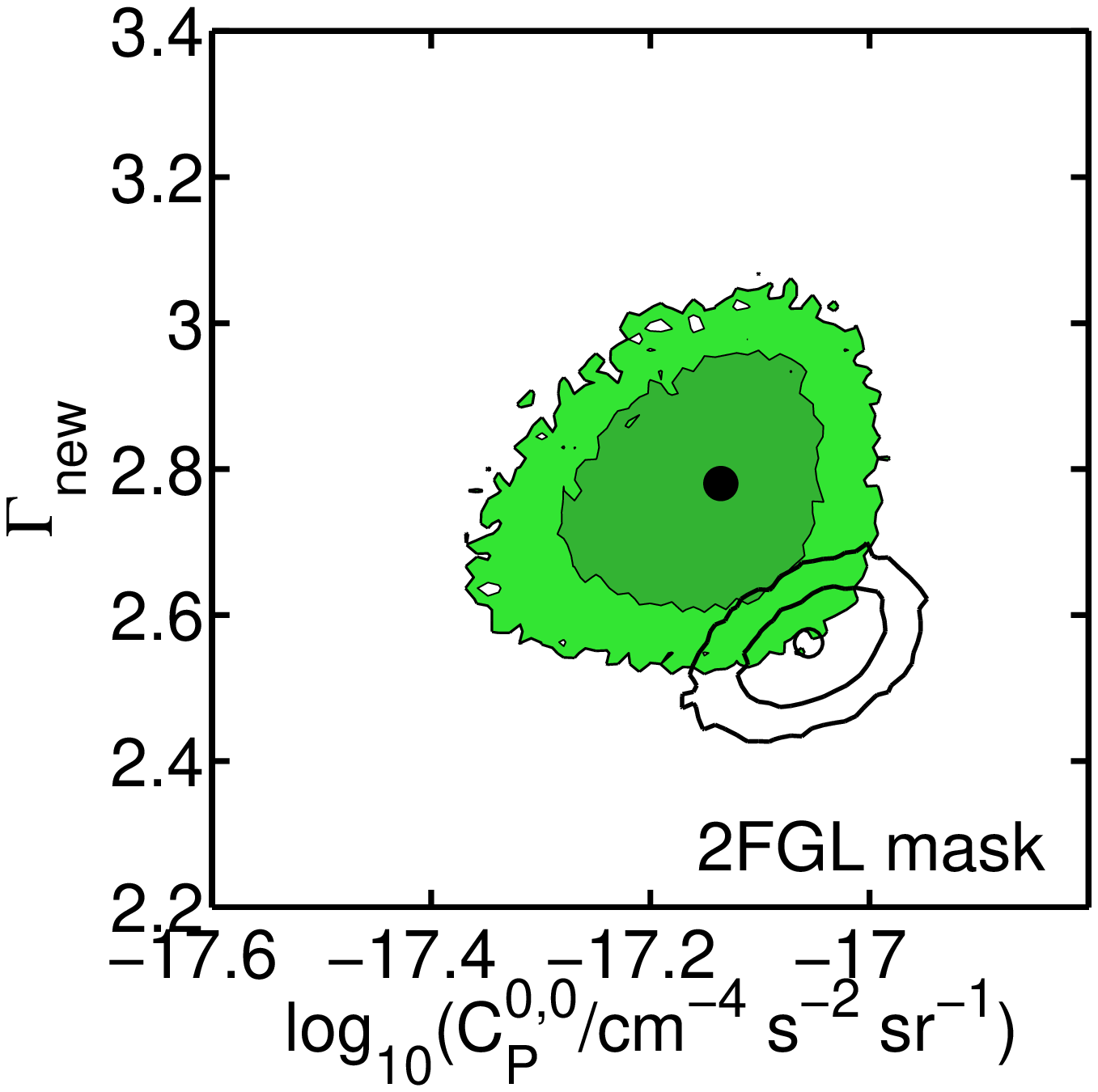}
\caption{\label{fig:2D_2FGL_new_class} Two-dimensional profile-likelihood contour plots for some combinations of parameters $A$, $\gamma_1$, $L_0$, $p_2^0$, $C_{\rm P}^{0,0}$ and $\Gamma_{\rm new}$. Filled contours and filled black dots refer to the scans performed by fitting only the auto- and cross-correlation APS from Ref.~\cite{Fornasa:2016ohl} in the case of the 2FGL mask, while empty contours and empty circles are for the fits to the APS and the source count distribution $dN/dF$ from the 3FGL catalogue. Inner contours mark the 68\% CL region and outer ones the 95\% CL region. The model predictions in the scans are computed with blazars and a new class of gamma-ray emitters.}
\end{figure*}

\begin{figure*}
\includegraphics[width=0.49\textwidth]{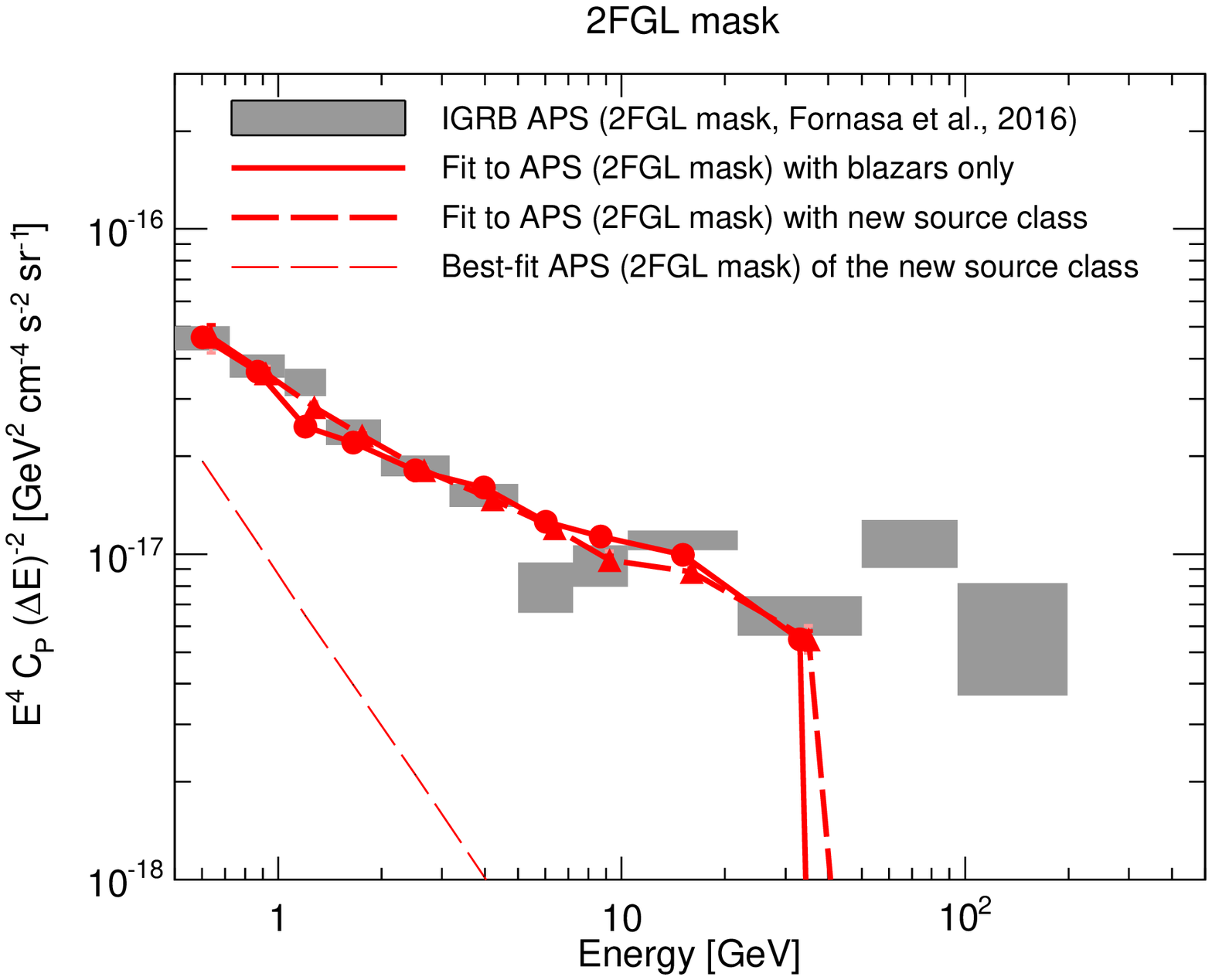}
\includegraphics[width=0.49\textwidth]{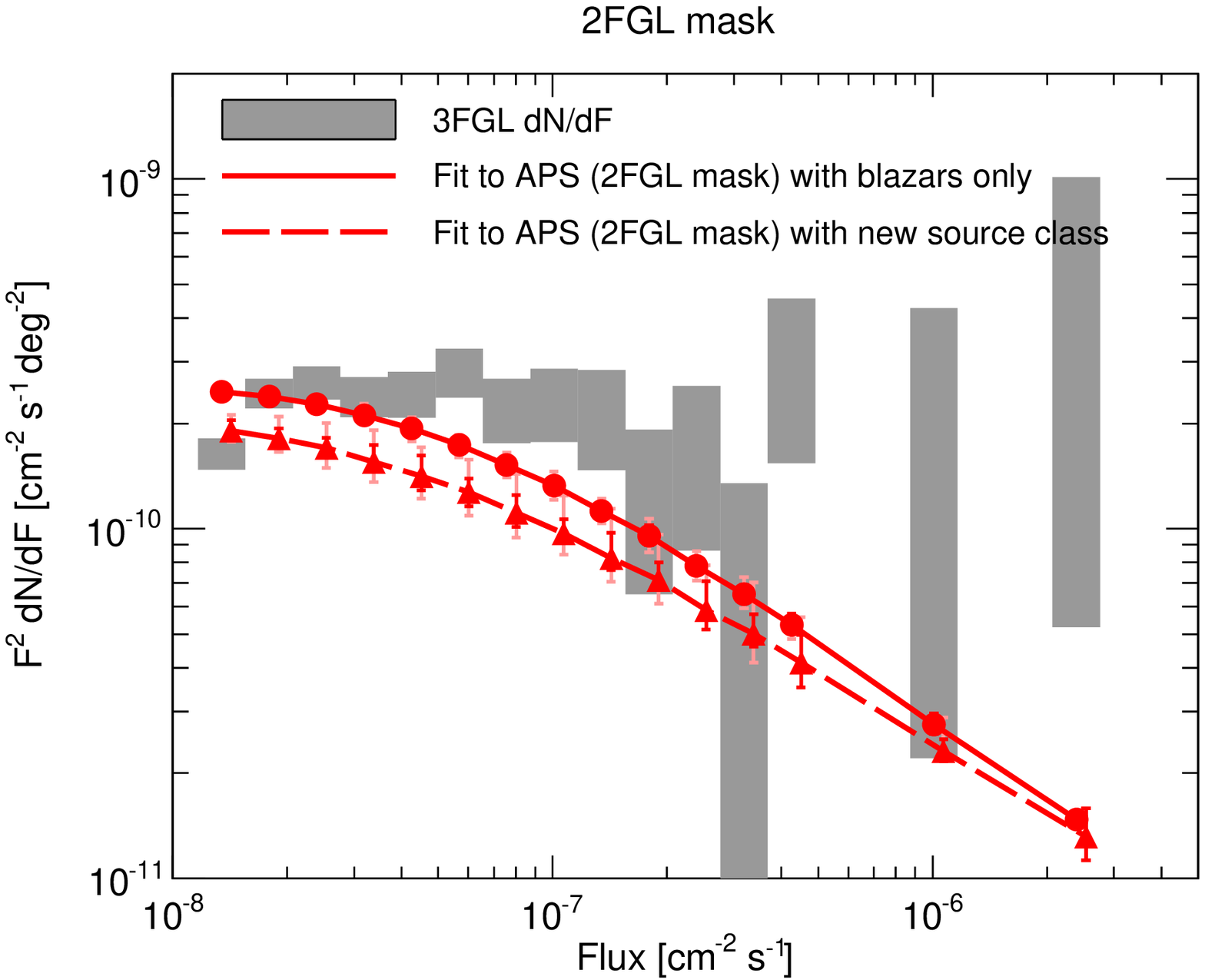}
\caption{\label{fig:bestfit_APS_dNdF_2FGL_onlyAPS} The gray boxes indicate the data points used in the scans, for the auto-correlation APS with the 2FGL mask (left panel) and the 3FGL $dN/dF$ (right panel). The solid red lines and red circles indicate the best-fit solution for the scan performed by fitting only the auto- and cross-correlation APS from Ref.~\cite{Fornasa:2016ohl} (2FGL mask) with a model including only blazars. The thicker dashed red lines and the red triangles denote the best-fit solutions for a scan fitting the same APS data but including a new source class (see text for details). The thinner dashed red line shows the auto-correlation APS of the new source class, separately. Around each red circle/triangle, the red (pink) vertical line shows the 68\% (95\%) CL uncertainty. Circles and triangles are slightly shifted with respect to each other to increase readibility.}
\end{figure*}

\begin{figure*}
\includegraphics[width=0.49\textwidth]{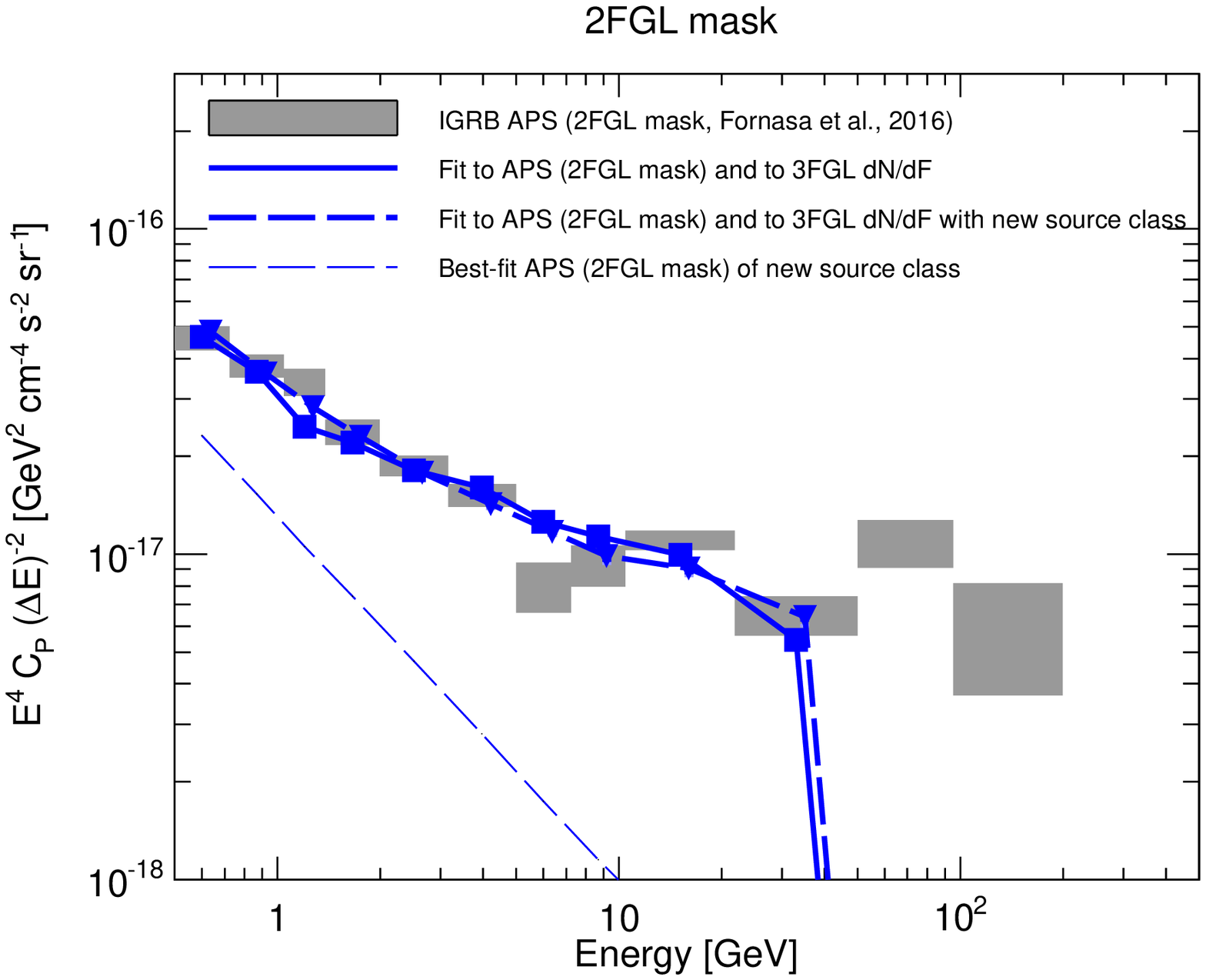}
\includegraphics[width=0.49\textwidth]{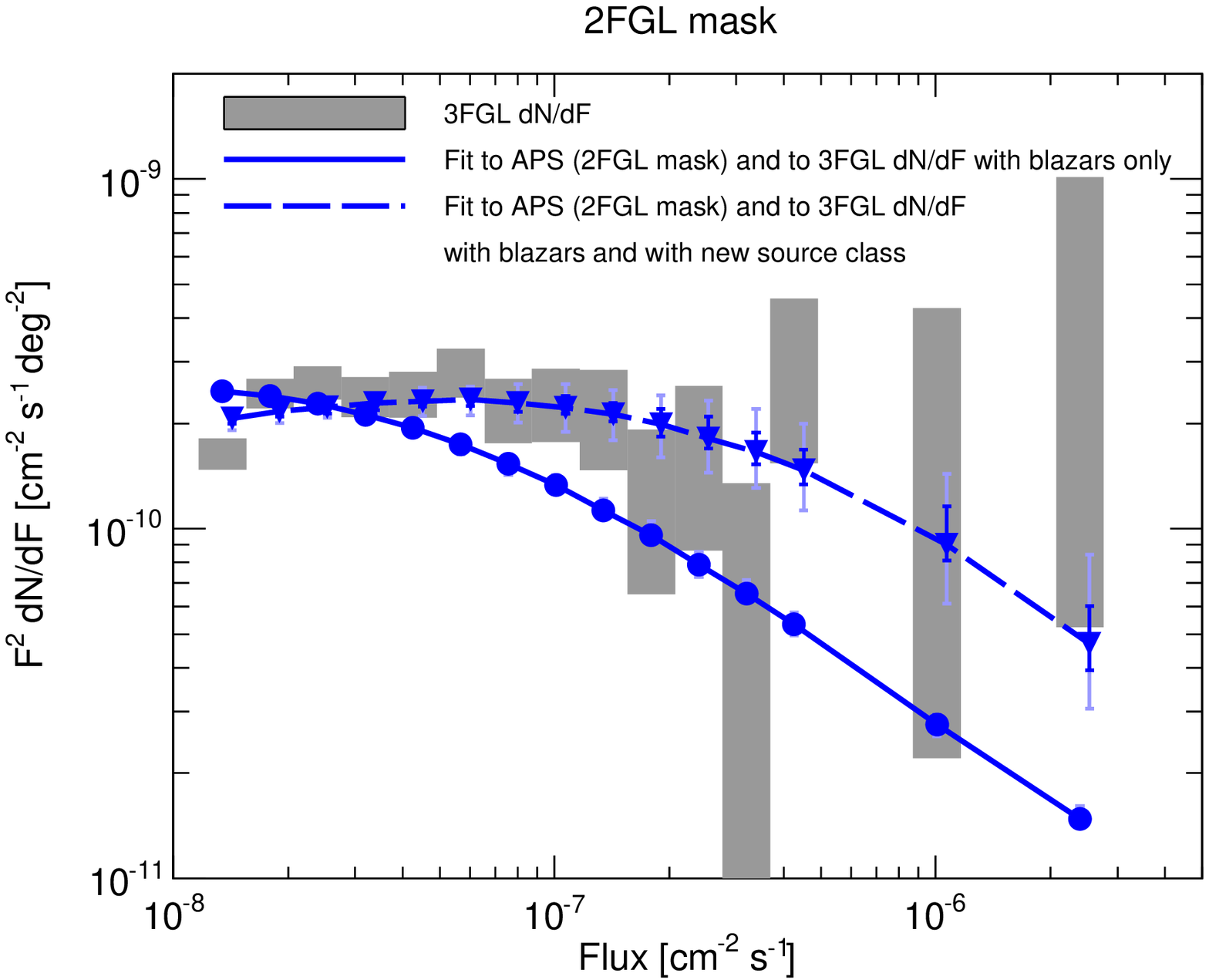}
\caption{\label{fig:bestfit_APS_dNdF_2FGL_APS_and_dNdF} The gray boxes indicate the data points used in the scans, for the auto-correlation APS with the 2FGL mask (left panel) and the 3FGL $dN/dF$ (right panel). The solid blue lines and blue squares indicate the best-fit solution for the scan performed by fitting the auto- and cross-correlation APS from Ref.~\cite{Fornasa:2016ohl} (2FGL mask) and the 3FGL $dN/dF$, with a model including only blazars. The thicker dashed blue lines and the blue triangles denote the best-fit solution for a scan fitting the same data but including a new source class (see text for details). The thinner dashed blue line shows the auto-correlation APS of the new class, separately. Around each blue square/triangle, the blue (light blue) vertical line shows the 68\% (95\%) CL uncertainty. Squares and triangles are slightly shifted with respect to each other to increase readibility.}
\end{figure*}

\subsection{Results for the 2FGL mask}
\label{sec:2FGL}
We repeat the scans discussed in the previous section but we now consider the
auto- and cross-APS obtained with the 2FGL mask. The solid red lines and
red circles in the first four panels of Fig.~\ref{fig:1D_2FGL} show the 
one-dimensional PL for the model parameters, in the case that the APS data are 
fitted only in terms of blazars (no new source class). Also, the blue
regions in Fig.~\ref{fig:2D_2FGL_onlyblazars} indicate their two-dimensional PL.
The contours approximately overlap with the red regions in 
Fig.~\ref{fig:2D_3FGL_onlyblazars}, confirming that the APS data point towards
the same source population, independent of the mask used. Indeed, for each
of the four model parameters, the best-fit solutions obtained by using the two
different masks (fitting only the APS and without additional source class) are 
in agreement with each other at 68\% CL. The main difference between 
Figs.~\ref{fig:2D_2FGL_onlyblazars} and \ref{fig:2D_3FGL_onlyblazars} is 
the size of the contours: for the 2FGL mask, the measurement of the auto- and 
cross-APS is characterized by smaller error bars than for the 3FGL mask and, 
therefore, a smaller portion of the parameter space can fit the data. 
Fig.~\ref{fig:bestfit_APS_dNdF_2FGL_onlyAPS} compares the APS data (left 
panel) and the 3FGL $dN/dF$ (right panel) with the best-fit solution (solid 
red line and red circles). Differently than in
Fig.~\ref{fig:bestfit_APS_dNdF_3FGL_onlyAPS}, there is no underestimation of 
the APS at low energies but few discrepancies (especially for the data point 
around 1~GeV and around 6~GeV) yield a best-fit $\chi^2$ of 117.24, with a 
$\chi^2$ per degree of freedom of 2.30, corresponding to a $p$-value of $4 
\times 10^{-7}$.\footnote{Allowing $p_2^0$ to be positive would improve the 
quality of the fit.}

The situation remains qualitatively unchanged if we include the 3FGL $dN/dF$
in the fit. The solid blue lines and blue squares in Fig.~\ref{fig:1D_2FGL} 
show the one-dimensional PL of the free parameters, while the empty contours in
Fig.~\ref{fig:2D_2FGL_onlyblazars} denote the two-dimensional PL. They follow
quite closely the case with only the APS data, apart from the PL of $\gamma_1$
that extends until 0.25 at 95\% CL. Thus, solid blue lines in 
Fig.~\ref{fig:bestfit_APS_dNdF_2FGL_APS_and_dNdF} are also similar to the
solid red lines in Fig.~\ref{fig:bestfit_APS_dNdF_2FGL_onlyAPS}, with no 
underestimate of the auto-correlation APS below 1~GeV but a systematic 
underestimate of the $dN/dF$ above 
3--$4 \times 10^{-8}~\mbox{cm}^{-2}~\mbox{s}^{-1}$. This indicates that the 
likelihood is dominated by the APS measurement. The $\chi^2$ of the best-fit 
solution is 166.44, corresponding to a $\chi^2$ per degree of freedom of 2.52 
and a $p$-value of $10^{-10}$.

As in the previous section, we extend our model by including an additional
source class, parametrized, as in Eq.~(\ref{eqn:APS_newpop}) by $C_{\rm P}^{0,0}$
and $\Gamma_{\rm new}$, for which we consider the same prior ranges as before.
The one-dimensional PL of the free parameters is shown in 
Fig.~\ref{fig:1D_2FGL} by dashed lines and triangles. Red lines are for the
fit to only the auto- and cross-APS and blue lines for the fit including the
3FGL $dN/dF$. The two-dimensional PL is shown in 
Fig.~\ref{fig:2D_2FGL_new_class} by the green regions (fit to APS data only)
and by the empty contours (fit to the APS and 3FGL $dN/dF$ data). The full
and empty circles indicate the best-fit points, respectively. As in the
previous section, the size of the contours increases with respect to 
Fig.~\ref{fig:2D_2FGL_onlyblazars} and they include the preferred regions 
in Fig.~\ref{fig:2D_2FGL_onlyblazars}. They are also in qualitative agreement 
with Fig.~\ref{fig:2D_3FGL_new_class}, confirming that the data sets 
corresponding to the two masks point towards the same blazar population. When 
the $dN/dF$ data are included (blue dashed lines in Fig.~\ref{fig:1D_2FGL} and 
empty contours in Fig.~\ref{fig:2D_2FGL_new_class}) there is a shift of the 
preferred regions to lower values of $A$ (and thus, larger values of 
$\gamma_1$). However, the two sets of contours are located along the same 
degeneracy. These regions correspond to solutions that are in a better 
agreement with the 3FGL $dN/dF$, as it can be seen by the blue dashed lines 
in Fig.~\ref{fig:bestfit_APS_dNdF_2FGL_APS_and_dNdF}: compared to the case 
with blazars only (blue solid lines in 
Fig.~\ref{fig:bestfit_APS_dNdF_2FGL_APS_and_dNdF}), the better agreement with
the APS data provided by the new source class increases the weight of the 
$dN/dF$ data in the likelihood. Thus, the scan is strongly driven towards 
configurations that also provide a good description to the source count 
distribution. 

When fitting the APS data alone 
(Fig.~\ref{fig:bestfit_APS_dNdF_2FGL_onlyAPS}), the best-fit $\chi^2$ is 74.51 
(best-fit $\chi^2$ per degree of freedom of 1.46 and $p$-value of 0.02). 
The likelihood-ratio test yields a $p$-value of $2 \times 10^{-10}$, 
corresponding to more than $10\sigma$ evidence in favour of the new source 
class. The strong preference is confirmed also within a Bayesian framework, 
with a $\ln B$ of 20.93. When we include the 3FGL $dN/dF$ data in the fit 
(Fig.~\ref{fig:bestfit_APS_dNdF_2FGL_APS_and_dNdF}), the best-fit $\chi^2$ is 
97.78, corresponding to a best-fit $\chi^2$ per degree of freedom of 1.53 and 
to a $p$-value of $4 \times 10^{-3}$. The $p$-value of the likelihood-ratio
test is $4 \times 10^{-16}$ and the Bayes factor $\ln B$ is 31.20. Both 
approaches strongly favour the presence of the new class.

\begin{figure*}
\includegraphics[width=0.49\textwidth]{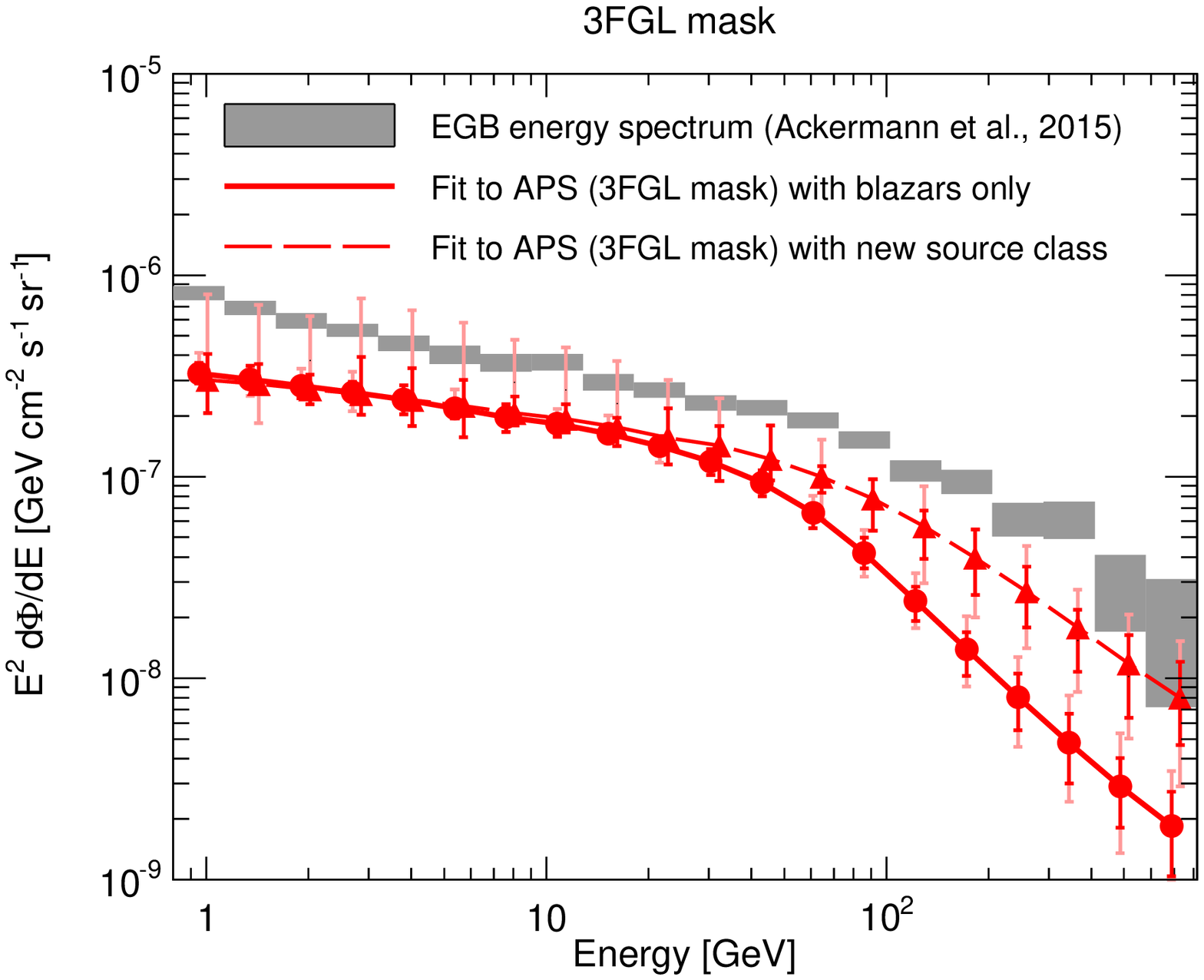}
\includegraphics[width=0.49\textwidth]{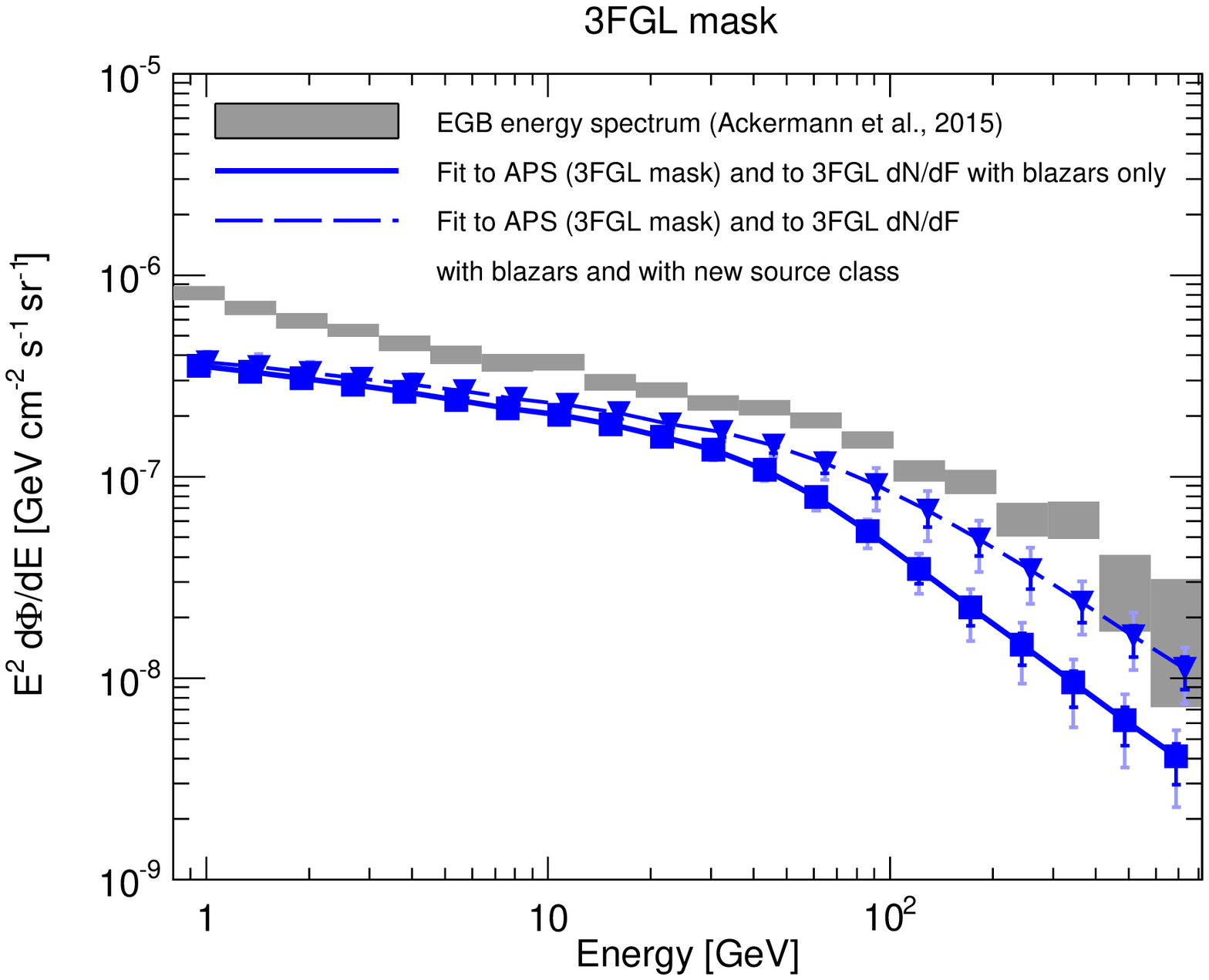}
\includegraphics[width=0.49\textwidth]{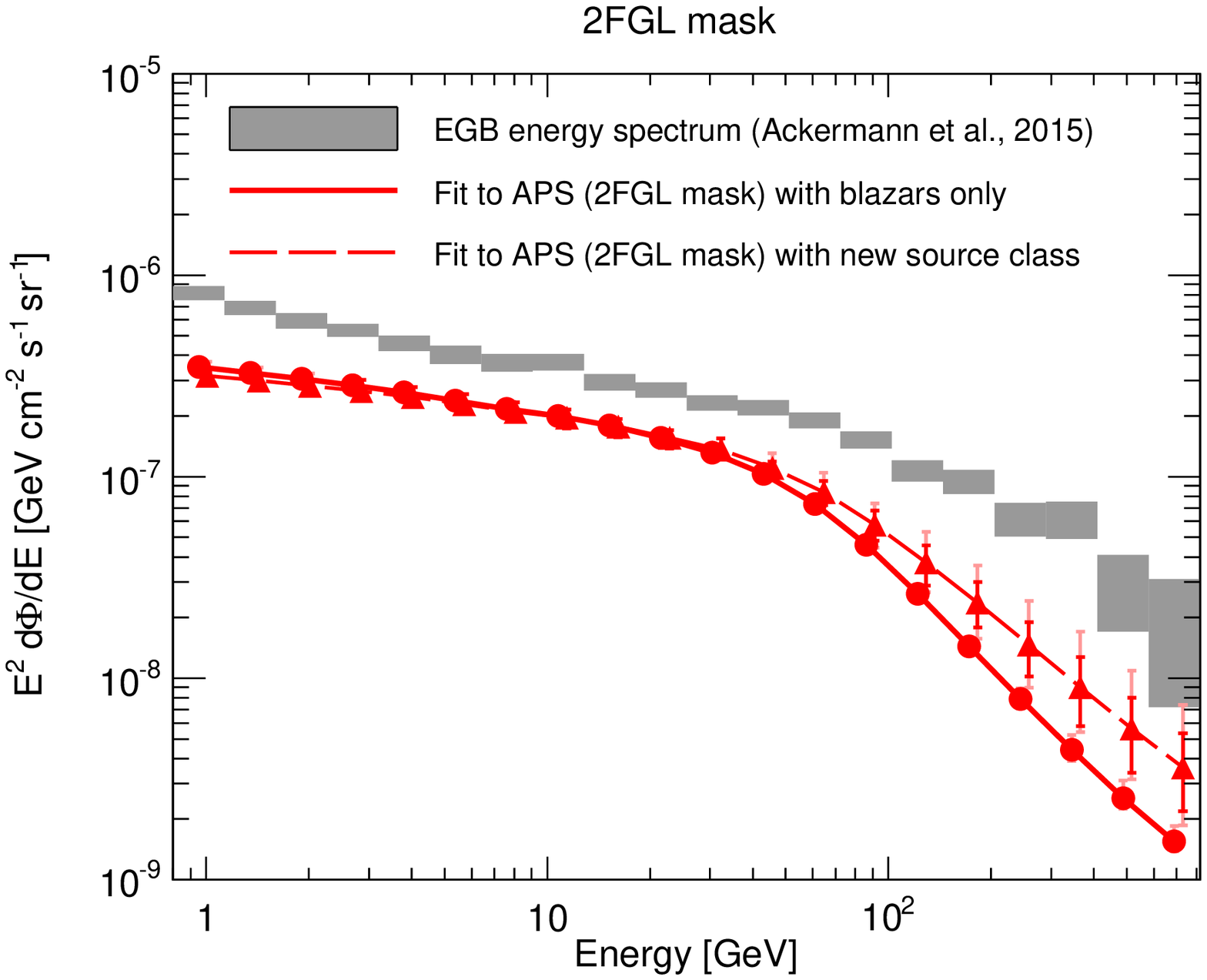}
\includegraphics[width=0.49\textwidth]{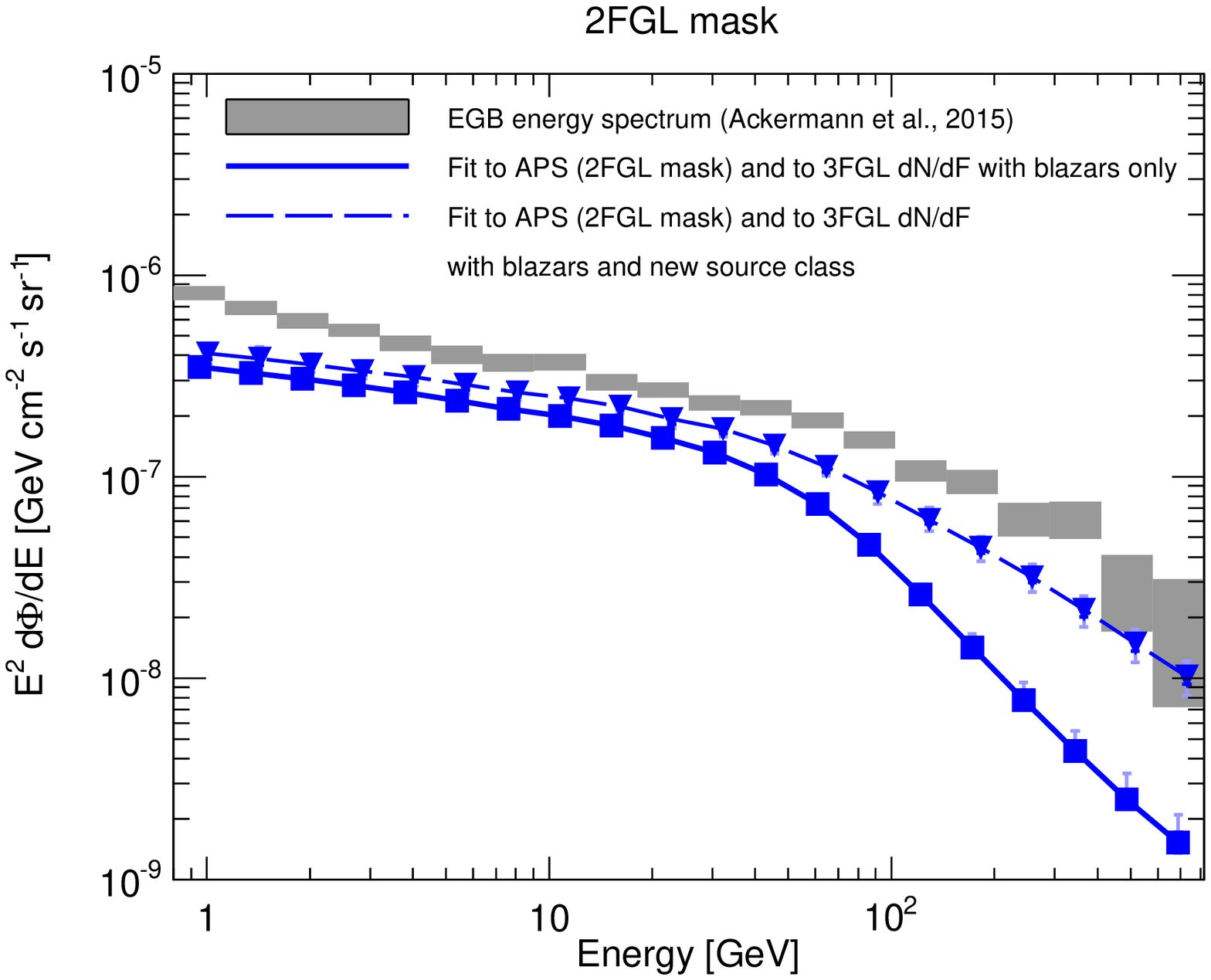}
\caption{\label{fig:bestfit_EGB} The gray boxes indicate the EGB intensity energy spectrum measured in Ref.~\cite{Ackermann:2014usa} (model A). (Top-left panel): The solid red line and red circles indicate the energy spectrum of the emission of all blazars (resolved and unresolved) for the best-fit solutions for the scan performed by fitting the auto- and cross-correlation APS from Ref.~\cite{Fornasa:2016ohl} (3FGL mask). The dashed red line and the red triangles denote the best-fit solutions for a scan performed fitting the same data but including a new source class (see text for details). (Top-right panel): The same quantities are plotted, as in the top-left panel, but for scans fitting the APS data (3FGL mask) and the 3FGL $dN/dF$. (Bottom panels): the same information as in the in the top panels is plotted, but for the case of the APS obtained with the 2FGL mask. In all panels, the red/blue bars around the markers indicate the 68\% CL uncertainty, while the pink/light blue ones the 95\% CL one. The two sets of data in each panel are slightly shifted with respect to each other to increase readibility.}
\end{figure*}

Note that the best-fit values of $C_{\rm P}^{0,0}$ and $\Gamma_{\rm new}$ (last
panel of Fig.~\ref{fig:2D_2FGL_new_class}) are in agreement (at 95\% CL) with 
the values obtained when fitting the APS with the 3FGL mask (with and without 
the 3FGL $dN/dF$ data, see Fig.~\ref{fig:2D_3FGL_new_params}). Indeed, the 
auto-correlation APS associated with the new source class (thin dash line in 
Fig.~\ref{fig:bestfit_APS_dNdF_3FGL_onlyAPS}, 
\ref{fig:bestfit_APS_dNdF_3FGL_APS_and_dNdF}, 
\ref{fig:bestfit_APS_dNdF_2FGL_onlyAPS} and 
\ref{fig:bestfit_APS_dNdF_2FGL_APS_and_dNdF}) is very similar independently
of the mask used. This might suggest that there are almost no members of
the new source class with fluxes between the sensitivities of the 3FGL and
of 2FGL catalogues.

\section{Discussion}
\label{sec:discussion}

\subsection{Comparison with the EGB intensity energy spectrum}
The gray boxes in Fig.~\ref{fig:bestfit_EGB} denote the EGB intensity energy 
spectrum from Ref.~\cite{Ackermann:2014usa} (model A). We compare it with
the energy spectrum of the emission produced by all blazars (resolved and
unresolved) according to the results of our scans. Top panels refer to the 
fits to the APS with the 3FGL mask (see Secs.~\ref{sec:APS_only_blazars_only}, 
\ref{sec:APS_and_dNdF_blazars_only}, and \ref{sec:with_new_source_class}), 
while the botton ones to the case of the 2FGL mask (see Sec.~\ref{sec:2FGL}). 
In the left panels we show results for fits to the APS data only, while in the
right ones we also include the 3FGL $dN/dF$. Each panel contains two data 
sets, one for the case in which only blazars are considered in the model 
predictions (solid lines) and one for the fit including the new source class 
(dashed lines). Each data point is surrounded by its 68\% CL (red or blue) and 
95\% CL (pink or light blue) estimated error. Note that, even when the new 
source class is included in the fit, Fig.~\ref{fig:bestfit_EGB} only shows the 
emission of blazars.

\begin{figure*}
\includegraphics[width=0.49\textwidth]{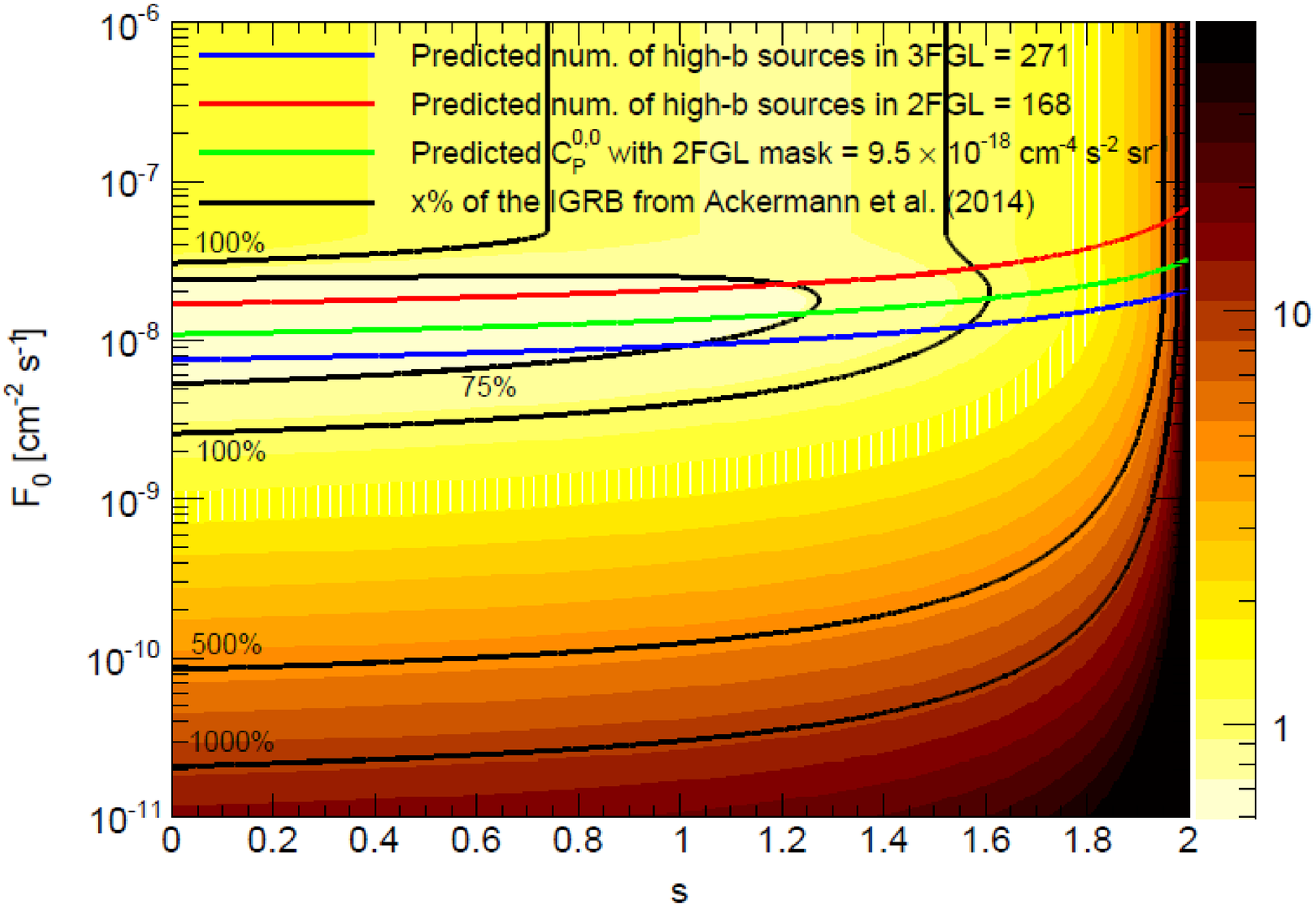}
\includegraphics[width=0.45\textwidth]{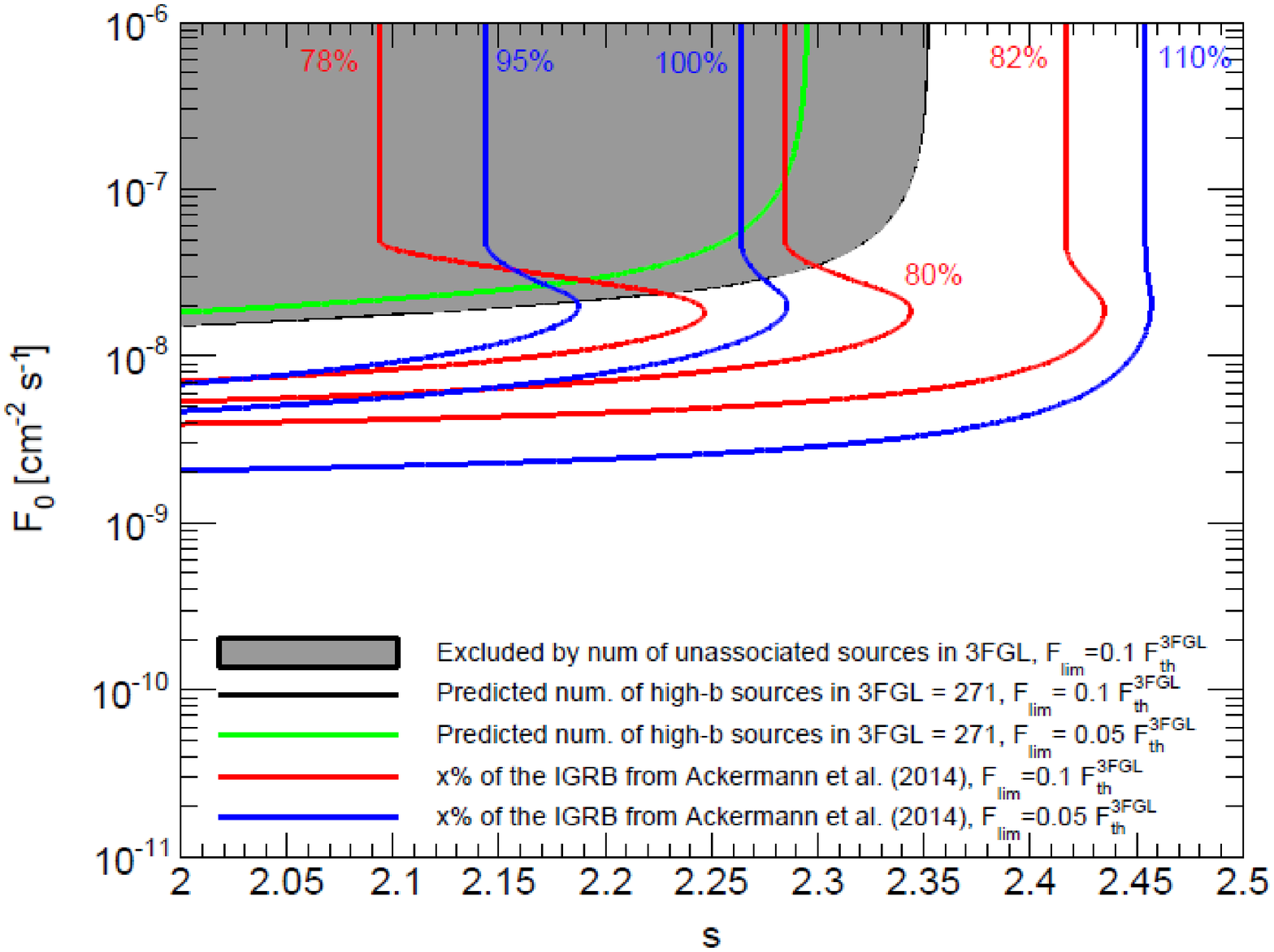}
\caption{\label{fig:new_population} (Left panel:) The colors indicate what fraction of the measured IGRB intensity can be ascribed to the new source class (above 100 MeV). The black lines mark specific levels, indicated in the labels. The blue (red) blue line shows what combination of $s$ and $F_0$ correspond to a number of sources equal to 271 (168) above the 3FGL (2FGL) sensitivity and for $|b|>30^\circ$. The green line indicates where the new source class predicts a $C_{\rm P}^{0,0}$ of $9.5 \times 10^{-18}~\mbox{cm}^{-4}~\mbox{s}^{-2}~\mbox{sr}^{-1}$, below the 2FGL threshold. (Right panel:) The red (blue) lines indicate what combinations of $s$ and $F_0$ correspond to specific fractions of the measured IGRB intensity above 100 MeV, for a $F_{\rm lim}=10\%$ ($F_{\rm lim}=5\%$) of the 3FGL sensitivity. The gray region is excluded because the new source class would overproduce the number of unassociated high-latitude emitters in the 3FGL catalogue. The black line denotes where we would have 271 sources in the 3FGL catalogue (with $|b|>30^\circ$) and the green one where we would have 168 sources in the 2FGL one (with $|b|>30^\circ$).}
\end{figure*}

As pointed out in the previous sections, all the scans performed in terms of 
blazars only (no new sources) are in qualitative agreement (independent of 
which APS data are used or if the $dN/dF$ data are included). Indeed, the four 
solid lines in Fig.~\ref{fig:bestfit_EGB} are very similar. Blazars are 
responsible for a fraction of the EGB above 800~MeV that goes from 45\% to 
49\%.

When we include the new source class (dashed lines), the emission is harder
and dashed lines deviate from the solid ones around 80-100~GeV. This is 
probably because, since the new class account for the low-energy regime, 
blazars are more tuned to reproduce the behaviour at high energies. In the 
best-fit scenarios, blazars account for between 43\% and 57\% of the EGB above 
800~MeV.

\subsection{Characterization of the new source class}
We start by noting that a slope $\Gamma_{\rm new}$ between approximately 2.5 
and 3.2 (depending on the scan considered) is too soft to be compatible
with the observed spectrum of star-forming galaxies~\cite{Ackermann:2012vca,
Linden:2016fdd} (at least at $\sim 1$ GeV) and of misaligned active galactic 
nuclei~\cite{DiMauro:2013xta,Hooper:2016gjy} (see, however, 
Ref.~\cite{Brown:2016sbl}). Also, according to Ref.~\cite{DiMauro:2014wha}, 
these two classes of gamma-ray emitters are not expected to give rise to APS 
as large as the one measured in Ref.~\cite{Fornasa:2016ohl} below 1 GeV.

Approximately one third of the sources in the 3FGL catalogue are 
unassociated~\cite{Ackermann:2015yfk}. These emitters are good candidates to 
play the role of the new source class uncovered in our analysis. However, 
these unassociated sources exhibit a harder energy spectra than what we find 
in our scan. The average slope for the 271 unassociated gamma-ray sources in 
the 3FGL catalogue detected at $|b|>30^\circ$ is 2.26.

The inferred energy spectrum of the new source class could be in agreement
with the gamma-ray emission expected from the annihilation or decay of Dark 
Matter in Galactic and/or extragalactic halos and subhalos. For example, a 
Dark Matter candidate with a mass of the order of few GeV and annihilating 
into $b$ quarks would give rise to the desired energy spectrum. Many works in 
the literature have estimated the level of anisotropies expected for the 
gamma-ray emission induced by Dark Matter \cite{Ando:2005xg,Ando:2006cr,
Ando:2006mt,Cuoco:2007sh,SiegalGaskins:2008ge,SiegalGaskins:2009ux,
Zavala:2009zr,Cuoco:2010jb,Fornasa:2012gu,Ando:2013ff,Fornasa:2016ohl}. 
Ref.~\cite{Fornasa:2016ohl} employed an hybrid method, based on the results of 
$N$-body simulations and complemented by analytical estimates. In that case, 
the anisotropy signal due to Dark Matter is dominated by the contribution of 
Galactic subhalos. However, it would be not be Poissonian, in contrast with 
the measured auto- and cross-APS used here.

In Ref.~\cite{Fornasa:2016ohl} the sub-GeV regime is where the measured auto-
and cross-APS is potentialy affected by systematics related to the subtraction 
of the Galactic foreground, to leakage outside the mask or to specific details 
of data selection (see Sec.~V-C of Ref.~\cite{Fornasa:2016ohl}). Even if it 
was tested that each of these effect cannot induce a deviation larger than 
$\sim 1\sigma$ from the final data set, maybe the simultaneous presence of
difference systamatics could have artificially enhanced the anisotropy 
expected from blazars to the level that is actually observed. This would 
reduce the need for the new class. The on-going measurement of gamma-ray 
anisotropies with Pass 8 {\it Fermi} LAT data will provide more information 
regarding this scenario.

Given the difficulty to associate the new source class with any known 
population of gamma-ray emitters, we attempt a phenomenological description. 
We assume that the new sources are well described by a broken-power-law source 
count distribution:
\begin{equation}
\frac{dN}{dF} =\left \{ 
\begin{array}{lr}
N_0 F^{-s} & \mbox{for } F < F_0 \\
N_0 F^{-2.5} & \mbox{for } F \geq F_0 \\
\end{array}.
\right.
\label{eqn:dNdF_new_class}
\end{equation}

The flux in Eq.~(\ref{eqn:dNdF_new_class}) is defined above 100 MeV. The index 
above the break $F_0$ is fixed to the Euclidean value, i.e. 2.5, typical of 
sources that are homogenously distributed in a local volume
\cite{Peacock,Ando:2017xcb}. This is particularly appropriate for rare 
emitters, as for the large-flux end of the distribution. On the contrary, the 
slope $s$ below the break is left free between 0 and 2.5. We determine the 
normalization $N_0$ by requiring that the corresponding auto-APS in the first
energy bins is equal to the measured APS (with the 3FGL mask)\footnote{We 
consider the best-fit value of $C_{\rm P}^{0,0}$ from the scan performed only 
with the APS data (3FGL mask), see Fig.~\ref{fig:1D_3FGL_new_class}. Also, 
since we want to reproduce the auto-correlation APS in the first energy bin 
from Ref.~\cite{Fornasa:2016ohl}, the integration variable is now the flux 
between 0.50 and 0.72 GeV. We translate the flux above 100 MeV used in 
Eq.~\ref{eqn:dNdF_new_class} into the flux between 0.50 and 0.72 GeV by 
assuming the best-fit value for $\Gamma_{\rm new}$ from 
Fig.~\ref{fig:1D_3FGL_new_class}.} As in Sec.~\ref{sec:model}, the APS is 
computed as the difference of $C_{\rm P,cov=1}$ (defined, in this case, as the
integral of $F^2 dN/dF$ below the 3FGL sensitivity from the last row of 
Tab.~\ref{tab:CP_3FGL}) and $C_{\rm P,cat}$ (i.e. the APS of the sources in the 
3FGL catalogue, see Tab.~\ref{tab:CP_3FGL}). Having determined $N_0$, we 
integrate $dN/dF$ for each value of $s$ and $F_0$ above the 3FGL sensitivity 
and above the 2FGL one to estimate the number of new sources expected in the 
3FGL and in the 2FGL catalogue, respectively. The two sensitivities are taken
from the last rows of Tab.~\ref{tab:CP_3FGL} and \ref{tab:CP_2FGL}. We also
compute the expected APS above the 2FGL threshold and between 0.50 and 0.72
GeV, and the integral of $F dN/dF$ above the 2FGL sensitivity. These 
quantities can be compared to the best-fit APS of the new source class (for
the 2FGL mask, see Sec.~\ref{sec:2FGL}) and to the IGRB intensity from 
Ref.~\cite{Ackermann:2014usa}.

For $0<s<2$, all the quantities mentioned above are well defined. We 
summarize the scenario in the left panel of Fig.~\ref{fig:new_population}: 
colors indicate the fraction of the observed IGRB (above 100~MeV) that can be 
explained by the new source class. Black lines indicate the regions where the 
fraction is 75\%, 100\%, 500\% and 1000\%. The region below the black line 
labelled ``100\%'' is not viable, disfavoring values of $F_0$ smaller than 
approximately $2-3 \times 10^{-9}~\mbox{cm}^{-2}~\mbox{s}^{-1}$ and value of $s$ 
between 1.5 and 2.0. The blue (red) line indicates the combinations of $s$ and 
$F_0$ that predict 271 (168) sources above the 3FGL (2FGL) sensitivity, 
respectively. Since these values correspond to the number of unassociated 
sources in the two catalogues for $|b|>30^\circ$, the regions above the blue 
and red lines are also excluded. Finally, the green line indicates the 
combinations of $s$ and $F_0$ that correspond to a $C_{\rm P}^{0,0}$ of 
$9.5 \times 10^{-18}~\mbox{cm}^{-4}~\mbox{s}^{-2}~\mbox{sr}^{-1}$ when computed for 
the 2FGL mask. This value is the upper bound of the 95\% CL interval for 
$C_{\rm P}^{0,0}$ in the fit to the APS data with the 2FGL mask (see 
Fig.~\ref{fig:2D_2FGL_new_class}). Thus, the region above the green line is 
also excluded as it would be incompatible at 95\% CL with the results of that 
fit. This leaves a narrow region around 
$F_{0} \sim 4 \times 10^{-9}~\mbox{cm}^{-2}~\mbox{s}^{-1}$ and $s<1.5$, in which 
the new source class dominates the measured IGRB intensity.

For $s \geq 2$, the computation of the IGRB intensity diverges for 
$F \rightarrow 0$. Thus, we introduce a cut $F_{\rm lim}$ below which we assume 
no source is present. We consider two benchmark cases for $F_{\rm lim}$: 10\% 
and 5\% of the 3FGL sensitivity from Tab.~\ref{tab:CP_3FGL}. The red (blue) 
lines in the right panel of Fig.~\ref{fig:new_population} indicate the region 
where the new source class accounts for certain fractions of the observed 
IGRB, for the higher (lower) $F_{\rm lim}$. If we consider the higher 
$F_{\rm lim}$, the new source class never overshoots the measured IGRB above
100~Mev, accounting for, at the most, $\sim 82$\% of the emission. However,
the gray region in the top left part of the panel is also not viable, as the 
new source class would predict too many sources in the 3FGL catalogue. For
the lower $F_{\rm lim}$, values of $F_{\rm 0}$ smaller than approximately 
$5 \times 10^{-9}~\mbox{cm}^{-2}~\mbox{s}^{-1}$ and of $s$ larger than 2.26 are 
excluded since the new gamma-ray emitters would overproduced the measured 
IGRB, above 100 MeV. The region above the green line is also not viable, as
it corresponds to too many unassociated sources at high latitude in the 3FGL
catalogue. The only allowed area is localized around 
$F_0=10^{-8}~\mbox{cm}^{-2}~\mbox{s}^{-1}$ and for $s<2.26$. 

\section{Conclusions}
\label{sec:conclusions}
In this work, we fit the recently published measurement of the IGRB 
anisotropy auto- and cross-APS with a physically-motivated model of blazars.
Ref.~\cite{Ajello:2015mfa} demonstrated that such a model provided a good 
description of the blazars observed by {\it Fermi} LAT. Here we use it to 
test whether blazars are able to reproduce the new APS measurement. A positive 
answer would confirm the result of Refs.~\cite{Cuoco:2012yf,Harding:2012gk} 
(based on the original 2012 APS data \cite{Ackermann:2012uf}), according to 
which the IGRB APS is compatible with being due entirely to blazars. On the 
other hand, a negative answer would corroborate the phenomenological analysis 
performed in Ref.~\cite{Fornasa:2016ohl}, establishing the need for more than 
one component to interpret IGRB anisotropies.

Our findings are summarized as follows:
\begin{itemize}
\item When fitting the new auto- and cross-APS (in the case of the 3FGL mask)
in terms of blazars only, our best-fit solution is in agreement at 68\% CL 
with the best fit obtained in Ref.~\cite{Ajello:2015mfa}, apart from our
predicted $p_2^0$, which is larger. Including the 3FGL $dN/dF$ in the fit
does not have a significant impact.

\item Blazars alone (with or without including the 3FGL $dN/dF$ in the fit)
underproduce the auto- and cross-APS observed with the 3FGL mask below
1~GeV. This suggests that a different class of gamma-ray emitters is needed to
reproduce the measured APS. Note that previous works analysing the IGRB 
anisotropies in terms of blazars (as, e.g., Refs.~\cite{Cuoco:2012yf,
Harding:2012gk}) could not be sensitive to this new source class as they were 
based on the 2012 APS measurement from Ref.~\cite{Ackermann:2012uf}, which did 
not extend below 1~GeV. Our result validates the findings of 
Ref.~\cite{Fornasa:2016ohl} and it suggests that sub-GeV anisotropies are due 
(at least in part) to gamma-ray emitters with a soft spectrum (with values of 
$\Gamma_{\rm new}$ ranging from 2.7 to 3.2). The properties of the new sources 
are consistent, whether we include the 3FGL $dN/dF$ data in the fit or not. 
By a likelihood-ratio test, the new source is preferred over the blazar-only 
scenario by, at least, $5\sigma$.

\item If we fit the APS obtained with the 2FGL mask with blazars only, our
best-fit solution does not underproduce the sub-GeV APS, as before. A new
source class still improves the fit to the data and it is preferred over the
blazar-only scenario by more than $10\sigma$. However, this new source class
is different than the one hinted at in Ref.~\cite{Fornasa:2016ohl}. In fact, 
in Ref.~\cite{Fornasa:2016ohl}, the APS below 2--3~GeV (with the 2FGL mask) is 
almost entirely due to the population with the lower energy break (in the 
scenario with the sources emitting as broken power laws, i.e. the description 
with the lowest $\chi^2$ per degree of freedom)\footnote{Private 
communication.}. While, in our Figs.~\ref{fig:bestfit_APS_dNdF_2FGL_onlyAPS} 
and \ref{fig:bestfit_APS_dNdF_2FGL_APS_and_dNdF} the new class is always 
subdominant. This indicates that the class that is responsible for the 
low-energy data in Ref.~\cite{Fornasa:2016ohl} is probably a mixture of 
different gamma-ray emitters, including blazars. This also attests the benefit 
of using a physically-motivated description of sources as we do here, instead 
of the phenomenological analysis performed in Ref.~\cite{Fornasa:2016ohl}.

\item When we include the new source class, the 95\% CL contours point towards
different regions of the parameter space, according to whether the 3FGL 
$dN/dF$ data are included or not in the fit. In particular, in order to
achieve a good description of the 3FGL source count distribution, $\gamma_1$
needs to be of the order of 0.75.

\item The auto- and cross-correlation APS predicted by the new source class is 
very similar, indipendent of whether the scan is performed with or without the 
3FGL $dN/dF$ data, with the 2FGL or 3FGL mask. It dominates the signal below 
1~GeV, in the case of the 3FGL mask and it plays a subdominant role for the 
2FGL mask. This implies that no many members of the new source class are 
present with a flux between the 3FGL and the 2FGL sensitivity. The slope of 
the energy spectrum $\Gamma_{\rm new}$ goes from 2.5 to 3.2.

\item The properties of the new class inferred from the fit to the APS data
disagree with the characteristics of known gamma-ray emitters, e.g.,
star-forming galaxies or misaligned active galactic nuclei. Also, unassociated 
sources in the 3FGL and 2FGL catalogues have, on average, a harder energy 
spectrum. Dark Matter halos and subhalos can reproduce the properties of the
new source class (especially for a Dark-Matter candidate with a mass of few
GeV and annihilating into $b$ quarks). However, the expected APS would 
probably not be Poissonian, as assumed here. Finally, the combination of 
different systematic effects (i.e. contamination from Galactic foreground, 
leakage outside the mask and data selection) could enhance the auto- and 
cross-APS predicted by blazars below $\sim 1$ GeV, improving the agreement 
with the data from Ref.~\cite{Fornasa:2016ohl} and reducing the need for the 
new source class.

\item We assume that the new gamma-ray emitters are characterized by a
source count distribution that follows a broken power law. We leave the 
position of the break $F_0$ and the low-flux index $s$ as free parameters.
In order to reproduce the APS measurement from Ref.~\cite{Fornasa:2016ohl}, 
without, at the same time, overshooting the number of unassociated sources in 
the 2FGL and 3FGL catalogue or the IGRB emission observed in 
Ref.~\cite{Ackermann:2014usa}, only values of $F_0$ around 
$4 \times 10^{-9}~\mbox{cm}^{-2}~\mbox{s}^{-1}$ (above 100 MeV) are allowed, for 
$s<1.5$. Alternatively, assuming that there is no source belonging to the new
class below a $F_{\rm lim}$ that is 10\% of the 3FGL threshold, all considered
values of $F_0$ and $s$ are allowed, apart from 
$F_0 > 3 \times 10^{-8}~\mbox{cm}^{-2}~\mbox{s}^{-1}$ and $2<s<2.35$. On the 
other hand, if we lower $F_{\rm lim}$ to 5\% of the 3FGL threshold, the only
viable region is around $F_0$ of $10^{-8}~\mbox{cm}^{-2}~\mbox{s}^{-1}$ and for
$2<s<2.27$. In all cases, the new source class would be the dominant 
component to the IGRB intensity.
\end{itemize}

The amount of information that we have been able to extract on the IGRB from
the new APS data attests the improvement that such a measurement represents,
with respect to the original 2012 one. However, it is very challenging to
achieve a coeherent and consistent description of the IGRB by employing only
one data set. We believe that the path to conclusively dissecting the
composition of the IGRB lays in the combination of multiple complementary
observables. Such a longer lever arm will also clarify the nature of the new
class of sources suggested by the present work.

\begin{acknowledgments}
SA and MF gratefully acknowledge support from the Netherlands Organization for 
Scientific Research (NWO) through a Vidi grant, and MF also thanks the project 
MultiDark CSD2009-00064. NF, MR and HSZ acknowledge support from the research 
grant Theoretical Astroparticle Physics number 2012CPPYP7 under the program 
PRIN 2012 funded by the Ministero dell'Istruzione, dell'Universit\`{a} e della 
Ricerca (MIUR) and from the research grant TAsP (Theoretical Astroparticle 
Physics) funded by the Istituto Nazionale di Fisica Nucleare (INFN). HSZ
gratefully acknowledges INFN for a post-doctoral fellowship in theoretical
physics on ``Astroparticle, Dark Matter and Neutrino Physics'', awarded under
the INFN Fellowship Programme 2015.
\end{acknowledgments}

\bibliography{Draft}
 
\end{document}